\documentclass[a4paper, 10pt]{LTJournalArticle}

\usepackage{amsmath}

\DeclareMathAlphabet\mathbfcal{OMS}{cmsy}{b}{n}

\runninghead{A systematic path to non-Markovian dynamics II}

\footertext{\textit{April} 2025}

\title{\large A systematic path to non-Markovian dynamics II: Probabilistic response
of nonlinear multidimensional systems to Gaussian colored noise excitation}

\author{G.A Athanassoulis \textsuperscript{1 \thanks{Corresponding author: \href{mailto:mathan@central.ntua.gr}{mathan@central.ntua.gr}, \href{mailto:makathan@gmail.com}{makathan@gmail.com}}} 
\orcid{0000-0001-8761-3560}, 
N.P Nikoletatos-Kekatos \textsuperscript{1}  \orcid{0009-0009-7265-6184}
and Konstantinos Mamis \textsuperscript{2}  \orcid{0000-0001-9772-810X} \\ 
\href{mailto:mathan@central.ntua.gr}{\small{mathan@central.ntua.gr}},
\href{mailto:nikoletatosnikos@central.ntua.gr}{\small{nikoletatosnikos@central.ntua.gr}},
\href{mailto:kmamis@uw.edu}{\small{kmamis@uw.edu}} 
}

\date{\footnotesize \textsuperscript{\textbf{1}}School of Naval Architecture \&
Marine Engineering, National Technical University of Athens,\\ 9 Iroon Polytechniou
st., 15780 Zografos, GREECE \\ \textsuperscript{\textbf{2}}Department of Applied
Mathematics, University of Washington, Seattle, WA 98195-3925 }

\renewcommand{\maketitlehookd}{
    \begin{abstract}
        \noindent
        The probabilistic characterization of non-Markovian responses to nonlinear dynamical systems under colored 
        excitation is an important issue, arising in many applications. Extending the Fokker-Planck-Kolmogorov equation, 
        governing the first-order response probability density function (pdf), to this case is a complicated task calling 
        for special treatment. In this work, a new pdf-evolution equation is derived for the response of nonlinear dynamical 
        systems under additive colored Gaussian noise. The derivation is based on the Stochastic Liouville equation (SLE), 
        transformed, by means of an extended version of the Novikov-Furutsu theorem, to an exact yet non-closed equation, 
        involving averages over the history of the functional derivatives of the non-Markovian response with respect to the 
        excitation. The latter are calculated exactly by means of the state-transition matrix of variational, time-varying 
        systems. Subsequently, an approximation scheme is implemented, relying on a decomposition of the state-transition
        matrix in its instantaneous mean value and its fluctuation around it. By a current-time approximation to the latter, 
        we obtain our final equation, in which the effect of the instantaneous mean value of the response is maintained, 
        rendering it nonlinear and non-local in time. Numerical results for the response pdf are provided for a bistable 
        Duffing oscillator, under Gaussian excitation. The pdfs obtained from the solution of the novel equation and a
        simpler small correlation time (SCT) pdf-evolution equation are compared to Monte Carlo (MC) simulations. The novel 
        equation outperforms the SCT equation as the excitation correlation time increases, keeping good agreement with
        the MC simulations.\\
    \end{abstract}

    \noindent
    \textbf{Keywords:} random differential equations; generalized Fokker-Planck-Kolmogorov
    equations; stochastic dynamical systems; response pdf-evolution equation;
    colored noise excitation; non-linear stochastic processes; \\

    \noindent
    \textbf{Mathematics Subject Classification} – MSC2020: 34F05, 60G65, 70L05 \\
                                }

\renewcommand{\theequation}{\arabic{section}.\arabic{equation}}
\counterwithin*{equation}{section}

\begin{document}

    \maketitle

    \setcounter{tocdepth}{3} 
    \tableofcontents 
    
    \newpage

\section{Introduction}\label{sec1}
Stochastic modeling of dynamical systems, initiated in the first decade of the 20th century in statistical physics 
\cite{Einstein_1905,Langevin_1908} and finance \cite{Bachelier_1900, Bachelier_2011}, has been constantly expanding, 
establishing a strong presence in science and engineering. In the 21-th century, stochastic modeling flourished further 
and infiltrated almost any mathematics-based discipline, from theoretical physics and engineering sciences to biology, 
climate, economics, and more.

In view of this vast landscape, we restrict our considerations to the area of macroscopic stochastic dynamics, for which 
a mathematical prototype is a system of nonlinear differential equations under random excitation. Such a system
will be abbreviated subsequently as RDE (Random Differential Equations) [Abbreviations are defined in first appearance. 
For reader’s convenience, an alphabetical list of abbreviations is given as \hyperref[secA1]{Appendix A}]. The random 
excitation may be Gaussian or non-Gaussian, delta-correlated or smoothly-correlated, and may appear in the RDE either 
additively or multiplicatively. All three dichotomies mentioned above are important, and any specific combination of them 
leads to specific difficulties and peculiarities. The dichotomy between delta-correlated and smoothly-correlated excitation, 
which is the main concern of the present paper, has profound implications on the probabilistic characterization of the
response, and on the theoretical and methodological background needed for treating the RDE.

The case of delta-correlated stochastic excitation (white noises) is the most intensively studied case in the current 
literature. In this case, it is well known that the response is Markovian, and the complete description of response’s 
probabilistic structure is given by the transition pdf, governed by the Fokker-Plank-Kolmogorov equation (FPKE) 
\cite{Moss_1989,Risken_1996}, \cite{Pugachev_2002} Sec. 5.6.6, \cite{Gardiner_1985} Ch. 5, \cite{Sun_2006}
Sec. 6.3, or by the Kolmogorov-Feller equation \cite{Pirrotta_2007,Zhu_2015,Tylikowski_1986}.

However, “the nature knows no delta-correlated processes. All actual processes and fields are characterized by a finite 
temporal correlation radius, and delta-correlated processes and fields result from asymptotic expansions
in terms of their temporal correlation radii” Klyatskin (2005) \cite{Klyatskin_2005}, p. 91. Even though smoothly-correlated 
(colored) excitation is a more realistic model, better suited to the mathematical modeling of stochastic dynamics
problems, remains poorly studied up to now, probably due to its increased difficulty. By assuming colored excitation, 
the response loses its Markovian character, and its probabilistic structure cannot be inferred by a single probability
density function (pdf). In that case, the complete probabilistic description of the response requires the determination 
of the infinite hierarchy of pdfs of various orders (one-time, two-times etc.), rendering the problem
unsolvable in the current state-of-the-art. The alternative formulation, using the characteristic functional and the 
Hopf’s equation (see \cite{Sapsis_2008}, Introduction, for a discussion and references), is promising but also remains
beyond the capabilities of the present research state; see e.g. \cite{Lumley_2001}, Sec. 2.1.5. A modest, feasible 
alternative is to formulate approximate equations, separately for each order. Such kind of equations for the first-order
(one-time) pdf have been developed in some cases, and are usually termed generalized FPKE (genFPKE). Their formulation 
is a difficult task since, in the non-Markovian case, all orders of pdfs are coupled and thus, an efficient, approximate
decoupling technique is required. In this paper, a systematic approach is developed for deriving genFPKE for the 
first-order, response pdf, corresponding to multidimensional, nonlinear, dynamical systems under additive Gaussian excitation. 
Needless to say, the scalar problem has been much more extensively studied \cite{Van_Kampen_1976,Hanggi_1978,Mamis_2019,
Sancho_1982, Hanggi_1985,Fox_1986,Hanggi_1994,Athanassoulis_2015,Bianucci_2024}. An extensive review of methods for 
constructing genFPKE for this case can be found in \cite{Sapsis_2008,Mamis_2019}. The present work is a sequel of 
\cite{Mamis_2019}, generalizing its methodology to the multidimensional case. The literature survey presented below mainly 
refers to the multidimensional problem, on which the present work is focused.

\subsection{Literature survey }
The importance of exploiting the colored noise excitation has been well appreciated in applications \cite{Zhang_2018, 
Zhou_2022,Wang_2018,Qi_2013,Mamis_2023,Mamis_2021, Venturi_2011}; besides, the complicacies that emerge from this modeling 
have been already pointed out \cite{Sapsis_2008,Hanggi_1994,Mamis_2019,Qi_2013,Venturi_2011,Mamis_2018,Mamis_phd,Luczka_2005} 
\cite{Van_kampen_2007} Sec IX.7. To this end, various approaches have been developed through time, shaping the methodological 
inventory for non-Markovian responses. First, and probably most known among engineering audience, is the filtering approach
\cite{Pugachev_2002}, Sec 5.10, \cite{Hanggi_1994,Chai_2015}, also called Markovianization method \cite{Kree_1985}, in 
which the RDE system is augmented by linear filters fed by white noises and producing as output an approximation of the
colored excitation. This approach allows the formulation of a FPKE (for the augmented system, including the filters), 
avoiding uncharted waters. However, the increment of the degrees of freedom, due to the presence of the filters, leads to 
an undesirable inflation of the spatial dimensions of the final FPKE.

Alternative lines of work, based on more fundamental considerations, initiated in the works of Lax \cite{LAX_1966}, 
Van Kampen \cite{Van_kampen_2007}, Fox \cite{Fox_1977}, and H{\"a}nggi \cite{Hanggi_1978}. Derivations of genFPKE corresponding 
to multidimensional RDE, usually under the assumption of Gaussian excitation, have been mainly developed by means of two 
generic methodologies: the functional calculus approach \cite{Mamis_2018,Cetto_1984,Dekker_1982,San_Miguel_1980}, and the 
cumulant expansion approach \cite{Garrido_1982,Fox_1983}. In both approaches the starting point is the Stochastic Liouville 
Equation (SLE) corresponding to the RDE. The SLE is exact, yet non-closed, and further considerations are required for 
the derivation of an approximate, solvable, genFPKE. The latter is more complicated than the standard FPKE obtained by 
Markovianization, but its spatial dimension coincides with the state-space dimension of the underlying system of RDE.

In the functional calculus approach, the Novikov-Furutsu (NF) theorem \cite{Novikov_1965,Athanassoulis_2019} is usually 
invoked to implement correlation splitting between response and excitation. The resulting equation contains the Volterra 
functional derivative of the response with respect to the excitation \cite{Mamis_phd,Van_Kampen_1976,Hanggi_1989}, 
conveying non-local features in the formulation. The treatment of the Volterra derivative is crucial
and has been addressed by two methods. The correlation time expansion method \cite{Dekker_1982,San_Miguel_1980}, based 
on the small correlation time (SCT) assumption, utilizes an asymptotic expansion of the derivative around the
current time, deriving approximate genFPKE, hierarchized by the correlation time orders. Alternatively, the Volterra 
derivative has been explicitly calculated without assuming SCT approximation \cite{Mamis_2018,Cetto_1984}, resulting
in a non-closed master equation \cite{Cetto_1984}, which requires a closure argument to become solvable \cite{Hanggi_1978}.

In the cumulant expansion method the NF theorem is not utilized. Instead, the SLE is transformed into the interaction 
picture \cite{Guenin_1966}, \cite{Sudarshan_1974}, Ch. 7, allowing the excitation’s generalized cumulants to be employed
\cite{Fox_1978} Sec. II.1, \cite{Bianucci_2020,Kubo_1962,Fox_1974,Fox_1975}. The resulting master equation is exact but 
non-closed, containing an infinite superposition of ordered cumulants. Under the assumption of weakly
correlated excitation, the equation can be truncated at a specific order and, using a current time approximation, genFPKE 
corresponding to each order can be formulated. Results obtained by Fox \cite{Fox_1983} using the cumulant expansion method, 
coincide with Dekker’s \cite{Dekker_1982} using the correlation expansion method. Further, the cumulant expansion method 
has also been applied to Poisson noise \cite{You_xin_1991}.

Two more methods for constructing genFPKE, mainly implemented for the scalar case, are the projection operator method 
\cite{Grigolini_1986,Terwiel_1974,Faetti_1988I,Faetti_1988II,Peacock_L_pez_1988} and the unified colored noise approximation 
(UCNA) \cite{Jung_1987}, \cite{Hanggi_1994} Sec. V.C. Extensions of these methods to the multidimensional case have recently 
been published; see \cite{Venturi_2014}, \cite{Duan_2020}.

\subsection{Structure of the present work }
In the present work we derive and validate new pdf-evolution equations governing the pdf dynamics of 
the response vector of a system of nonlinear and non-autonomous RDEs, under additive Gaussian excitation. In
\hyperref[sec2]{Section 2}, we first formulate the underlying dynamical problem and the corresponding
SLE. Then, we briefly describe the methodology used for the derivation of the new genFPKE and summarize the main results 
of this work. The detailed exposition of the derivation starts in \hyperref[sec3]{Section 3}, where we utilize an extended
form of the NF theorem \cite{Athanassoulis_2019} to obtain a novel version of the SLE, called herein transformed SLE, 
which is exact but non-closed. In \hyperref[sec4]{Section 4}, we present some simple applications of the transformed SLE. 
We consider three simple problems: linear systems of RDEs under general Gaussian excitation, nonlinear systems of RDEs under 
white noise excitation, and nonlinear systems of RDEs under colored noise excitation of small correlation time. In the
first two cases the closure is automatic. Especially in the second problem we (re)derive the classical FPKE in a simple 
and elegant way. In the third problem we derive a SCT approximate version of genFPKE, again in a surprisingly easy
and straightforward manner, which does not seem to have been given explicitly in the literature. The main original 
contribution of this paper is developed in \hyperref[sec5]{Section 5}, where a new closure technique is applied to the 
transformed SLE, permitting us to derive a novel pdf-evolution equation which outperforms the SCT model. In \hyperref[sec6]{Section 6}, 
first numerical results are presented for a bistable Duffing oscillator. The obtained response pdf, both in the
transient and the steady-state regime, are nicely compared with results obtained by using Monte Carlo simulation. 
Finally, in \hyperref[sec7]{Section 7}, we provide a critical discussion of the theory developed in this paper, 
pointing out its main features, the approximations utilized to obtain the novel genFKE, and indicating how it can be improved.


\section{Formulation of the problem and main results}\label{sec2}

\subsection{Formulation of the problem and preliminary discussion}\label{sec2.1}

The problem we are dealing with in this paper is the determination of the one-time probabilistic structure of the response 
of a multidimensional, nonlinear system of RDEs, under generic, colored, Gaussian, additive excitation. The system of
RDEs can be written in the form

\begin{subequations} \label{eqs:2.1}
    \begin{align}
         & \dot X_{n}(t ; \theta) = h_{n}({\bfv{X}}(t;\theta),t) + \Xi_{n}(t;\theta ), \label{eq:2.1a} \\
         & {X_{n}}({t_{0}};\theta ) = X_{n}^{0}(\theta ), \ n = 1, 2, \ldots ,N, \label{eq:2.1b}
    \end{align}
\end{subequations}
where the overdot denotes the differentiation with respect to time $t$, $\theta$ denotes the stochastic argument, and the 
deterministic functions $h_{n}({\bfv{x}},t)$ are assumed to be Lipschitz continuous in ${\bfv{x}}$ and continuous in $t$. 
The random excitations $\Xi_{n}(t,\theta)$ and the random initial values $X_{n}^{0}(\theta)$ are the data of the system, 
and thus they are assumed to be known. We shall follow the usual (useful and convenient) assumption that the random data 
are Gaussian. More precisely, the vector ${\bfv{X}}^{0}(\theta) = \left( X_{n}^{0}(\theta) \right)_{n=1}^{N}$ and the 
vector function $\mathbf{\Xi}(t,\theta) = \left( \Xi_{n}(t,\theta) \right)_{n=1}^{N}$ are considered jointly Gaussian. 
Their probabilistic structure is completely defined by means of their mean vectors ${\bfv{m}}_{{\bfv{X}}^{0}}$ and
${\bfv{m}}_{\mathbf{\Xi}}(t)$, the autocovariance matrices ${\bfv{C}}_{{\bfv{X}}^{0}{\bfv{X}}^{0}}$,
${\bfv{C}}_{\mathbf{\Xi}\mathbf{\Xi}}$, and the cross-covariance matrix ${\bfv{C}}_{{\bfv{X}}^{0}\mathbf{\Xi}}$. Let it 
be noted that, to the best of our knowledge, the case where excitations are correlated to the initial values has not been 
studied before, with the exception of papers \cite{Athanassoulis_2019,Mamis_phd,Mamis_2018}, by the present research group, 
dealing with the scalar problem.

The starting point of the probabilistic study of system (\ref{eq:2.1a}) is the representation of the response pdf as an 
averaged random delta function \cite{Hanggi_1978, Mamis_2018, Sancho_1982, Fox_1986, Cetto_1984,Dekker_1982, Sancho_1980, Wang_2013},
also called pdf method in the theory of turbulence \cite{Lundgren_1967,Pope_1985} or delta projection method in 
\cite{Mamis_2019,Mamis_phd}. Employing the notation $f_{{\bfv{X}}(t)}({\bfv{x}}) = f_{{X}_{1}(t){X}_{2}(t)\cdots {X}_{N}(t)}({x}_{1},{x}_{2},\ldots ,{x}_{N})$
for the one-time response pdf, the delta projection method is based on the representation:

\begin{equation}
    {{f}_{\bfv{X}(t)}}(\bfv{x})={{\mathbb{E}}^{\theta }}\left[ \delta (\bfv{x}
        -\bfv{X}(t;\theta ) ) \right], \label{eq:2.2}
\end{equation}
where
\begin{equation}
    \delta (\bfv{x}-\bfv{X}(t;\theta ))= \delta ({{x}_{1}}-{{X}_{1}}(t;\theta
    )) \times \cdots \times \delta ({{x}_{N}}-{{X}_{N}}(t;\theta )), \label{eq:2.3}
\end{equation}
is a multidimensional random delta function. Differentiating both sides of equation
(\ref{eq:2.3}) with respect to time, employing the identity
\begin{equation*}
    \frac{\partial \delta (\bfv{x}-\bfv{X}(t;\theta ))}{\partial t}=-\frac{\partial
        \delta (\bfv{x}-\bfv{X}(t;\theta ))}{\partial \bfv{x}}\cdot \bfv{\dot{X}}
    (t;\theta ),
\end{equation*}
or, equivalently,
\begin{equation*}
    \frac{\partial \delta (\bfv{x}-\bfv{X}(t;\theta ))}{\partial t}=-\sum\limits
    _{n=1}^{N}{{{{\dot{X}}}_{n}}(t;\theta ){\delta }'({{x}_{n}}-{{X}_{n}}(t;\theta ))\prod\limits_{\begin{smallmatrix}j=1 \\ j\ne n\end{smallmatrix}}^{N}{\delta ({{x}_{j}}-{{X}_{j}}(t;\theta ))}}
    ,
\end{equation*}
and substituting the resulting time derivative of the response vector by using equation (\ref{eq:2.1a}), we obtain the 
corresponding Stochastic Liouville equation (SLE) (after averaging):
\begin{equation}
    {{\partial }_{t}}{{f}_{\bfv{X}(t)}}(\bfv{x})+ \sum\limits_{n=1}^{N}{\frac{\partial }{\partial {{x}_{n}}}\left( {{h}_{n}}\left( \bfv{x},t \right){{f}_{\bfv{X}(t)}}(\bfv{x}) \right)}
    =-\sum\limits_{n=1}^{N}{\frac{\partial }{\partial {{x}_{n}}}{{\mathbb{E}}^{\theta }}\left[ \delta (\bfv{x}-\bfv{X}(t;\theta )){{\Xi }_{n}}(t;\theta ) \right]}
    . \label{eq:2.4}
\end{equation}
Note that, in equation (\ref{eq:2.4}), the ensemble average ${{\mathbb{E}}^{\theta }}\left[ \cdot \right]$ should be 
taken with respect to the joint response-excitation probability measure ${{P}_{\bfv{X}\mathbf{\Xi} }}$. If the external 
excitation is zero, $\mathbf{\mathbf{\Xi}}(t;\theta )=0$, the above equation is reduced to the classical Liouville-Gibs 
equation \cite{Saaty_1981}. The terminology Stochastic Liouville equation was introduced by Kubo in 1963 \cite{Kubo_1963}. 
Even though equation (\ref{eq:2.4}) is exact, deterministic, and its left-hand side has the form of a first order PDE 
with respect to ${{f}_{\bfv{X}(t)}}(\bfv{x})$, the averages:
\begin{align}
    \mathcal{N}_{n}^{\bfv{X}\mathbf{\Xi} }(\bfv{x},t)\equiv \mathcal{N}_{n}^{\bfv{X}\mathbf{\Xi} }: & ={{\mathbb{E}}^{\theta }}\left[ \delta (\bfv{x}-\bfv{X}(t;\theta ) ){{\Xi }_{n}}(t;\theta ) \right] = \nonumber                          \\
                                                                                                    & =\int_{\mathbb{R}}{{{\xi }_{n}}{{f}_{\bfv{X}(t){{\Xi }_{n}}(t)}}(\bfv{x},{{\xi }_{n}})}d{{\xi }_{n}}, \,\, n=1,\ldots ,N, \label{eq:2.5}
\end{align}
in the right-hand side, involve the joint response-excitation pdfs ${{f}_{\bfv{X}(t){{\Xi }_{n}}(t)}}(\bfv{x},{{\xi }_{n}})$, 
making the equation non-closed. Aiming to derive a closed evolution equation for ${{f}_{\bfv{X}(t)}}(\bfv{x})$, we employ 
the extended (multidimensional) Novikov-Furutsu (NF) Theorem \cite{Athanassoulis_2019,Mamis_phd}, to eliminate the 
dependence of $\mathcal{N}_{n}^{\bfv{X}\mathbf{\Xi} }$ on the corresponding stochastic excitation $\Xi_{n}$. An important 
implication of this procedure is that the terms $\mathcal{N}_{n}^{\bfv{X}\mathbf{\Xi} }$ become non-local in time.

Starting with equation (\ref{eq:2.4}), in \hyperref[sec3]{Section 3}, we reformulate it by using the extended NF Theorem 
\cite{Athanassoulis_2019}, obtaining a transformed SLE which contains averages over the derivatives of $\bfv{X}(t;\theta )$ 
with respect to the initial values $X_{n}^{0}(\theta )$ and the excitation functions ${{\Xi }_{n}}(s;\theta )$, $s<t$, 
called herein variational derivatives. The second ones are Volterra functional derivatives \cite{volterra_1930,Hanggi_1989}, 
\cite{Athanassoulis_2019} Appx. A. These derivatives satisfy linear, time-varying ODEs (Ordinary Differential Equations), 
namely, the variational problems associated with equations (\ref{eqs:2.1}); see \hyperref[sec3.3.2]{Section 3.3.2}.
The treatment of these terms is critical for deriving a “good” pdf-evolution equation. In contrast to all other approaches, 
in this paper we solved the variational problems exactly, in terms of the state-transition matrices. The latter contains 
the time history of response function $\bfv{X}(t;\theta )$, retaining the non-local and non-closed character of the 
equation. Nevertheless, the exact character of solutions for the variational derivatives permits us to perform a highly 
improved treatment of the corresponding terms, leading to efficient, approximate pdf-evolution equations. The main steps 
towards obtaining improved closed equations are the separation of the (instantaneous) mean value of the response from its 
random fluctuations, and the use of the Magnus expansion to obtain handy, explicit formulae for the state transition
matrices. This permits us to keep intact and treat exactly the effect of the instantaneous mean values of the response 
(modeled as time-integrals of some generalized response moments), restricting the eventual current time approximation to 
the random fluctuations only. To the best of our knowledge, this treatment is innovative and applied for the first time 
to this kind of problem. It leads to new closed, nonlinear, and non-local pdf-evolution equations, significantly improved 
in comparison of those based on the SCT approximation schemes. Our equations share features with mean-field nonlinear FPKE, as
presented by \cite{Frank_2005}, but it also exhibits non-locality in time, a feature reflecting the non-Markovian 
character of the response. The form of these equations is presented in the next subsection, along with a brief comparison
with the classical FPKE.

\subsection{Main results: A new, closed, solvable pdf- evolution equation
    for non-Markovian response} \label{sec2.2}

The main result of the present paper, being derived in \hyperref[sec5]{Section 5}, is the following pdf-evolution equation, 
governing the one-time response pdf of the nonlinear system (\ref{eqs:2.1}):
\begin{align}
     & \frac{\partial {{f}_{\bfv{X}(t)}}(\bfv{x})}{\partial t}+
    \sum\limits_{n=1}^{N}{\frac{\partial }{\partial {{x}_{n}}}\left[ \left( {{h}_{n}}(\bfv{x},t)+\underline{{m_{\Xi_{n}}}(t)} \right){{f}_{\bfv{X}(t)}}(\bfv{x}) \right]}= \nonumber                                                                                                                                                                                                                                                                   \\
     & =\sum\limits_{n=1}^{N}{\sum\limits_{\nu =1}^{N}{\frac{\partial }{\partial {{x}_{n}}}\frac{\partial }{\partial {{x}_{\nu }}}}}\left[ \left( \underline{\mathcal{D}_{\nu n}^{{{X}_{0}}\mathbf{\Xi} }\left[ {{f}_{\bfv{X}(\cdot )}}(\cdot );\bfv{x},t \right]}+\underline{\underline{\mathcal{D}_{\nu n}^{\mathbf{\Xi} \mathbf{\Xi} }\left[ {{f}_{\bfv{X}(\cdot )}}(\cdot );\bfv{x},t \right]}}\right){{f}_{\bfv{X}(t)}}(\bfv{x}) \right]. \label{eq:2.6}
\end{align}
This equation exhibits both similarities and differences with respect to the classical FPKE. The simply underlined terms 
model the effects of the (non-zero) mean value of the excitation (${{m}_{{{\Xi }_{n}}}}$) and of the (possible)
dependance of the initial values with the excitation ($\mathcal{D}_{\nu n}^{{{X}_{0}}\mathbf{\Xi}}$), which usually do 
not appear in such kind of equations. Nevertheless, these features are not the main source of novelty of the equation
(\ref{eq:2.6}). Neglecting these two terms, we see an equation structurally similar to the classical FPKE. However, this 
similarity with FPKE is only apparent. The most profound difference of equation (\ref{eq:2.6}) from FPKE 
(and other similar ones) lies within the doubly underlined diffusion coefficients $\mathcal{D}_{\nu n}^{\mathbf{\Xi} \mathbf{\Xi} }$. 
The latter are not simply functions of the state vector $\bfv{x}$ and the current time $t$ (as usually), but they are, in addition, 
dependent on the time history of some response moments, namely
\begin{equation}
    {{R}_{ij}}(t)\equiv{{R}_{ij}}[{{f}_{\bfv{X}(t)}}(\cdot ),t]=\int_{{{\mathbb{R}}^{N}}}
    {\frac{\partial {{h}_{i}}(\bfv{x},t)}{\partial {{x}_{j}}}}{{f}_{\bfv{X}(t)}}
    (\bfv{x})d\bfv{x}. \label{eq:2.7}
\end{equation}
The diffusion coefficient matrix has the form
\begin{equation}
    \mathcal{D}^{\mathbf{\Xi} \mathbf{\Xi} }\left[{{f}_{\bfv{X}(\cdot )}}
        ( \cdot ); \bfv{x},t \right]=\int\limits_{{{t}_{0}}}^{t}{\bfv{K}(\bfv{x},t,s)\Phi [\bfv{R}](t,s)\mathbf{C}_{\mathbf{\Xi} \mathbf{\Xi} }(t,s)ds}
    , \label{eq:2.8}
\end{equation}
where $\bfv{K}(\bfv{x},t,s)$ is an exponential matrix, dependent on the approximation scheme, and $\Phi [\bfv{R}](t,s)$ 
is a state-transition, matrix solving the time-dependent matrix differential equation $\bfv{\dot{Y}}(t)=\bfv{R}(t)\bfv{Y}(t)$, 
$\bfv{Y}(s)=\bfv{I}$. The components of matrix $\bfv{R}$ are given by equation (\ref{eq:2.7}). 

Equations (\ref{eq:2.7}) and (\ref{eq:2.8}) explain the nonlocal and nonlinear character of equation (\ref{eq:2.6}), 
which makes it essentially different from any other existing model equations for the same problem. These features are 
the overhead for taking into account the correlation time of the excitation and the implied non-Markovian character of the 
response. Nevertheless, they reward us with an equation richer than other existing ones (recovering them after simplifications),
having larger range of validity with respect to the correlation time. The last statement has been confirmed by numerical 
results, validated by Monte Carlo simulations. 

\begin{remark}The somewhat unusual notation ${{f}_{\bfv{X}(\cdot )}}(\cdot )$  
used in the diffusion coefficients, expresses the fact that the corresponding dependence on the pdf ${{f}_{\bfv{X}(t)}}(\bfv{x})$ 
occurs by means of integration over the time history as well as over the entire state space. This becomes clear by the 
equations (\ref{eq:2.7}) and (\ref{eq:2.8}).
\end{remark}


\section{Transformation of SLE via the extended NF Theorem}\label{sec3}
\subsection{A second form of the SLE}\label{sec3.1}

According to equations (\ref{eqs:2.1}), the response path function $\bfv{X}(t;\theta )$ (for each $\theta$) is uniquely 
determined at any specific time $t$, by means of the initial value ${{\bfv{X}}_{0}}(\theta )$ and the history of the excitation 
$\mathbf{\Xi}(s,\theta )$, for all $s\in [{{t}_{0}},t]$. To formalize and exploit this dependence, we claim that $\bfv{X}(t;\theta )$ 
is a function-functional (abbreviated to $\operatorname{FF} \ell$) on the data ${{\bfv{X}}_{0}}(\theta )$ and $\mathbf{\Xi}(s,\theta )$, 
$s\in [{{t}_{0}},t]$, and we write
\begin{equation}
    \bfv{X}(t;\theta )=\mathbfcal{X}\left[{{\bfv{X}}^{0}}(\theta );\mathbf{\Xi}
        (\left. \cdot \right |_{{{t}_{0}}}^{t};\theta ) \right] \equiv \left({\mathcal{X}_{n}}
    \left[{{\bfv{X}}^{0}}(\theta );\mathbf{\Xi}(\left. \cdot \right|_{{{t}_{0}}}
            ^{t};\theta ) \right] \right)_{n=1}^{N}. \label{eq:3.1}
\end{equation}
That is, $\mathbfcal{X} \left[ \cdot ;\cdot \right]$ denotes the vector function-functional representation of $\bfv{X}(t;\theta )$, 
while ${\mathcal{X}_{n}}\left[ \cdot ;\cdot \right]$ are the components of $\mathbfcal{X}\left[ \cdot ;\cdot \right]$, 
representing ${{X}_{n}}(t;\theta)$. The symbol $\mathbf{\Xi}(\left. \cdot \right|_{{{t}_{0}}}^{t};\theta )$ denotes the 
history $\left\{ \Xi (s,\theta ),s\in [{{t}_{0}},t] \right\}$ of the excitation. The simpler symbol $\mathbf{\Xi}(\cdot ;\theta )$ 
could, in principle, be used, but in the subsequent analysis we have to consider cases where the history of the excitation 
must be taken over different time intervals, e.g. the interval $(s,t]$ or $(s,\tau ]$, where ${{t}_{0}}\le s<\tau\le t$. 
Thus, keeping the endpoints of the time interval in the notation, as in $\mathbf{\Xi}(\left. \cdot \right|_{s}^{t};\theta )$ or 
$\mathbf{\Xi}(\left. \cdot \right|_{s}^{\tau };\theta )$, is a convenient way to keep track on the correct dependence in each case.

Next, we extend the $\operatorname{FF}\ell$ point of view to the random delta function $\delta (\bfv{x}-\bfv{X}(t;\theta ))$, 
appearing in equation (\ref{eq:2.4}), writing
\begin{align} \label{eq:3.2}
    \mathcal{F}\left[{{\bfv{X}}^{0}}(\theta );\mathbf{\Xi}(\left. \cdot \right
    |_{{{t}_{0}}}^{t};\theta ) \right] & =
    \delta \left( \bfv{x}-\bfv{X} \left[{{\bfv{X}}^{0}}
            (\theta );\mathbf{\Xi}(\left. \cdot \right|_{{{t}_{0}}}^{t};\theta ) \right
    ] \right) = \nonumber                                                                                                                                                                                        \\
                                       & = \prod\limits_{n=1}^{N}{\delta \left( {{x}_{n}}-{{X}_{n}}\left[ {{\bfv{X}}^{0}}(\theta );\mathbf{\Xi} (\left. \cdot \right|_{{{t}_{0}}}^{t};\theta ) \right] \right)}.
\end{align}
The symbol $\mathcal{F}\left[ \cdot ;\cdot \right]$ is used to emphasize the function-functional structure of the dependence 
on the random data, suppressing temporarily its exact form. Using this notation, the average terms $\mathcal{N}_{n}^{\bfv{X}\mathbf{\Xi} }$, equation (\ref{eq:2.5}), take the form:
\begin{equation}
    \mathcal{N}_{n}^{\bfv{X}\mathbf{\Xi} }={{\mathbb{E}}^{\theta }}\left[ \mathcal{F}
        \left[{{\bfv{X}}^{0}}(\theta );\mathbf{\Xi}(\left. \cdot \right|_{{{t}_{0}}}
            ^{t};\theta ) \right]{{\Xi }_{n}}(t;\theta ) \right].\label{eq:3.3}
\end{equation}
Substituting (\ref{eq:3.1}) in the SLE (\ref{eq:2.4}), we obtain the following second form of the SLE, namely:
\begin{align} \label{eq:3.4}
    {{\partial }_{t}}{{f}_{\bfv{X}(t)}}(\bfv{x})+ & \sum\limits_{n=1}^{N}{\frac{\partial }{\partial {{x}_{n}}}\left( {{h}_{n}}\left( \bfv{x},t \right){{f}_{\bfv{X}(t)}}(\bfv{x}) \right)}= \nonumber                                                                                                              \\
                                                  & =-\sum\limits_{n=1}^{N}{\frac{\partial }{\partial {{x}_{n}}}\left( {{\mathbb{E}}^{\theta }}\left[  \mathcal{F} \left[ {{\bfv{X}}^{0}}(\theta );\mathbf{\Xi} (\left. \cdot \right|_{{{t}_{0}}}^{t};\theta ) \right]{{\Xi }_{n}}(t;\theta ) \right] \right)}, 
\end{align}
Equation (\ref{eq:3.3}) (and (\ref{eq:3.4})) reveals an interesting feature of the non-closed terms 
$\mathcal{N}_{n\cdot }^{\bfv{X}\mathbf{\Xi} }$ They contain averages of specific (yet unknown) $\operatorname{FF}\ell$s 
of the random data, multiplied by the components of the Gaussian excitation ${{\Xi }_{n}}(t;\theta )$. This structure 
facilitates the application of the NF Theorem.

\subsection{Application of the extended NF Theorem}\label{sec3.2}

In \cite{Athanassoulis_2019,Mamis_phd} Athanassoulis and Mamis developed various extensions of the classical Novikov-Furutsu 
theorem \cite{Novikov_1965,Furutsu_1963}. The one which is given (and applied) below refers to random $\operatorname{FF}\ell$s 
of the form $\mathcal{F}[{{\bfv{X}}^{0}}(\theta );\mathbf{\Xi}(\left. \cdot \right|_{{{t}_{0}}}^{t};\theta )]$ (abbreviated as
$\mathcal{F}[\cdot \cdot \cdot ]$ for conciseness), whose arguments ${{\bfv{X}}^{0}}(\theta )$, $\mathbf{\Xi}(\cdot ;\theta )$ 
are jointly Gaussian. The theorem provides the following reduction of dependance (correlation splitting):
\begin{align} \label{eq:3.5}
    \mathbb{E}^{\theta }\left[ \mathcal{F}[\cdots ]{{\Xi }_{n}}(t;\theta ) \right] = & m_{\Xi_{n}}(t)\mathbb{E}^{\theta }\left[ \mathcal{F}[\cdots ] \right]+ \nonumber                                                                                                                                                      \\
                                                                                    & +\sum\limits_{{{n}_{1}}=1}^{N}{C_{X_{{{n}_{1}}}^{0}{{\Xi }_{n}}}(t)\mathbb{E}^{\theta }\left[ \frac{\partial \mathcal{F}[\cdots ]}{\partial X_{{{n}_{1}}}^{0}(\theta )} \right]}+ \nonumber
    \\
                                                                                    & +\sum\limits_{{{n}_{2}}=1}^{N}{\int\limits_{{{t}_{0}}}^{t}{C_{{{\Xi }_{n}}{{\Xi }_{{{n}_{2}}}}}(t,s)\mathbb{E}^{\theta }\left[ \frac{\delta \mathcal{F}[\cdots ]}{\delta {{\Xi }_{{{n}_{2}}}}(s;\theta )} \right]ds}}. 
\end{align}
In equation (\ref{eq:3.5}), $\delta \cdot \mathbf{/}\delta{{\Xi }_{{{n}_{2}}}}(s;\theta )$ denotes the Volterra 
functional derivative with respect to the excitation function ${{\Xi }_{{{n}_{2}}}}(\cdot ;\theta )$, at the point 
(time instant) $s$. Setting $m_{\Xi_{n}}=0$ and $C_{X_{{{n}_{1}}}^{0}{{\Xi }_{n}}}(t,s)= 0$, equation (\ref{eq:3.5}) 
reduces to the classical NF theorem \cite{Novikov_1965,Furutsu_1963}. This classical form has been extensively used 
before for the formulation of genFPKE \cite{Sancho_1982, Dekker_1982, San_Miguel_1980, Cetto_1984, Venturi_2011,Fox_1986}. 
The extended form of the NF theorem, equation (\ref{eq:3.5}), has been used in \cite{Mamis_2019} for the study of a 
scalar RDE, and in \cite{Mamis_2018,Mamis_phd} in a prelimina ry analysis of the multidimensional problem. 
To implement the NF theorem (\ref{eq:3.5}), we need to elaborate on the derivatives 
$\partial \mathcal{F}[\cdot \cdot \cdot ]\mathbf{/}\partial X_{{{n}_{1}}}^{0}(\theta )$ and 
$\delta \mathcal{F}[\cdot \cdot \cdot ]\mathbf{/}\delta{{\Xi }_{{{n}_{2}}}}(s;\theta)$.
Using equation (\ref{eq:3.2}) and keeping in mind equation (\ref{eq:3.1}), we find
\begin{subequations} \label{eqs:3.6}
    \begin{align}
        \frac{\partial \mathcal{F}[\cdot \cdot \cdot ]}{\partial X_{{{n}_{1}}}^{0}(\theta )} & = \frac{\partial }{\partial X_{{{n}_{1}}}^{0}(\theta )}\delta (\bfv{x}-\bfv{X}(t;\theta )) = \nonumber                                                                                                                                                                                                                                                                                                 \\
                                                                                             & = -\sum\limits_{\nu =1}^{N}{\frac{\partial \delta ({{x}_{\nu }}-{{X}_{\nu }}(t;\theta ))}{\partial {{x}_{\nu }}}\frac{\partial {{X}_{\nu }}(t;\theta )}{\partial X_{{{n}_{1}}}^{0}(\theta )}\prod\limits_{\begin{smallmatrix}{{\nu }_{1}}=1 \\ {{\nu }_{1}}\ne \nu\end{smallmatrix}}^{N}{\delta ({{x}_{{{\nu }_{1}}}}-{{X}_{{{\nu }_{1}}}}(t;\theta ))}}, \label{eq:3.6a}
    \end{align}
    \begin{align}
        \frac{\delta \mathcal{F}[\cdot \cdot \cdot ]}{\delta {{\Xi }_{{{n}_{2}}}}(s;\theta )} & = \frac{\delta }{\delta {{\Xi }_{{{n}_{2}}}}(s;\theta )}\delta (\bfv{x}-\bfv{X}(t;\theta )) = \nonumber                                                                                                                                                                                                                                                                                                  \\
                                                                                              & = -\sum\limits_{\nu =1}^{N}{\frac{\partial \delta ({{x}_{\nu }}-{{X}_{\nu }}(t;\theta ))}{\partial {{x}_{\nu }}}\frac{\delta {{X}_{\nu }}(t;\theta )}{\delta {{\Xi }_{{{n}_{2}}}}(s;\theta )}\prod\limits_{\begin{smallmatrix}{{\nu }_{1}}=1 \\ {{\nu }_{1}}\ne \nu\end{smallmatrix}}^{N}{\delta ({{x}_{{{\nu }_{1}}}}-{{X}_{{{\nu }_{1}}}}(t;\theta ))}}. \label{eq:3.6b}
    \end{align}
\end{subequations}
Note that, in equation (\ref{eq:3.6b}), we apply the chain rule for the Volterra derivative with respect to excitation 
${{\Xi }_{{{n}_{2}}}}(\cdot ;\theta )$. Further, we see that the essential part for the calculation of the derivatives 
$\partial \mathcal{F} [\cdot \cdot \cdot ]\mathbf{/}\partial X_{{{n}_{1}}}^{0}(\theta)$
and
$\delta \mathcal{F}[\cdot \cdot \cdot ]\mathbf{/}\delta{{\Xi }_{{{n}_{2}}}}(s \theta)$ 
is the calculation of the following variational derivatives:
\begin{subequations} \label{eqs:3.7}
    \begin{align}
         & V_{\nu {{n}_{1}}}^{{{\bfv{X}}^{0}}}(t;\theta )=\frac{\partial {{X}_{\nu }}(t;\theta )}{\partial X_{{{n}_{1}}}^{0}(\theta )}\equiv \frac{\partial{{X}_{\nu }}\left[{{\bfv{X}}^{0}}(\theta );\mathbf{\Xi}(\left. \cdot \right|_{{{t}_{0}}}^{t};\theta ) \right]}{\partial X_{{{n}_{1}}}^{0}}, \label{eq:3.7a}
        \\
         & V_{\nu {{n}_{2}}}^{\mathbf{\Xi} (s)}(t;\theta )=\frac{\delta {{X}_{\nu }}(t;\theta )}{\delta {{\Xi }_{{{n}_{2}}}}(s;\theta )}\equiv \frac{\delta{{X}_{\nu }}\left[{{\bfv{X}}^{0}}(\theta );\mathbf{\Xi}(\left. \cdot \right|_{{{t}_{0}}}^{t};\theta ) \right]}{\delta {{\Xi }_{{{n}_{2}}}}(s;\theta )}. \label{eq:3.7b}
    \end{align}
\end{subequations}
Derivatives (\ref{eqs:3.7}) express the rate of change of the response components ${{X}_{\nu }}(t;\theta )$ with respect 
to the components of the initial values $X_{{{n}_{1}}}^{0}(\theta )$, and the components of the excitation 
${{\Xi }_{{{n}_{2}}}}(t;\theta )$, ${{n}_{2}}=1,\ldots,N$, respectively. This is why they are usually
called variational derivatives of the system (\ref{eqs:2.1}). The
variational derivatives (\ref{eqs:3.7}) satisfy linear initial value problems (IVPs), formulated
and solved in \hyperref[sec3.3]{Section 3.3}. Using equations (\ref{eqs:3.6}) and (\ref{eqs:3.7}), in conjunction with
the NF theorem (\ref{eq:3.5}), we readily reformulate the SLE (\ref{eq:3.4}) in the form:
\begin{align} \label{eq:3.8}
    \frac{\partial{{f}_{\bfv{X}(t)}}(\bfv{x})}{\partial t} & + \sum\limits_{n=1}^{N}{\frac{\partial }{\partial {{x}_{n}}}\left[ \left( {{h}_{n}}(\bfv{x}, t) + m_{{{\Xi }_{n}}}(t) \right){{f}_{\bfv{X}(t)}}(\bfv{x}) \right]}= \nonumber                                                                                                                                                              \\
                                                           & +\sum\limits_{n=1}^{N}{\sum\limits_{\nu =1}^{N}{\frac{\partial }{\partial {{x}_{n}}}\frac{\partial }{\partial {{x}_{\nu }}}\sum\limits_{{{n}_{1}}=1}^{N}{C_{X_{{{n}_{1}}}^{0}{{\Xi }_{n}}}(t)\mathbb{E}^{\theta }\left[ \delta (\bfv{x}-\bfv{X}(t;\theta ))V_{\nu {{n}_{1}}}^{{{\bfv{X}}^{0}}}(t;\theta ) \right]}}}+ \nonumber                   \\
                                                           & +\sum\limits_{n=1}^{N}{\sum\limits_{\nu =1}^{N}{\frac{\partial }{\partial {{x}_{n}}}\frac{\partial }{\partial {{x}_{\nu }}}\sum\limits_{{{n}_{2}}=1}^{N}{\int\limits_{{{t}_{0}}}^{t}{C_{{{\Xi }_{n}}{{\Xi }_{{{n}_{2}}}}}(t,s)\mathbb{E}^{\theta }\left[ \delta (\bfv{x}-\bfv{X}(t;\theta ))V_{\nu {{n}_{2}}}^{\Xi (s)}(t;\theta ) \right]ds}}}}.
\end{align}
Equation (\ref{eq:3.8}) will hereafter be referred to as the transformed SLE. It is exact but non-closed due to the 
averages over the variational derivatives, which are dependent on the whole history of excitation. Equation (\ref{eq:3.8}) 
is a generalization of the usually derived SLE \cite{Cetto_1984,Dekker_1982,San_Miguel_1980,Hernandez_Machado_1983} to 
the case of non-zero mean excitations and excitations correlated with the initial data.

\subsection{ Calculation of the variational derivatives }\label{sec3.3}

The calculation of the variational derivatives will be carried out by formulating and solving the corresponding variational 
problems, obtained by differentiating the system of RDEs (\ref{eqs:2.1}) with respect to the data (initial values
and excitation functions). Differentiations will be caried out path-wisely, i.e., for each $\theta$ separately. Thus, 
the presence of the stochastic argument $\theta$ is not essential herein, and it is omitted in this section.
We start by rewriting equations (\ref{eqs:2.1}) in the following form
\addtocounter{equation}{1}
\begin{equation} \label{eqs:3.9}
    {{\dot{X}}_{\nu }}(t)={{h}_{\nu }}\left( \bfv{X}(t),t \right)+{\Xi}_{\nu }
    (t), \ \ \
    {{X}_{\nu }}({{t}_{0}})=X_{\nu }^{0}, \ \ \nu=1,2,\ldots ,N. \tag{3.9a,b}
\end{equation}

\subsubsection{Calculation of the variational derivatives
\texorpdfstring{$V_{\nu {{n}_{1}}}^{{{\bfv{X}}^{0}}}$}{with respect to initial value} } \label{sec3.3.1}

Applying the operator $\partial \cdot /\partial X_{{{n}_{1}}}^{0}$ to both sides of the equations (\ref{eqs:3.9}), and 
recalling that each path-function ${{\Xi }_{\nu }}(t)$ is not functionally dependent on $X_{{{n}_{1}}}^{0}$,${{n}_{1}}\in \{1,\ldots ,N\}$, 
we get the following IVPs:
\begin{subequations} \label{eqs:3.10}
    \begin{align}
         & \frac{d}{dt}V_{\nu {{n}_{1}}}^{{{\bfv{X}}^{0}}}(t)=\sum\limits_{{{\nu }_{1}}=1}^{N}{\frac{\partial {{h}_{\nu }}\left( \bfv{X}(t),t \right)}{\partial X_{{{\nu }_{1}}}(t)}V_{{{\nu }_{1}}{{n}_{1}}}^{{{\bfv{X}}^{0}}}(t)}+\underbrace{\frac{\partial {{\Xi }_{\nu }}(t)}{\partial X_{{{n}_{1}}}^{0}}}_{=0}, \ \ {{t}_{0}}<t, \label{eq:3.10a} \\
         & V_{\nu {{n}_{1}}}^{{{\bfv{X}}^{0}}}({{t}_{0}})=\frac{\partial X_{\nu }^{0}}{\partial X_{{{n}_{1}}}^{0}}={{\delta }_{\nu {{n}_{1}}}}, \  \ \nu =1,\ldots ,N, \label{eq:3.10b}
    \end{align}
\end{subequations}
where ${{\delta }_{\nu {{n}_{1}}}}$ denote the Kronecker delta. Note that in equation (\ref{eq:3.10a}) we have assumed 
the interchangeability of derivative operators $\partial \cdot /\partial X_{{{n}_{1}}}^{0}$ and $d\cdot/dt$. Under the
assumption that $\partial{{h}_{\nu }}\left( \bfv{X}(t),t \right)/\partial X_{{{\nu }_{1}}}(t)$ are continuous functions of 
$t$, the solutions of the above time-varying, linear systems of ODEs can be written as $\left({{n}_{1}}=1,\ldots,N \right)$:
\begin{equation}
    V_{\nu {{n}_{1}}}^{{{\bfv{X}}^{0}}}(t)=\sum\limits_{{{\nu }_{1}}=1}^{N}{\Phi _{\nu {{\nu }_{1}}}^{{{\bfv{X}}_{0}}}(t;{{t}_{0}})}
    {{\delta }_{{{\nu }_{1}}{{n}_{1}}}}=\Phi_{\nu {{n}_{1}}}^{{{\bfv{X}}_{0}}}
    (t;{{t}_{0}}), \ \ v=1,\ldots ,N, \label{eq:3.11}
\end{equation}
where $\Phi_{\nu {{n}_{1}}}^{{{\bfv{X}}_{0}}}(t;{{t}_{0}})$ denote the state-transition matrix, \cite{Brockett_1970}, of 
the IVP (\ref{eqs:3.10}).

\subsubsection{ Calculation of the variational derivatives \texorpdfstring{$V_{m\ell }^{\mathbf{\Xi}(s)}$}{with respect to excitation} } \label{sec3.3.2}

In analogy with the above procedure, we calculate the variational derivative $\delta{{\dot{X}}_{\nu }}(t)/\delta{{\Xi }_{{{n}_{2}}}}(s)$ 
by applying the Volterra derivative operator $\delta\cdot / \delta{{\Xi }_{{{n}_{2}}}}(s)$, $s\in [t_{0},t]$, to both 
sides of equations (\ref{eqs:3.9}). Assuming that the chain rule holds true for the Volterra derivatives, we derive the 
following problems:
\begin{subequations} \label{eqs:3.12}
    \begin{align}
         & \frac{\delta {{{\dot{X}}}_{\nu }}(t)}{\delta {{\Xi }_{{{n}_{2}}}}(s)}=\sum\limits_{{{\nu }_{1}}=1}^{N}{\frac{\partial {{h}_{\nu }}(\bfv{X}(t),t)}{\partial {{X}_{{{\nu }_{1}}}}(t)}\frac{\delta {{X}_{{{\nu }_{1}}}}(t)}{\delta {{\Xi }_{{{n}_{2}}}}(s)}}+\frac{\delta {{\Xi }_{\nu }}(t)}{\delta {{\Xi }_{{{n}_{2}}}}(s)}, \ {{t}_{0}}<s,t \label{eq:3.12a} \\
         & \frac{\delta {{X}_{\nu }}({{t}_{0}})}{\delta {{\Xi }_{{{n}_{2}}}}(s)}=0, \ \nu =1,\ldots ,N. \label{eq:3.12b}
    \end{align}
\end{subequations}
Under the assumption that the derivative operators $\delta \cdot /\delta{{\Xi }_{{{n}_{2}}}}(s)$ and $d\cdot /dt$ are
interchangeable, and using the well-known result $\delta{{\Xi }_{\nu }}(t)/\delta {{\Xi }_{{{n}_{2}}}}(s)={{\delta }_{\nu {{n}_{2}}}}\delta(t-s)$, 
equation (\ref{eq:3.12a}) takes the form:
\begin{equation}\label{eq:3.12a'}
    \frac{d}{dt}\frac{\delta {{X}_{\nu }}(t)}{\delta {{\Xi }_{{{n}_{2}}}}(s)}
    =\sum\limits_{{{\nu }_{1}}=1}^{N}{\left( \frac{\partial {{h}_{\nu }}\left( \bfv{X}(t),t \right)}{\partial {{X}_{{{\nu }_{1}}}}(t)} \right)}
    \frac{\delta {{X}_{{{\nu }_{1}}}}(t)}{\delta {{\Xi }_{{{n}_{2}}}}(s)}+{{\delta }_{\nu {{n}_{2}}}}
    \delta (t-s), \ \ {{n}_{2}}=1,\ldots ,N. \tag{3.12a'}
\end{equation}

\begin{remark}
    It can be shown that the assumptions made in the derivation of equations (\ref{eq:3.12a'}) and (\ref{eq:3.10a}) 
    are valid under reasonable technical conditions on the analytical structure of the studied equations and the excitations. 
    In the present work we focus on the essential steps of the derivation, assuming that the appropriate technical 
    conditions are satisfied. 
\end{remark}

Equation (\ref{eq:3.12a'}) is not in a form convenient for further study and solution. It can be reformulated as a linear 
system in the standard form by invoking a causality argument, as in \cite{Dekker_1982,Cetto_1984}. Any variation of the
excitation, $\delta{{\Xi }_{{{n}_{2}}}}(s)$, at the time $s$, cannot result in a variation of the response, 
$\delta{{X}_{{{\nu }_{1}}}}(t)$ , at any previous time $t<s$, implying $\delta{{X}_{\nu }}(t)/\delta{{\Xi }_{{{n}_{2}}}}(s)=0$
for $t<s$ and any ${{n}_{2}},{{\nu }_{1}}\in \{1,\ldots ,N\}$. Thus, we have
\begin{equation} \label{eq:3.13}
    V_{\nu {{n}_{2}}}^{\mathbf{\Xi} (s)}(\tau )=\frac{\delta {{X}_{\nu }}(\tau
    )}{\delta {{\Xi }_{{{n}_{2}}}}(s)}\equiv \frac{\delta{{X}_{\nu }}\left[{{\bfv{X}}^{0}}(\theta);\mathbf{\Xi}(\left.
        \cdot \right|_{s}^{\tau };\theta ) \right]}{\delta{{\Xi }_{{{n}_{2}}}}(s)}
    =0, \ for \ any \ \tau \in [{{t}_{0}},s).
\end{equation}
Based on equation (\ref{eq:3.13}), we reformulate equation (\ref{eq:3.12a'}) in integral form, integrating with respect 
to time over the interval $[s-\varepsilon ,t]$, for small $\varepsilon>0$ :
\begin{equation} \label{eq:3.12a''}
    \frac{\delta {{X}_{\nu }}(t)}{\delta {{\Xi }_{{{n}_{2}}}}(s)}=\sum\limits
    _{{{\nu }_{1}}=1}^{N}{\int_{s-\varepsilon }^{t}{\left( \frac{\partial {{h}_{\nu }}\left( \bfv{X}(\tau ),\tau \right)}{\partial {{X}_{{{\nu }_{1}}}}(\tau )} \right)\frac{\delta {{X}_{{{\nu }_{1}}}}(\tau )}{\delta {{\Xi }_{{{n}_{2}}}}(s)}}d\tau }
    +{{\delta }_{\nu {{n}_{2}}}}\underbrace{\int_{s-\varepsilon }^{t}{\delta (\tau -s)d\tau }}
    _{=1}.\tag{3.12a''}
\end{equation}
Integrating from $s-\varepsilon$ (up to $t$) is consistent with equation (\ref{eq:3.13}), and helps us to formulate a 
well-defined integral of the delta function (equals to 1), avoiding having the singularity point at the lower integration limit.
By taking the limit as $\varepsilon \to 0$, we obtain a Volterra integral equation of the second kind, which for ${{n}_{2}}\in \{1,\ldots ,N\}$ 
is equivalent to the following IVP \cite{Polyanin_2008}:
\begin{subequations} \label{eqs:3.14}
    \begin{align}
         & \frac{d}{dt}V_{\nu {{n}_{2}}}^{\mathbf{\Xi} (s)}(t)= \sum\limits_{{{\nu }_{1}}=1}^{N}{\left( \frac{\partial {{h}_{\nu }}\left( \bfv{X}(t),t \right)}{\partial {{X}_{{{\nu }_{1}}}}(t)} \right)}V_{{{\nu }_{1}}{{n}_{2}}}^{\mathbf{\Xi} (s)}(t), \ \ t>s,
        \\
         & V_{\nu {{n}_{2}}}^{\mathbf{\Xi} (s)}(s)={{\delta }_{\nu {{n}_{2}}}}, \ \ \nu =1,\ldots,N,
    \end{align}
\end{subequations}
System (\ref{eqs:3.14}), being in the standard form, permits us to write its solution in the form:
\begin{equation}
    V_{\nu {{n}_{2}}}^{\mathbf{\Xi} (s)}(t)=\sum\limits_{{{\nu }_{1}}=1}^{N}{\Phi _{\nu {{\nu }_{1}}}^{\mathbf{\Xi} }(t;s)}
    {{\delta }_{{{\nu }_{1}}{{n}_{2}}}}=\Phi_{\nu {{n}_{2}}}^{\mathbf{\Xi} }(
    t;s), \ \nu ,{{n}_{2}}=1,\ldots ,N, \label{eq:3.15}
\end{equation}
where $\Phi_{\nu {{n}_{2}}}^{\mathbf{\Xi} }(t;s)$ is the state-transition matrix of the IVP (\ref{eqs:3.14}).

\subsection{Final form of the transformed SLE and further discussion} \label{sec3.4}

To formulate the final form of the SLE, we restore the stochastic argument and substitute in the transformed SLE (\ref{eq:3.8}) 
the solutions $V_{\nu {{n}_{1}}}^{{{\bfv{X}}^{0}}}(t;\theta)$, $V_{\nu {{n}_{2}}}^{\mathbf{\Xi}(s)}(t;\theta )$, given 
by equations (\ref{eq:3.11}) and (\ref{eq:3.15}), respectively. Then we obtain the following equation:
\begin{equation}
    \begin{aligned}
        \frac{\partial{{f}_{\bfv{X}(t)}}(\bfv{X})}{\partial t} & +\sum\limits_{n=1}^{N}{\frac{\partial }{\partial {{x}_{n}}}\left[ \left( {{h}_{n}}\left( \bfv{x} \right)+m_{{{\Xi }_{n}}}(t) \right){{f}_{\bfv{X}(t)}}(\bfv{x}) \right]}=                                                                                                                                                                                                     \\
                                                               & =\sum\limits_{n=1}^{N}{\sum\limits_{\nu =1}^{N}{\frac{\partial }{\partial {{x}_{n}}}\frac{\partial }{\partial {{x}_{\nu }}}\sum\limits_{{{n}_{1}}=1}^{N}{C_{X_{{{n}_{1}}}^{0}{{\Xi }_{n}}}(t)\mathbb{E}^{\theta }\left[ \delta (\bfv{x}-\bfv{X}(t;\theta ))\Phi _{\nu {{n}_{1}}}^{{{\bfv{X}}^{0}}}(t;{{t}_{0}},\theta ) \right]}}}+                                           \\
                                                               & +\sum\limits_{n=1}^{N}{\sum\limits_{\nu =1}^{N}{\frac{\partial }{\partial {{x}_{n}}}\frac{\partial }{\partial {{x}_{\nu }}}\sum\limits_{{{n}_{2}}=1}^{N}{\int\limits_{{{t}_{0}}}^{t}{C_{{{\Xi }_{n}}{{\Xi }_{{{n}_{2}}}}}(t,s)\mathbb{E}^{\theta }\left[ \delta (\bfv{x}-\bfv{X}(t;\theta ))\Phi _{\nu {{n}_{2}}}^{\mathbf{\Xi} }(t;s,\theta ) \right]ds}}}}. \label{eq:3.16}
    \end{aligned}
\end{equation}
Equation (\ref{eq:3.16}) is structurally identical to equation (\ref{eq:3.8}). The difference is that the variational 
derivatives (\ref{eqs:3.7}) in (\ref{eq:3.8}) have been expressed by using the state-transition matrices which realize 
the solutions of the variational problems (\ref{eqs:3.10}) and (\ref{eqs:3.14}). The explicit calculation (or a good analytic
approximation) of the state-transition matrices is an essential step to move forward. The treatment of the latter varies 
among authors, see e .g. \cite{Cetto_1984,Dekker_1982}. To that end, we prioritize the analytic treatment of the state-transition
matrices, as expressed by the Peano-Baker \cite{Brockett_1970} and Magnus \cite{Blanes_2009,Magnus_1954} series representations, 
see \hyperref[secB1]{Appendix B}. This perspective permits us to derive new genFPKE for nonlinear systems of RDE, outperforming 
existing ones (see \hyperref[sec5]{Section 5}). However, before proceeding to the derivation of our novel pdf-evolution equation, 
it seems appropriate to present some simple applications of
the transformed SLE.


\section{First, simple applications of the transformed SLE} \label{sec4}

In this Section, we shall first derive an equation for the evolution of the response pdf of a linear system, under colored 
Gaussian excitation. The obtained model equation encompasses the complete (transient and long-term) probabilistic behavior
of the system, is analytically solvable, and provides us with a simple benchmark for testing numerical schemes.

\subsection{Pdf-evolution equation for linear systems, under colored noise excitation} \label{sec4.1}

Assuming that the ${{h}_{n}}$ functions in equations (\ref{eqs:2.1}), are linear in $\bfv{X}$,
\begin{equation}
    {{h}_{n}}\left( \bfv{X}(t;\theta ),t \right)=\sum\limits_{m=1}^{N}{{{\eta }_{nm}}(t){{X}_{m}}(t;\theta )}
    , \ \ n=1,2,\ldots,N, \label{eq:4.1}
\end{equation}
the corresponding system of RDE becomes linear. In order to specify the general equation (\ref{eq:3.16}) to the present 
case (as well as to any of the cases studied subsequently), we have to calculate the average terms of its right-hand side,
which are concisely written as follows:
\begin{subequations} \label{eqs:4.2}
    \begin{align}
         & {{\mathcal{G}}_{\nu {{n}_{1}}}}(t;{{t}_{0}})=\mathbb{E}^{\theta }\left[ \delta (\bfv{x}-\bfv{X}(t;\theta ))\Phi_{\nu {{n}_{1}}}^{{{\bfv{X}}^{0}}}(t;{{t}_{0}},\theta ) \right], \label{eq:4.2a} \\
         & {{\mathcal{G}}_{\nu {{n}_{2}}}}(t;s)=\mathbb{E}^{\theta }\left[ \delta (\bfv{x}-\bfv{X}(t;\theta ))\Phi_{\nu {{n}_{2}}}^{\mathbf{\Xi} }(t;s,\theta ) \right]. \label{eq:4.2b}
    \end{align}
\end{subequations}

To calculate the above averages, first we need to determine the state-transition matrices $\Phi^{{{\bfv{X}}_{0}}}$ and $\Phi^{\mathbf{\Xi} }$, 
which are defined path-wisely (and thus, in general, they depend on the stochastic argument $\theta$). In the linear case, 
the common system matrix of the variational problems (\ref{eqs:3.10}) and (\ref{eqs:3.14}),
$\partial{{h}_{n}}(\bfv{X}(t;\theta ),t)/\partial X_{m}={{\eta }_{nm}}(t)$, becomes independent of $\bfv{X}(t;\theta )$. 
Thus, the variational problems (\ref{eqs:3.10}) and (\ref{eqs:3.14}) become deterministic, time-varying 
linear systems of ODEs. This allows us to calculate the averages (\ref{eqs:4.2}) directly, since the matrices
$\Phi^{{{\bfv{X}}_{0}}}$ and $\Phi^{\mathbf{\Xi} }$ are factored out of the expected value operator, obtaining the expressions:
\addtocounter{equation}{1}
\begin{equation}
    {{\mathcal{G}}_{\nu {{n}_{1}}}}(t;{{t}_{0}})=\Phi_{\nu {{n}_{1}}}^{{{\bfv{X}}_{0}}}
    (t;{{t}_{0}}){{f}_{\bfv{X}(t)}}(\bfv{x}), \ \ \
    {{\mathcal{G}}_{\nu {{n}_{2}}}}
    (t;s)=\Phi_{\nu {{n}_{2}}}^{\mathbf{\Xi} }(t;s){{f}_{\bfv{X}(t)}}(\bfv{x}
    ). \tag{4.3a,b} \label{eqs:4.3a,b}
\end{equation}
Then, by substituting equations (\ref{eqs:4.3a,b}) into equation (\ref{eq:3.16}), we obtain the following linear pdf-evolution equation:
\begin{align}
    \frac{\partial {{f}_{\bfv{X}(t)}}(\bfv{x})}{\partial t}+ & \sum\limits_{n=1}
    ^{N}{\frac{\partial }{\partial {{x}_{n}}}\left[ \left( {{h}_{n}}(\bfv{x},t)+{{m}_{{{\Xi }_{n}}}}(t) \right){{f}_{\bfv{X}(t)}}(\bfv{x}) \right]}
    = \nonumber
    \\
                                                             & =\sum\limits_{n=1}^{N}{\sum\limits_{\nu =1}^{N}{\left( \mathcal{D}_{\nu n}^{{{X}_{0}}\mathbf{\Xi} }(t)+\mathcal{D}_{\nu n}^{\mathbf{\Xi} \mathbf{\Xi} }(t) \right)\frac{\partial }{\partial {{x}_{n}}}\frac{\partial }{\partial {{x}_{\nu }}}}}
    {{f}_{\bfv{X}(t)}}(\bfv{x}), \label{eq:4.4}
\end{align}
where the diffusion coefficients on the right-hand side of equation (\ref{eq:4.4}),
are given by:
\begin{subequations} \label{eqs:4.5}
    \begin{align}
         & \mathcal{D}_{\nu n}^{{{X}_{0}}\mathbf{\Xi} }(t;{{t}_{0}})=\sum\limits_{{{n}_{1}}=1}^{N}{C_{X_{{{n}_{1}}}^{0}{{\Xi }_{n}}}(t)\Phi _{\nu {{n}_{1}}}^{{{\bfv{X}}_{0}}}(t;{{t}_{0}})},\label{eq:4.5a}                     \\
         & \mathcal{D}_{\nu n}^{\mathbf{\Xi} \mathbf{\Xi} }(t)=\sum\limits_{{{n}_{2}}=1}^{N}{\int\limits_{{{t}_{0}}}^{t}{C_{{{\Xi }_{n}}{{\Xi }_{{{n}_{2}}}}}(t,s)\Phi _{\nu {{n}_{2}}}^{\mathbf{\Xi} }(t;s)ds}}.\label{eq:4.5b}
    \end{align}
\end{subequations}
Note that explicit closed-form expression for the matrices $\Phi^{{{\bfv{X}}_{0}}}$ and $\Phi^{\mathbf{\Xi} }$ are not 
generally available for time-varying variational problems. A discussion on the main representations of state-transition
matrices and their approximations is presented in \hyperref[secB1]{Appendix B}. In the case of a system of RDE with constant 
coefficients, i.e. ${{\eta }_{nm}}(t)={{\eta }_{nm}}$, the common system matrix of the variational problems becomes constant, 
namely ${{({{\mathbf{J}}^{\bfv{h}}})}_{nm}}:= \partial{{h}_{n}}(\bfv{X}(t;\theta ))/\partial X_{m} = {{\eta }_{nm}}$, 
and the corresponding state-transition matrices $\Phi^{{{\bfv{X}}_{0}}}$ and $\Phi^{\mathbf{\Xi}}$ take the form of matrix 
exponentials (\cite{Brockett_1970}, Section 1.5), namely,
\begin{equation} \label{eqs:4.6a,b}
    \Phi_{\nu {{n}_{1}}}^{{{\bfv{X}}_{0}}}(t;{{t}_{0}})={{\left( \exp \left( \left( t-{{t}_{0}} \right){{\mathbf{J}}^{\bfv{h}}} \right) \right)}_{\nu {{n}_{1}}}}
    , \ \ \
    \Phi_{\nu {{n}_{2}}}^{\mathbf{\Xi} }(t;s)={{\left( \exp \left( \left( t-s \right){{\mathbf{J}}^{\bfv{h}}} \right) \right)}_{\nu {{n}_{2}}}}
    . \tag{4.6a,b} 
\end{equation}
\addtocounter{equation}{1}
\noindent
Equation (\ref{eq:4.4}) is a linear transport-diffusion PDE for the evolution of ${{f}_{\bfv{X}(t)}}(\bfv{x})$, which 
has also been presented in \cite{Mamis_2018}, for the case of a system with constant coefficients. This equation,
supplemented by the initial condition:
\begin{equation} \label{eq:4.7a}
    {{f}_{\bfv{X}({{t}_{0}})}}(\bfv{x})={{f}_{0}}(\bfv{x})\left( \text{=}\text{general Gaussian}\right),\tag{4.7a} 
\end{equation}
and the additional data (coefficient functions)
\addtocounter{equation}{1}
\begin{equation}
    {{\mathbf{m}}_{\mathbf{\Xi} }}(t),{{\mathbf{C}}_{{{\bfv{X}}^{0}}\mathbf{\Xi} }}
    (t),\tag{4.7b,c}\label{eqs:4.7b,c}
\end{equation}
can be solved analytically via order reduction, using Fourier transform and the method of characteristics. Its solution 
is presented in \cite{Mamis_phd}. On the other hand, it is well known that a linear system under Gaussian excitation yields
a Gaussian response, which permits us to construct the response pdf by solving the corresponding moment problem. 
The direct analytic solution of PDE (\ref{eq:4.4}), for system with constant coefficients, coincides with the solution obtained
by means of the moments \cite{Mamis_phd}, a fact that provides a first justification of the present approach.

Numerical results for the underdamped linear oscillator
\begin{equation*}
    \ddot{X}(t;\theta )+2\zeta{{\omega }_{0}}\dot{X}(t;\theta )+\omega_{0}^{2}
    X(t;\theta )=\Xi(t;\theta ),
\end{equation*}
under colored Gaussian excitation are presented, as a benchmark problem, in the Supplementary Material. The linear 
pdf-evolution equation (\ref{eq:4.4}), corresponding to this case, is specified and then solved numerically by means of 
a partition of unity finite element scheme. The approximated 2D response and 1D marginals densities are calculated and 
nicely compared with the known analytic solutions, and MC simulations. This is a first test of validity for the main 
elements of the numerical scheme, which is also used for the solution of the nonlinear case in \hyperref[sec6]{Section 6}. 
It also allows us to assess the MC simulation procedure, as the analytical form of the response pdf is known in this case.

\subsection{The case of white-noise excitation. An alternative derivation of the classical FPKE} \label{sec4.2}

\begin{figure} 
    \centering
    \includegraphics[width=0.5\textwidth]{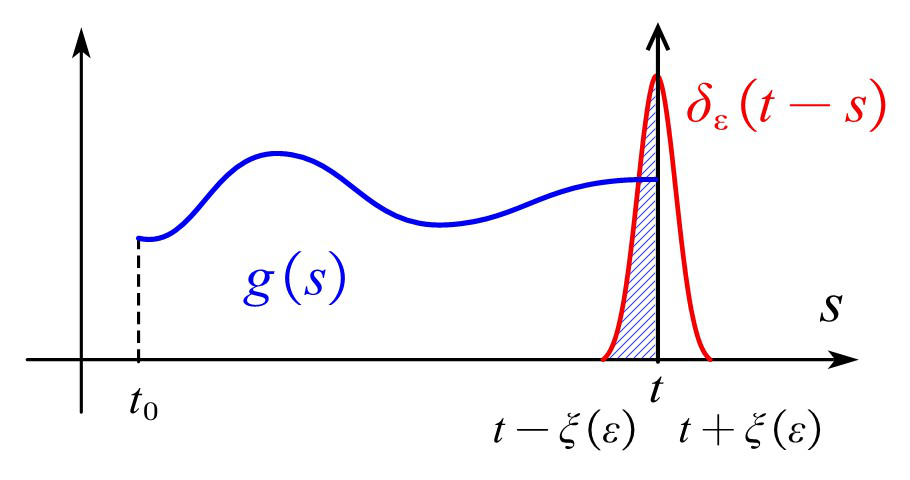}
    \caption{ \centering Geometrical interpretation of the integral $\underset{\varepsilon \,\,\downarrow \,\,0}{\mathop{\lim }}\,\,\,\int_{{{t}_{0}}}^{t}{{{\delta }_{\varepsilon }}\,(\,t-s\,)\,\,g\,(\,s\,)\,\,ds}$.} \label{fig:1}
\end{figure}

Under the assumption of a white noise Gaussian excitation, we have that:
\addtocounter{equation}{1}
\begin{equation} \label{eqs:4.8}
    {{m}_{\mathbf{\Xi} }}(t)=0, \ \ \  \mathbf{C}_{\mathbf{\Xi}\mathbf{\Xi} }^{\text{W}\text{N}}(t,s)=2D(t)\delta (t-s), \tag{4.8a,b}
\end{equation}
where $D(t)$ is a positive-definite matrix which is called the noise intensity matrix. Assuming further
\begin{equation}
    {{\mathbf{C}}_{{{\bfv{X}}_{0}}\mathbf{\Xi} }}(t)=0, \tag{4.8c}
\end{equation}
and substituting equations (\hyperref[eqs:4.8]{4.8}) in equation (\ref{eq:3.16}), the later simplifies
in the following form:
\addtocounter{equation}{1}
\begin{align*}\label{eq:4.9a}
    \frac{\partial {{f}_{\bfv{X}(t)}}(\bfv{x})}{\partial{t}} & +\sum\limits_{n=1}^{N}{\frac{\partial }{\partial {{x}_{n}}}\left( {{h}_{n}}(\bfv{x},t){{f}_{\bfv{X}(t)}}(\bfv{x}) \right)}=                                                                                                                                                           \\
                                                             & =\sum\limits_{n=1}^{N}{\sum\limits_{\nu =1}^{N}{\frac{{{\partial }^{2}}}{\partial {{x}_{n}}\partial {{x}_{\nu }}}\sum\limits_{{{n}_{2}}=1}^{N}{2{{D}_{n{{n}_{2}}}}(t)\mathbb{E}^{\theta }\left[ \delta (\bfv{x}-\bfv{X}(t;\theta )){{B}_{\nu {{n}_{2}}}}(t) \right]}}}, \tag{4.9a}
\end{align*}
where
\begin{equation} \label{eq:4.9b}
    {{B}_{\nu {{n}_{2}}}}(t)=\int_{{{t}_{0}}}^{t}{\delta (t-s)\Phi _{\nu {{n}_{2}}}^{\mathbf{\Xi} }(t;s,\theta )ds}. \tag{4.9b}
\end{equation}
To finish the derivation of the classical FPKE, we have to calculate the integral ${{B}_{\nu {{n}_{2}}}}(t)$. 
A technical difficulty in this calculation arises from the fact that the integrand contains a delta function whose singular
point coincides with the upper limit of integration. Such a calculation does not have a standard mathematical meaning 
and should be performed as a limiting calculation, guided by the underlying physics. In this connection, we regularize 
the integral, representing the delta function by a delta family of smooth functions ${{\delta }_{\varepsilon }}(t-s)$, 
which recovers the delta function as the weak limit for $\varepsilon \downarrow 0$:
\begin{equation} \label{{eq:4.10}}
    \delta (t-s)=\underset{\varepsilon \downarrow 0}{\mathop{\lim }}{{\delta }_{\varepsilon }}
    (t-s)=\underset{\varepsilon \downarrow 0}{\mathop{\lim }}\frac{1}{\varepsilon
    }q\left( \frac{t-s}{\varepsilon }\right),
\end{equation}
where $q(\cdot )$ is any non-negative, normalized smooth function (kernel), supported around zero, \cite{Cheney_2009}. 
Under these considerations, equation (\hyperref[eqs:4.8]{4.8b}) is written as
\begin{equation} \label{eq:4.8b'}
    \mathbf{C}_{\mathbf{\Xi} \mathbf{\Xi} }^{\text{W}\text{N}}
    (t,s)=\underset{\varepsilon \downarrow 0}{\mathop{\lim }}\mathbf{C}_{\mathbf{\Xi}
    \mathbf{\Xi} }^{\text{(}\varepsilon \text{)}}(t,s)=2D(t)\underset
    {\varepsilon \downarrow 0}{\mathop{\lim }}\frac{1}{\varepsilon }q\left( \frac{t-s}{\varepsilon
    }\right). \tag{4.8b'}
\end{equation}
Since any stationary autocovariance 
$\mathbf{C}_{\mathbf{\Xi} \mathbf{\Xi} }^{(\varepsilon)}(t,s)=\mathbf{C}_{\mathbf{\Xi} \mathbf{\Xi} }^{(\varepsilon)}(t-s)$
is an even function, the kernel function $q(\cdot )$, defining the delta function in this case, must be an even function 
as well. This requirement implies, for any continuous function $g(\cdot )$, that:
\begin{equation} \label{eq:4.11}
    \underset{\varepsilon \downarrow 0}{\mathop{\lim }}\int\limits_{t_0}^{t}{{\delta}_{\varepsilon }}(t-s)g(s)ds = 
    \underset{\varepsilon \downarrow 0}{\mathop{\lim }}\left( g(\tilde{t}(\varepsilon ))\int\limits_{t-\xi(\varepsilon)}^{t}{{\delta }_{\varepsilon }}(t-s)ds \right)=\frac{1}{2}
    g(t),
\end{equation}
where $t-\xi(\varepsilon )<\tilde{t}(\varepsilon)<t$. The first equality in the above equation is implied by the mean 
value theorem, and the second one by the fact that $\underset{\varepsilon \downarrow 0}{\mathop{\lim }}g(\tilde{t}(\varepsilon))=g(t)$ 
and the restriction of the integration of the symmetric ${{\delta }_{\varepsilon }}(t-s)$ to the half-interval, see Figure \ref{fig:1}.
Applying equation (\ref{eq:4.11}), the calculation of the integral ${{B}_{\nu {{n}_{2}}}}(t)$ is now straightforward. 
Indeed, the state-transition matrix $\Phi (t,s;\theta )$ is a continuous function with respect to both arguments $t,s$, 
and $\underset{s\to t}{\mathop{\lim }}\Phi (t,s;\theta )=\bfv{I}$, where $\bfv{I}$ indicates the identity matrix. Thus, 
we readily obtain ${{B}_{\nu {{n}_{2}}}}(t)=(1/2){{\delta }_{\nu {{n}_{2}}}}$. Finally, by substituting ${{B}_{\nu {{n}_{2}}}}(t)$ 
in (\ref{eq:4.9a}), using the identity 
$\sum\limits_{{{n}_{2}}=1}^{N}{{{D}_{n{{n}_{2}}}}(t){{\delta }_{\nu {{n}_{2}}}}}={{D}_{n\nu }}(t)$, and recalling equation 
(\ref{eq:2.2}), we derive
\begin{equation} \label{eq:4.12}
    \frac{\partial {{f}_{\bfv{X}(t)}}(\bfv{x})}{\partial t}+\sum\limits_{n=1}
    ^{N}{\frac{\partial }{\partial {{x}_{n}}}\left[ {{h}_{n}}(\bfv{x},t){{f}_{\bfv{X}(t)}}(\bfv{x}) \right]}
    =\sum\limits_{n=1}^{N}{\sum\limits_{\nu =1}^{N}{{{D}_{n\nu }}(t)\frac{{{\partial }^{2}}{{f}_{\bfv{X}(t)}}(\bfv{x})}{\partial {{x}_{n}}\partial {{x}_{\nu }}}}},
\end{equation}
which is the classical Fokker-Planck-Kolmogorov (FPK) equation \cite{Moss_1989,Risken_1984}, corresponding to a 
nonlinear system of RDEs, excited by additive Gaussian white noise.
\begin{remark}
    It is interesting to note that the above derivation of the FPKE is straightforward, concise, and clear. This is due 
    to the explicit consideration of the non-locality of the problem, via the SLE. The present approach permits us to 
    circumvent the need to resort to the (somewhat artificial) infinite-dimensional kinetic equation (Kramer-Moyal expansion;
    \cite{Paul_2013} pp. 43-47, \cite{soong_1973} Sec. 7.2.3), or to the purely analytic treatment of Kolmogorov 
    (see, e.g., \cite{Paul_2013} pp. 47-49, \cite{Gardiner_1985} Sec. 3.4-3.5.2). Here, the standard FPKE results naturally
    from the transformed SLE (after using the NF Theorem) via the assumption of delta correlation for the excitation, by 
    only using equation (\ref{eq:4.11}). This apparently exact result can also be considered as an automatic current-time 
    approximation, since it forcefully excludes (by its delta-type definition) any memory effects carried over by the 
    transition matrix $\Phi^{\Xi }$; see equation (\ref{eq:4.9b}).
\end{remark}

\subsection{The small correlation time (SCT) pdf-evolution equation} \label{sec4.3}

In this section, we derive a SCT pdf-evolution equation for a nonlinear system of RDE, under additive Gaussian coloured 
noise excitation. The meaning of the SCT approximation is that the significant effects of $C_{X_{{{n}_{1}}}^{0}{{\Xi }_{{{n}_{2}}}}}(t)$ 
and $C_{{{\Xi }_{{n}_{1}}}{{\Xi }_{{{n}_{2}}}}}(t,s)$ are concentrated near ${{t}_{0}}$ and near the current time $t$, 
respectively. The derivation presented herein is easy and straightforward, due to the analytic approximations of the 
state-transition matrices $\Phi^{{{\bfv{X}}_{0}}}$ and $\Phi^{\mathbf{\Xi} }.$

Approximating $\Phi^{\mathbf{\Xi} }$ by using the first two terms of the Peano-Baker series (see \hyperref[secB1]{Appendix B}), 
we obtain
\begin{equation}\label{eq:4.13}
    \Phi^{\mathbf{\Xi} }(t;s,\theta )\approx \bfv{I}+\int_{s}^{t}{{{\mathbf{J}}^{\bfv{h}}}\left( \bfv{X}(u;\theta ),u \right)}
    du.
\end{equation}
Introducing a current-time approximation for the integral in the above expression, we get
\begin{equation} \label{eq:4.14}
    \Phi^{\mathbf{\Xi} }(t;s,\theta )\approx \bfv{I}+{{\mathbf{J}}^{\bfv{h}}}
    \left( \bfv{X}(t;\theta ),t \right)(t-s).
\end{equation}
The same treatment applies to the matrix $\Phi^{{{\bfv{X}}^{0}}},$ and the corresponding final approximation is given by 
equation (\ref{eq:4.14}) with $s={{t}_{0}}$. By using these approximations for the matrices $\Phi^{{{\bfv{X}}^{0}}}$
and $\Phi^{\mathbf{\Xi} }$, the expected value operators in equations (\ref{eqs:4.2}) can be easily calculated, resulting 
in their localization at the current time $t$:
\begin{subequations} \label{eqs:4.15}
    \begin{align}
         & {{\mathcal{G}}_{\nu {{n}_{1}}}}(t;{{t}_{0}})={{I}_{\nu {{n}_{1}}}}{{f}_{\bfv{X}(t)}}(\bfv{x})+{{\left( {{\mathbf{J}}^{\bfv{h}}}\left( \bfv{x},t \right) \right)}_{\nu {{n}_{1}}}}{{f}_{\bfv{X}(t)}}(\bfv{x})(t-{{t}_{0}}), \label{eq:4.15a} \\
         & {{\mathcal{G}}_{\nu {{n}_{2}}}}(t;s)={{I}_{\nu {{n}_{2}}}}{{f}_{\bfv{X}(t)}}(\bfv{x})+{{\left( {{\mathbf{J}}^{\bfv{h}}}\left( \bfv{x},t \right) \right)}_{\nu {{n}_{2}}}}{{f}_{\bfv{X}(t)}}(\bfv{x})(t-s). \label{eq:4.15b}
    \end{align}
\end{subequations}
Via the above expressions, the transformed SLE (\ref{eq:3.16}) takes the form:
\begin{align} \label{eq:4.16}
    \frac{\partial {{f}_{\bfv{X}(t)}}(\bfv{x})}{\partial t} + & \sum\limits_{n=1}^{N}{\frac{\partial }{\partial {{x}_{n}}}\left[ \left( {{h}_{n}}\left( \bfv{x}, t \right)+m_{{{\Xi }_{n}}}(t) \right){{f}_{\bfv{X}(t)}}(\bfv{x}) \right]}= \nonumber                                                                                                                                                        \\
                                                            = & \sum\limits_{n=1}^{N}{\sum\limits_{\nu =1}^{N}{\frac{\partial }{\partial {{x}_{n}}}\frac{\partial }{\partial {{x}_{\nu }}}\left[ \left( \mathcal{D}_{\nu n}^{{\mathbf{X}_{0}}\mathbf{\Xi} }\left[ \bfv{x},t \right]+\mathcal{D}_{\nu n}^{\mathbf{\Xi}\mathbf{\Xi} }\left[ \bfv{x},t \right] \right){{f}_{\bfv{X}(t)}}(\bfv{x}) \right]}},
\end{align}
where
\begin{subequations} \label{eqs:4.17}
    \begin{align}
         & \mathcal{D}_{\nu n}^{{{X}_{0}}\mathbf{\Xi} }\left[ \bfv{x},t \right]=\sum\limits_{{{n}_{1}}=1}^{N}{C_{X_{{{n}_{1}}}^{0}{{\Xi }_{n}}}(t)\left( {{I}_{\nu {{n}_{1}}}}+{{\left( {{\mathbf{J}}^{\bfv{h}}}\left( \bfv{x},t \right) \right)}_{\nu {{n}_{1}}}}(t-{{t}_{0}}) \right)},\label{eq:4.17a}                        \\
         & \mathcal{D}_{\nu n}^{\mathbf{\Xi\Xi} }\left[ \bfv{x},t \right]=\sum\limits_{{{n}_{2}}=1}^{N}{\int\limits_{{{t}_{0}}}^{t}{C_{{{\Xi }_{n}}{{\Xi }_{{{n}_{2}}}}}(t,s)\left( {{I}_{\nu {{n}_{2}}}}+{{\left( {{\mathbf{J}}^{\bfv{h}}}\left( \bfv{x},t \right) \right)}_{\nu {{n}_{2}}}}(t-s) \right)ds}}. \label{eq:4.17b}
    \end{align}
\end{subequations}
To compare equation (\ref{eq:4.16}) with other existing ones, first we have to impose the assumptions of zero-mean 
excitations, $m_{{{\Xi }_{n}}}(t)=0$, and $\mathcal{D}_{\nu n}^{{{X}_{0}}\Xi }\left[ \bfv{x},t \right]=0$ (initial
data uncorrelated with the excitation). Terms containing the above two functions do not appear in existing literature, 
probably because the extended NF theorem, equation (\ref{eq:3.5}), was not available. Further, a comparison of equation 
(\ref{eq:4.16}) with existing genFPKE is not straightforward since in the literature the results are usually given for 
the multiplicative excitation, leaving the additive case as a special one (obtained after some analytical manipulations). 
It can be shown that the additive excitation versions of the pdf-evolution equation derived by San Miguel and Sancho in 
1980 \cite{San_Miguel_1980}, and Dekker in 1982 \cite{Dekker_1982}, degenerates to equation (\ref{eq:4.16}) . 
In the latter work, the equation has been derived in a more laborious way, via the correlation time expansion method. 
Moreover, equation (\ref{eq:4.16})  can be also deduced from the results of references \cite{Fox_1983,Garrido_1982} 
obtained by the ordered cumulant expansion method. In comparison with our derivation, the main methodological difference 
lies in the use of the analytic approximation of the Volterra derivatives $V_{\nu {{n}_{2}}}^{\Xi (s)}(t)$. It seems that 
our approach provides an easier derivation for equation (\ref{eq:4.16}), pertaining to the additive excitation by colored noise.

\section{A novel pdf-evolution equation for N-dimensional nonlinear systems} \label{sec5}

In this section we present the derivation of our novel response pdf-evolution equation, announced in equation (\ref{eq:2.6}), 
which is based on the separation of the (instantaneous) mean value of the response, from its random fluctuations.
This separation is reflected at the level of the state-transition matrices $\Phi^{{{\bfv{X}}_{0}}}$ and $\Phi^{\mathbf{\Xi} }$, 
via an appropriate matrix decomposition. To motivate the method and facilitate its understanding, it is useful to highlight 
the main steps first in the case of a scalar RDE (one-dimensional case). The latter case has been presented in \cite{Mamis_2019}, 
by using a more complicated procedure.

\subsection{ Motivation: Applying the novel closure to a scalar RDE} \label{sec5.1}

In the scalar case, the transition matrix $\Phi^{\Xi }$ has the form of an exponential function, thus $\mathcal{G}(t;s)$, 
equation (\ref{eq:4.2b}), is now a scalar quantity, which takes the form:
\begin{equation}
    \mathcal{G}(t;s)\equiv{{\mathbb{E}}^{\theta }}\left[ \delta (x-X(t;\theta
    ))\exp \left( \int_{s}^{t}{{h}'(X(u;\theta ),u)du}\right) \right]. \label{eq:5.1}
\end{equation}
That is, the exponential function plays the role of the one-dimensional state-transition matrix, $\Phi^{\Xi }(t;s,\theta )$. 
The main idea is to decompose ${h}'$ (which now is a scalar function) into its mean and fluctuating (around the mean value) part:
\begin{equation}
    {h}'(X(t;\theta ),t)=\mathbb{E}^{\theta }\left[{h}'\left( X(t;\theta
        ),t \right) \right]+{{\Delta }_{{h}'}}\left( X(t;\theta ),t \right). \label{eq:5.2}
\end{equation}
where 
${{\Delta }_{{h}'}}\left( X(t;\theta ),t \right)={h}'\left( X(t;\theta),t \right)-\mathbb{E}^{\theta }\left[{h}'\left( X(t;\theta),t \right) \right]$. 
We note that the instantaneous mean value of ${h}'$ engages a generalized response moment,
\addtocounter{equation}{1}
\begin{equation}
    \mathbb{E}^{\theta }\left[{h}'\left( X(t;\theta ),t \right) \right]=
    \int_{\mathbb{R}}{{h}'\left( x,t \right){{f}_{X(t)}}(x)dx}, \tag{5.3a}\label{eq:5.3a}
\end{equation}
which subsequently treated without any further simplification. This quantity is explicitly dependent on the unknown pdf 
at the time instant $t$. To highlight this dependence, we introduce the notation,
\begin{equation}
    {{R}_{{h}'}}\left[{{f}_{X(t)}}(\cdot ),t \right]:=\mathbb{E}^{\theta
    }\left[{h}'\left( X(t;\theta ),t \right) \right]. \tag{5.3b}\label{eq:5.3b}
\end{equation}
Substituting the decomposition (\ref{eq:5.2}) into the exponential function in equation
(\ref{eq:5.1}), we obtain,
\begin{align} \label{eq:5.4}
     & \exp \left( \int_{s}^{t}{{h}'(X(u;\theta ),u)du}\right)= \nonumber \\
     & =\exp \left( \int_{s}
    ^{t}{{{R}_{{h}'}}\left[ {{f}_{X(u)}}(\cdot ),u \right]du}\right)\exp
    \left( \int_{s}^{t}{{{\Delta }_{{h}'}}\left( X(u;\theta ),u \right)du}\right).
\end{align}
Introducing a current-time approximation in the second integral of the right-hand side of equation (\ref{eq:5.4}), 
which contains only the fluctuating part ${{\Delta }_{{h}'}}\left( X(t;\theta ),t \right)$, we obtain
\begin{equation}
    \exp \left( \int_{s}^{t}{{{\Delta }_{{h}'}}\left( X(u;\theta ),u \right)du}
    \right)\approx \exp \left({{\Delta }_{{h}'}}\left( X(t;\theta ),t \right)
    (t-s) \right). \label{eq:5.5}
\end{equation}
Substituting equation (\ref{eq:5.4}) into equation (\ref{eq:5.1}), utilizing equation (\ref{eq:5.5}), and factoring 
out of the expected value operator the deterministic exponential term, we obtain
\begin{align} \label{eq:5.6}
    \mathcal{G}(t;s) & =\exp \left( \int_{s}^{t}{{{R}_{{{h}'}}}[{{f}_{X(u)}}(\cdot ),u]du}\right) \times  & \nonumber \\
                     & \ \quad \times   {{\mathbb{E}}^{\theta }}\left[ \exp \left({{\Delta }_{{h}'}}\left( X(t;\theta ),t \right)(t-s) \right)\delta (x-X(t;\theta )) \right]= \nonumber  \\
                     & =\int\limits_{\mathbb{R}}{\delta (x-u)\exp\left(\Delta\left( u,t \right)(t-s) \right){{f}_{X(t)}}(u)du} = \nonumber \\
                     & =\exp \left( \int_{s}^{t}{{{R}_{{{h}'}}}[{{f}_{X(u)}}(\cdot ),u]du}\right)\exp \left({{\Delta }_{{h}'}}\left( x,t \right)(t-s) \right){{f}_{X(t)}}(x). 
\end{align}
According to the second equality in the above equation, $\Delta(x,t)$ is the function obtained by $\Delta (X(t;\theta),t)$ 
when replacing the random process $X(t;\theta )$ by its corresponding state variable $x$.The term $\mathcal{G}(t;{{t}_{0}})$ 
is obtained by setting $s={{t}_{0}}$ in equation (\ref{eq:5.6}). Simplifying the SLE (\ref{eq:3.16}) for $N=1$, and 
substituting $\mathcal{G}(t;{{t}_{0}})$ and $\mathcal{G}(t;s)$ in the latter, we obtain equation (\ref{eq:2.6}), for 
$n=\nu =1$. The coefficients $\mathcal{D}^{{{X}_{0}}\Xi (\cdot)}$ and $\mathcal{D}^{\Xi \Xi }$, take now the form:
\begin{subequations} \label{eqs:5.7}
    \begin{align}
         & \mathcal{D}^{{{X}_{0}}\Xi }\left[{{f}_{X(\cdot )}}(\cdot );x,t \right]= \nonumber                                                                                                                         \\
         & = C_{X^{0}\Xi }(t)\exp \left( \int_{{{t}_{0}}}^{t}{{{R}_{{{h}'}}}[{{f}_{X(u)}}(\cdot ),u]du}\right)\exp \left({{\Delta }_{{h}'}}\left( x,t \right)(t-{{t}_{0}}) \right), \label{eq:5.7a}                  \\
         & \mathcal{D}^{\Xi \Xi }\left[{{f}_{X(\cdot )}}(\cdot );x,t \right] = \nonumber                                                                                                                       \\
         & =\int\limits_{{{t}_{0}}}^{t}{C_{\Xi \Xi }(t,s)\exp \left( \int_{s}^{t}{{{R}_{{{h}'}}}[{{f}_{X(u)}}(\cdot ),u]du} \right)\exp \left( {{\Delta }_{{h}'}}\left( x,t \right)(t-s) \right)ds}. \label{eq:5.7b}
    \end{align}
\end{subequations}
Equation (\ref{eq:2.6}) (with coefficients (\ref{eqs:5.7})) for the one-dimensional problem, has been first derived in 
\cite{Mamis_2019}, by using a more laborious methodology, based on Volterra series expansions. It is worthwhile to note 
that it has been extensively tested numerically \cite{Mamis_2019} Sec. 4, \cite{Athanassoulis_2018}, proving itself a 
very capable model for calculating the response pdf of scalar RDEs, far beyond the SCT excitation regime, up to 
non-dimensional correlation time 3 or more.

\begin{figure} 
    \centering
    \includegraphics[width=0.7\textwidth]{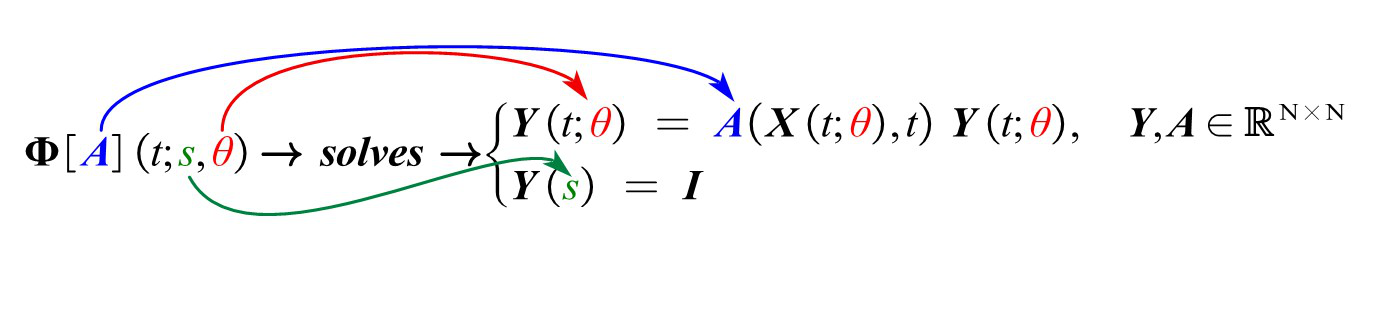}
    \caption{ \centering Visual explanation of the notation
        $\Phi [\mathbf{A}](t;s,\theta )$}
    \label{fig:2}
\end{figure}

The extension of the above approach to a system of RDEs requires manipulations of the state-transition matrices $\Phi^{{{\bfv{X}}_{0}}}$ 
and $\Phi^\mathbf{\Xi}$ and, also, the introduction of additional state-transition matrices associated with IVPs different than 
the variational problems (\ref{eqs:3.10}) and (\ref{eqs:3.14}). It is, therefore, expedient to enrich the notation used 
for the state-transition matrix $\Phi =\Phi (t;{{t}_{0}},\theta ),$ associated with the IVP
\begin{equation*}
    {{\mathbf{\dot{y}}}_{[N\times 1]}}(t;\theta )={{\mathbf{A}}_{[N\times N]}}
    (\bfv{X}(t;\theta ),t){{\mathbf{y}}_{[N\times 1]}}(t;\theta ), \ \ \  \mathbf{y}
    ({{t}_{0}};\theta )={{\mathbf{y}}^{0}}(\theta ), \tag{P}\label{eq:P}
\end{equation*}
as follows: $\Phi =\Phi [\mathbf{A}](t;{{t}_{0}},\theta )$. Now, the underlying system-matrix $\mathbf{A}$ of the IVP is 
also indicated in the notation. The function $\mathbf{y}(t;\theta )=\Phi [\mathbf{A}](t;{{t}_{0}},\theta){{\mathbf{y}}^{0}}(\theta )$ 
solves problem (P) and, thus, by definition, the matrix $\mathbf{Y}(t;\theta )=\Phi [\mathbf{A}](t;{{t}_{0}},\theta )$
satisfies the matrix differential equation $\mathbf{\dot{Y}}(t;\theta )=\mathbf{A}(\bfv{X}(t;\theta ),t)\mathbf{Y}(t;\theta )$, 
$\mathbf{Y}({{t}_{0}})=\bfv{I}$. A visual interpretation of the dependencies in the notation $\Phi [\mathbf{A}](t;s,\theta )$ 
is shown in \hyperref[fig:2]{Figure 2}.
Using the above notation, equation (\ref{eq:5.4}), for the scalar case, can be rephrased, in terms of one-dimensional 
state-transition matrices, in the form
\begin{equation*}
    \Phi^{\Xi }[{h}'](t;s,\theta )=\Phi [{{R}_{{{h}'}}}](t;s,\theta )\Phi
    [{{\Delta }_{{h}'}}](t;s,\theta ).
\end{equation*}
This equation implies a decomposition of the state-transition matrix, associated with the additive decomposition (\ref{eq:5.2}) 
and the exponential property ${{e}^{a(t)+b(t)}}={{e}^{a(t)}}{{e}^{b(t)}}.$ In the multidimensional case, the corresponding 
decomposition for exponential matrices does not hold in general 
(${{e}^{\mathbf{A}(t)+\mathbf{B}(t)}}\ne{{e}^{\mathbf{A}(t)}}{{e}^{\mathbf{B}(t)}}$) and further considerations should be made.

\subsection{Generalization of the novel closure to the N-dimensional case}\label{sec5.2}

In the multidimensional case, the system matrix of both variational problems, (\ref{eqs:3.10}) and (\ref{eqs:3.14}), is 
the Jacobian matrix of the vector function $\bfv{h}(\bfv{X}(t;\theta ),t)=$ $({{h}_{1}},\cdots ,{{h}_{N}})(\bfv{X}(t;\theta ),t)$,
see equation (\hyperref[eq:2.1a]{2.1a}), denoted as
\begin{equation}
    {{\mathbf{J}}^{\bfv{h}}}:={{\mathbf{J}}^{\bfv{h}}}\left( \bfv{X}(t;
        \theta ),t \right)=\nabla \bfv{h}(\bfv{X}(t;\theta ),t). \label{eq:5.8}
\end{equation}
In this case, we again decompose the matrix ${{\mathbf{J}}^{\bfv{h}}}(\bfv{X}(t;\theta ),t)$, in its mean value and its 
random fluctuations (around the mean value):
\begin{equation}
    {{\mathbf{J}}^{\bfv{h}}}\left( \bfv{X}(t;\theta ),t \right):=\mathbf{R}
    \left[{{f}_{\bfv{X}(t)}}(\cdot ),t \right]+\mathbf{\Delta }\left( \bfv{X}
        (t;\theta ),t \right), \label{eq:5.9}
\end{equation}
where
\begin{align}
     & \mathbf{R}(t)\equiv \mathbf{R}\left[{{f}_{\bfv{X}(t)}}(\cdot ),t \right]={{\mathbb{E}}^{\theta }}\left[{{\mathbf{J}}^{\bfv{h}}}\left( \bfv{X}(t;\theta ),t \right) \right]
     =\int_{{{\mathbb{R}}^{N}}}{{{\mathbf{J}}^{\bfv{h}}}\left( x,t \right){{f}_{\bfv{X}(t)}}(\bfv{x})d\bfv{x}}, \label{eq:5.10} \\
     & \mathbf{\Delta }\left( \bfv{X}(t;\theta ),t \right)={{\mathbf{J}}^{\bfv{h}}}\left( \bfv{X}(t;\theta ),t \right)-\mathbf{R}\left[{{f}_{\bfv{X}(t)}}(\cdot ),t \right]. \label{eq:5.11}
\end{align}
Following the steps explained in the scalar case, we need to reflect the decomposition (\ref{eq:5.9}) on the level of 
the state-transition matrix $\Phi^{\bfv{\Xi} }[{{\mathbf{J}}^{\bfv{h}}}]$. To this end, we seek for a decomposition of 
$\Phi^{\mathbf{\Xi}}[{{\mathbf{J}}^{\bfv{h}}}]$, in terms of the transition-matrix
\begin{equation}
    \Phi [\mathbf{R}](t;s)\equiv \Phi [\mathbf{R}[{{f}_{\bfv{X}(\cdot )}}(\cdot
            ),\cdot ]](t;s). \label{eq:5.12}
\end{equation}
Herein we will make use of the simpler notation $\Phi [\mathbf{R}](t;s)$ whenever the dependence of $\mathbf{R}$ on the 
density field ${{f}_{\bfv{X}(\cdot )}}(\cdot )$ is not important to be pointed out. It can be proved (see, e.g., \cite{Frazer_1938}
p. 219) that, under the additive decomposition (\ref{eq:5.9}), the following multiplicative decomposition holds true:
\addtocounter{equation}{1}
\begin{equation}
    \Phi^{\mathbf{\Xi} }[{{\mathbf{J}}^{\bfv{h}}}](t;s,\theta )=\Phi [\mathbf{R}
    ](t;s)\Phi [\mathbf{B}](t;s,\theta ), \tag{5.13a} \label{eq:5.13a}
\end{equation}
where the matrix $\mathbf{B}(t,s;\theta )$ is expressed as:
\begin{equation}
    \mathbf{B}(t,s;\theta )={{\Phi }^{-1}}[\mathbf{R}](t;s)\mathbf{\Delta }\left
    ( \bfv{X}(t;\theta ),t \right)\Phi [\mathbf{R}](t;s). \tag{5.13b}\label{eq:5.13b}
\end{equation}
The decomposition (\ref{eq:5.13a}) is the analog to equation (\ref{eq:5.4}), in the multidimensional case. By substituting 
equation (\ref{eq:5.13a}) into the non-local term (\ref{eq:4.2b}) and factoring the deterministic matrix $\Phi [\mathbf{R}](t;s)$ 
out of the mean-value operator, we obtain
\begin{equation}
    \mathcal{G}(t;s)=\Phi [\mathbf{R}](t;s)\mathbb{E}^{\theta }\left[ \delta
        (\bfv{x}-\bfv{X}(t;\theta ))\Phi [\mathbf{B }](t;s,\theta ) \right].
    \label{eq:5.14}
\end{equation}
In equation (\ref{eq:5.14}), $\mathcal{G}(t;s)$ represents the matrix whose components ${{\mathcal{G}}_{\nu {{n}_{2}}}}(t;s)$ 
are given by equation (\ref{eq:4.2b}). To implement our approximation scheme, we need an explicit representation of the matrix
$\Phi [\mathbf{B}](t;s,\theta )$. To that end, we use the Magnus expansion of $\Phi [\mathbf{B }](t;s,\theta )$ 
(see \hyperref[secB1]{Appendix B}, equations (\ref{eq:B3}) and (\ref{eq:B4})), keeping only its first term, obtaining
\addtocounter{equation}{1}
\begin{equation}
    \Phi [\mathbf{B}](t;s,\theta )\cong \exp \left( \int_{s}^{t}{\mathbf{B}(u,s;\theta )du}
    \right). \tag{5.15a}\label{eq:5.15a}
\end{equation}
Higher order terms of Magnus series can also be considered, leading to more complicated approximations. This issue will 
be further studied in future work. Combining equations (\ref{eq:5.13b}) and (\ref{eq:5.15a}), we get
\begin{equation}
    \Phi [\mathbf{B}](t;s,\theta )\cong \exp \left( \int_{s}^{t}{{{\Phi }^{-1}}[\mathbf{R}](t;s)\mathbf{\Delta }\left( \bfv{X}(u;\theta ),u \right)\Phi [\mathbf{R}](t;s)du}
    \right). \tag{5.15b}\label{eq:5.15b}
\end{equation}
Taking into account that the elements of the matrix $\Phi [\mathbf{B }](t;s,\theta )$ are eventually multiplied by the
elements of $\mathbf{C}_{\mathbf{\Xi\Xi}}(t,s)$ (see equation (\ref{eq:3.16})), we understand that we need an approximation 
of the integral, in the right-hand side of equation (\ref{eq:5.15b}), which should become exact as $s$ tends to $t$. The
simplest approximation of this kind is the current-time approximation of the integrand (at $t$), under which we perform the 
integration, obtaining
\begin{align}
    \Phi [\mathbf{B}](t;s,\theta ) & \cong \exp \left({{\Phi }^{-1}}[\mathbf{R}](t;s)\mathbf{\Delta }\left( \bfv{X}(t;\theta ),t \right)\Phi [\mathbf{R}](t;s)(t-s) \right)= \nonumber   \\
                                   & ={{\Phi }^{-1}}[\mathbf{R}](t;s)\exp \left( \mathbf{\Delta }\left( \bfv{X}(t;\theta ),t \right)(t-s) \right)\Phi [\mathbf{R}](t;s). \label{eq:5.16}
\end{align}
Again, better approximation can be considered by using Taylor expansion. Now, by substituting equation (\ref{eq:5.16}) 
into equation (\ref{eq:5.14}), we obtain
\begin{align}
    \mathcal{G}(t;s) & =\mathbb{E}^{\theta }\left[ \delta (\bfv{x}-\bfv{X}(t;\theta ))\exp \left( \mathbf{\Delta }\left( \bfv{X}(t;\theta ),t \right)(t-s) \right) \right]\Phi [\mathbf{R}](t;s)=\nonumber \\
                     & =\exp \left( \mathbf{\Delta }\left( \bfv{x},t \right)(t-s) \right)\Phi [\mathbf{R}](t;s){{f}_{\bfv{X}(t)}}(\bfv{x}), \label{eq:5.17}
\end{align}
where the second equality is justified as follows:
\begin{align}
    \mathbb{E}^{\theta } & \left[ \delta (\bfv{x}-\bfv{X}(t;\theta ))\exp \left( \mathbf{\Delta}\left( \bfv{X}(t;\theta ),t \right)(t-s) \right) \right]= \nonumber
    \\
                         & =\int\limits_{{{\mathbb{R}}^{N}}}{\delta (\bfv{x}-\mathbf{u})\exp \left( \mathbf{\Delta }\left( \mathbf{u},t \right)(t-s) \right){{f}_{\bfv{X}(t)}}(\mathbf{u})d\mathbf{u}} = \nonumber \\
                         & = \exp \left( \mathbf{\Delta }\left( \bfv{x},t \right)(t-s) \right){{f}_{\bfv{X}(t)}}(\bfv{x}). \label{eq:5.18}
\end{align}
That is, $\mathbf{\Delta }\left( \bfv{x},t \right)$ is the matrix obtained by $\mathbf{\Delta }\left( \bfv{X}(t;\theta ),t \right)$ 
when replacing the random vector $\bfv{X}(t;\theta )$ by a deterministic vector $\bfv{x}$. The matrix $\mathcal{G}(t;{{t}_{0}})$, 
whose components appear in equation (\ref{eq:4.2a}), is obtained by equation (\ref{eq:5.17}), after setting $s={{t}_{0}}$. 
Note that equation (\ref{eq:5.17}) is the multidimensional analog of equation (\ref{eq:5.6}), of the scalar case. 
Last, by substituting $\mathcal{G}(t;s)$ and $\mathcal{G}(t;{{t}_{0}})$ into the transformed SLE (\ref{eq:3.16}), we 
obtain the new pdf-evolution equation (\ref{eq:2.6}), along with the exact forms of the diffusion coefficients, namely
\begin{subequations} \label{eqs:5.19}
    \begin{align}
         & \mathcal{D}_{\nu n}^{{{X}_{0}}\mathbf{\Xi} }\left[{{f}_{\bfv{X}(\cdot )}}(\cdot );x,t \right]= \nonumber
        \\
         & =\sum\limits_{{{n}_{1}}=1}^{N}{\sum\limits_{k=1}^{N}{C_{X_{{{n}_{1}}}^{0}{{\Xi }_{n}}}(t){{\Phi }_{k{{n}_{1}}}}[\mathbf{R}[{{f}_{\bfv{X}(\cdot )}}(\cdot ),\cdot ]](t;{{t}_{0}}){{\left( \exp \left( \mathbf{\Delta }(\bfv{x},t)(t-{{t}_{0}}) \right) \right)}_{\nu k}}}}, \label{eq: 5.19a}                       \\
         & \mathcal{D}_{vn}^{\mathbf{\Xi\Xi} }\left[{{f}_{\bfv{X}(\cdot )}}(\cdot );x,t \right] = \nonumber
        \\
         & =\sum\limits_{{{n}_{2}}=1}^{N}{\sum\limits_{k=1}^{N}{\int\limits_{{{t}_{0}}}^{t}{C_{{{\Xi }_{n}}{{\Xi }_{{{n}_{2}}}}}(t,s){{\Phi }_{k{{n}_{2}}}}[\mathbf{R}[{{f}_{\bfv{X}(\cdot )}}(\cdot ),\cdot ]](t;s)}{{\left( \exp \left( \mathbf{\Delta }(\bfv{x},t)(t-s) \right) \right)}_{\nu k}}\,ds}}. \label{eq: 5.19b}
    \end{align}
\end{subequations}
Equations (\ref{eqs:5.19}) specify the expressions of the diffusion coefficients for our novel one-time pdf-evolution 
equation (\ref{eq:2.6}). 

\subsection{Comparison with the SCT approximation}\label{sec5.3}

In this section we compare our novel pdf-evolution equation, which is (\ref{eq:2.6}) with diffusion coefficients given 
by equations (\ref{eqs:5.19}) (called novel genFPKE, or ngFPK, for conciseness), with the SCT approximation, which is equation 
(\ref{eq:2.6}) with diffusion coefficients given by equations (\ref{eqs:4.17}). In this conjunction, we consider
zero-mean excitation, $m_{{{\Xi }_{n}}}(t)=0$, and uncorrelated excitation to the initial value, 
$\mathcal{D}_{\nu n}^{{{X}_{0}}\mathbf{\Xi} }\left[ \bfv{x},t \right]=0$. Then, the two equations have identical drift 
part and differ only in the diffusion coefficients matrix $\mathcal{D}^{\mathbf{\Xi\Xi}}$. To compare the two cases, it 
is expedient to rewrite the components $\mathcal{D}_{vn}^{\mathbf{\Xi\Xi}}$ of the diffusion matrix, equations (\ref{eq: 5.19b}) 
and (\ref{eq:4.17b}), in the following form:
\begin{equation}
    \mathcal{D}_{vn}^{\mathbf{\Xi\Xi}}\left[{{f}_{\bfv{X}(\cdot )}}(
        \cdot );\bfv{x},t \right]=\sum\limits_{{{n}_{2}}=1}^{N}{\int\limits_{{{t}_{0}}}^{t}{ C_{{{\Xi }_{n}}{{\Xi }_{{{n}_{2}}}}}(t,s) \left\{ \begin{aligned}&D_{\nu {{n}_{2}}}^{\operatorname{ngFPK}}[{{f}_{\bfv{X}(\cdot )}}(\cdot );\bfv{x},t,s] \\&D_{\nu {{n}_{2}}}^{\operatorname{SCT}}(\bfv{x},t,s) \\\end{aligned} \right\} } }
    ds, \label{eq:5.20}
\end{equation}
where
\addtocounter{equation}{1}
\begin{equation}
      D_{\nu {{n}_{2}}}^{\operatorname{ngFPK}}({{f}_{\bfv{X}(\cdot )}}(\cdot );
    \bfv{x},t,s) = \nonumber                                                                                                                                                                              \\
      \sum\limits_{k=1}^{N}{{{\left( \exp \left( \mathbf{\Delta }\left( \bfv{x},t \right)(t-s) \right) \right)}_{\nu k}}{{\Phi }_{k{{n}_{2}}}}[\mathbf{R}[{{f}_{\bfv{X}(\cdot )}}(\cdot ),\cdot ]](t;s)}
    , \tag{5.21a}\label{eq:5.21a}
\end{equation}
and
\begin{equation}
    D_{\nu {{n}_{2}}}^{\operatorname{SCT}}(\bfv{x},t,s)={{I}_{\nu {{n}_{2}}}}
    +{{\left( {{\mathbf{J}}^{\bfv{h}}}\left( \bfv{x},t \right) \right)}_{\nu {{n}_{2}}}}
    (t-s). \tag{5.21b}\label{eq:5.21b}
\end{equation}

For both ngFPKE and SCT, the diffusion coefficients are of the same integral form (\ref{eq:5.20}), differing only 
ith respect to the matrix in bracket $\{\}$, on the right-hand side. We will now show how the matrix 
$\mathbf{D}^{\operatorname{ngFPK}}$ is reduced to the corresponding matrix $\mathbf{D}^{\operatorname{SCT}}$, by means 
of a series of approximations. The first step is to approximate $\Phi [\mathbf{R}[{{f}_{\bfv{X}(\cdot )}}(\cdot ),\cdot ]](t;s)$ 
(in equations (\ref{eq:5.21a})) via the first two terms of its Peano-Baker series expansion, namely,
\addtocounter{equation}{1}
\begin{equation}
    \Phi [\mathbf{R}[{{f}_{\bfv{X}(\cdot )}}(\cdot ),\cdot ]](t;s)\approx \bfv{I}
    +\int_{s}^{t}{\mathbf{R}[{{f}_{\bfv{X}(u)}}(\cdot ),u]du}.\tag{5.22a}\label{eq:5.22a}
\end{equation}
Then, we introduce a current-time approximation to the integral in the above equation, obtaining
\begin{equation}
    \Phi [\mathbf{R}[{{f}_{\bfv{X}(\cdot )}}(\cdot ),\cdot ]](t;s)\approx \bfv{I}
    +\mathbf{R}[{{f}_{\bfv{X}(t)}}(\cdot ),t](t-s). \tag{5.22b}\label{eq:5.22b}
\end{equation}
Since in the SCT derivation, the decomposition (\ref{eq:5.9}) of the matrix ${{\mathbf{J}}^{\bfv{h}}}$ is not used, we 
have to come back to equation (\ref{eq:5.17}) and repeat the averaging by using equation (\ref{eq:5.11}):
\begin{align*}
     & \mathbb{E}^{\theta }\left[ \delta (\bfv{x}-\bfv{X}(t;\theta ))\exp \left( \mathbf{\Delta }\left( \bfv{X}(t;\theta ),t \right)(t-s) \right) \right]= \nonumber                                                                               \\
     & =\mathbb{E}^{\theta }\left[ \delta (\bfv{x}-\bfv{X}(t;\theta ))\exp \left( \left({{\mathbf{J}}^{\bfv{h}}}\left( \bfv{X}(t;\theta ),t \right)-\mathbf{R}\left[{{f}_{\bfv{X}(t)}}(\cdot ),t \right] \right)(t-s) \right) \right] =  \nonumber \\
     & =\exp \left( \left({{\mathbf{J}}^{\bfv{h}}}\left( \bfv{x},t \right)-\mathbf{R}\left[{{f}_{\bfv{X}(t)}}(\cdot ),t \right] \right)(t-s) \right){{f}_{\bfv{X}(t)}}(\bfv{x}). \label{eq:ex}
\end{align*}
This means that, in all manipulations made after calculating the averaging [in equation (\ref{eq:5.17})], it is 
permissible to use the relation
\begin{equation*}
    \exp \left( \mathbf{\Delta }\left( \bfv{x},t \right)(t-s) \right)=\exp \left
    ( \left({{\mathbf{J}}^{\bfv{h}}}(\bfv{x},t)-\mathbf{R}\left[{{f}_{\bfv{X}(t)}}
        (\cdot ),t \right] \right)(t-s) \right).
\end{equation*}
It should be noted that this equation cannot be inferred by the decomposition (\ref{eq:5.9}) [or (\ref{eq:5.11})]. 
It is a consequence of the specific localization (via the delta function $\delta (\bfv{x}-\bfv{X}(t;\theta ))$) performed in
equation (\ref{eq:5.17}) via (\ref{eq:5.18}).

Now, by expanding the exponential matrix $\exp \left( \mathbf{\Delta }\left(\bfv{x},t \right)(t-s) \right)$ in power series 
keeping only the two first terms, and exploiting equation (\ref{eq:5.22b}), we calculate the corresponding approximation 
of matrix $\mathbf{D}^{\operatorname{ngFPK}}$, equation (\ref{eq:5.21a}), as follows
\begin{align} \label{eq:5.23}
     & \mathbf{D}^{\operatorname{ngFPK}}({{f}_{\bfv{X}(\cdot )}}(\cdot );\bfv
    {x},t,s)\approx \bfv{I} +{{\mathbf{J}}^{\bfv{h}}}\left( \bfv{x},t \right
    )(t-s) + \nonumber                                                          \\
     & + \mathbf{R}[{{f}_{\bfv{X}(t)}}(\cdot ),t]\left({{\mathbf{J}}^{\bfv{h}}}
    \left( \bfv{x},t \right)-\mathbf{R}[{{f}_{\bfv{X}(t)}}(\cdot ),t] \right)
    {{(t-s)}^{2}}.
\end{align}
Neglecting the second order term with respect to $(t-s)$, equation (\ref{eq:5.23}) becomes equation (\ref{eq:5.21b}), 
which means that the novel genFPKE reduces to the SCT equation.

\begin{remark}
    It is interesting to notice that, to derive the SCT approximate equation from our novel genFPKE, we have to neglect all 
    memory effects included in $\Phi [\mathbf{R}[{{f}_{\bfv{X}(\cdot )}}(\cdot ),\cdot ]]$, to simplify the exponential 
    matrix $\exp \left( \mathbf{\Delta }\left( \bfv{x},t \right)(t-s) \right)$, and to neglect a second-order term, 
    remaining after these simplifications. These facts are indicative of the amount of additional information included 
    in the novel genFPK equation in comparison with the classical SCT approximation.

\end{remark}

\begin{remark}
    Further, it is easy to see that the novel genFPKE equation can be reduced to the pdf-evolution equation (\ref{eq:4.4}), in the 
    case of linear systems, and to the standard FPKE (\ref{eq:4.12}), in the case of white noise excitation.
\end{remark}


\section{First numerical results}\label{sec6}

To test the validity of the proposed novel genFPKE (\ref{eq:2.6}), numerical simulations have been performed and briefly 
presented in this section. We consider the case of a randomly excited Duffing oscillator, and we compare its response pdf 
as obtained by solving the SCT equation, the novel genFPKE and Monte Carlo (MC) simulations. The numerical method used for 
solving genFPKE employs: i) a partition of unity finite element method (PUFEM) for the discretization in the state space 
\cite{Babuska_1996,Babuska_1997}, ii) a Bubnov-Galerkin technique for deriving ODEs governing the evolution of the pdf and iii) a 
Crank-Nicolson scheme for solving the obtained ODEs in the time domain. Similar numerical methods have been used for 
solving the stationary FPKE \cite{Kumar_2006,Kumar_2009} and the transient FPKE \cite{Kumar_2010,Sun_2015}. Details on 
the numerical treatment will be presented elsewhere. An additional test of validity of our numerical scheme and
simulation procedure is presented in the Supplementary Material, where results concerning the solution of the pdf-evolution 
equation corresponding to a linear oscillator are given. Since, in this case, analytical solution of the pdf evolution 
is available, we are able to test the efficiency of the corresponding MC simulations and the quality of the obtained 
PUFEM approximation, giving a first justification of our numerical treatment.

\subsection{A Duffing oscillator and its excitation}\label{sec6.1}
Consider the randomly excited Duffing oscillator:
\addtocounter{equation}{1}
\begin{equation}
    m\ddot{X}(t;\theta )+b\dot{X}(t;\theta )+{{\eta }_{1}}X(t;\theta )+{{\eta }_{3}}{{X}^{3}}(t;\theta )=\Xi (t;\theta ), \tag{6.1a}\label{eq:6.1a}
\end{equation}
\begin{equation}
    X({{t}_{0}};\theta )={{X}^{0}}(\theta ), \ \ \ \dot{X}({{t}_{0}};\theta )={{\dot{X}}^{0}}(\theta ). \tag{6.1b,c}\label{eq:6.1b,c}
\end{equation}
In equations (6.1), we assume $b>0$, ${{\eta }_{1}}<0$ (bistable case), and ${{\eta }_{3}}>0$, to ensure global stability 
of the oscillator. The excitation is written in the form $\Xi (t;\theta )={{\xi }_{0}}{{\tilde{\Xi }}_{0}}(t;\theta ) +$
${{m}_{\Xi }}(t)$, where the coefficient ${{\xi }_{0}}$ has dimensions of force, leaving ${{\tilde{\Xi }}_{0}}$ a 
dimensionless Gaussian excitation with zero-mean, ${{m}_{{{{\tilde{\Xi }}}_{0}}}}(t)=0$ and unitary variance 
$\sigma_{{{{\tilde{\Xi }}}_{0}}}^{2}=1$. In this way, the variance of the excitation $\Xi$ is equal to $\sigma_{\Xi }^{2}=\xi_{0}^{2}$.
Introducing the dimensionless variables $\tilde{t}=t\sqrt{{{\eta }_{1}}/m}$, $\tilde{X}=X\sqrt{{{\eta }_{3}}/|{{\eta }_{1}}|}$, 
${{\tilde{X}}_{0}}={{X}_{0}}\sqrt{{{\eta }_{3}}\mathbf{/}|{{\eta }_{1}}|}$ and ${{\dot{\tilde{X}}}_{0}}={{\dot{X}}_{0}}
\sqrt{m{{\eta }_{3}}}\mathbf{/}|{{\eta }_{1}}|$, equations (\hyperref[eq:6.1a]{6.1a,b,c}) take the following non-dimensional form:
\addtocounter{equation}{1}
\begin{equation}
    \ddot{\tilde{X}}(\tilde{t};\theta )+2\zeta \dot{\tilde{X}}(\tilde{t};\theta)-\tilde{X}(\tilde{t};\theta )+{{\tilde{X}}^{3}}(\tilde{t};\theta )=\tilde{\Xi }(\tilde{t};\theta ), \tag{6.2a}\label{eq:6.2a}
\end{equation}
\begin{equation}
    \tilde{X}({{t}_{0}};\theta )={{\tilde{X}}_{0}}(\theta ), \ \ \ \dot{\tilde{X}}({{t}_{0}};\theta )={{\dot{\tilde{X}}}_{0}}(\theta ), \tag{6.2b,c}\label{eq:6.2b,c}
\end{equation}
where $\zeta =b/\left( 2\sqrt{m|{{\eta }_{1}}|}\right)$ is the dimensionless
damping ratio, $\tilde{\Xi }$ is the normalized excitation, expressed as
\addtocounter{equation}{1}
\begin{equation}
    \tilde{\Xi }(\tilde{t};\theta )={{\Pi }_{\Xi }}{{\tilde{\Xi }}_{0}}(\tilde{t};\theta )+{{m}_{\tilde{\Xi }}}(\tilde{t}), 
    \ \ \ {{m}_{\tilde{\Xi }}}(\tilde{t})={{({{\eta }_{3}}\mathbf{/}|{{\eta }_{1}}{{|}^{3}})}^{1/2}}{{m}_{\Xi }}(\tilde{t}), \tag{6.3a,b}\label{eq:6.3b,c}
\end{equation}
and
${{\Pi }_{\Xi }}={{\xi }_{0}}\sqrt{{{\eta }_{3}}\mathbf{/}|{{\eta }_{1}}{{|}^{3}}}$
is the dimensionless forcing coefficient, which also plays the role of a nonlinearity parameter. Indeed, under the above 
choice of characteristic quantities, the effects of nonlinearity and the noise intensity are both included in the
parameter ${{\Pi }_{\Xi }}$ and studied jointly. Last, we specify ${{\tilde{\Xi }}_{0}}(t;\theta )$ to be a Gaussian 
Filter (GF) excitation, with autocovariance function
\begin{equation}
    {{C}_{{{{\tilde{\Xi }}}_{0}}{{{\tilde{\Xi }}}_{0}}}}(t,s)={{C}_{{{{\tilde{\Xi }}}_{0}}{{{\tilde{\Xi }}}_{0}}}}(t-s)=
    \exp \left( -{{a}^{2}}{{\left( t-s \right)}^{2}}\right)\cos ({{\omega }_{{{{\tilde{\Xi }}}_{0}}}}(t-s)), \label{eq:6.4}
\end{equation}
where the parameter ${{\omega }_{{{{\tilde{\Xi }}}_{0}}}}$ denotes the peak frequency of the excitation spectrum, and 
the parameter $a$ calibrates the correlation time of the excitation, ${{\tau }_{cor}}$. The latter is defined in terms of 
the autocovariance function, \cite{Yaglom_1987} Chap. 2, Sec.9. In the present work we adopt the following definition of 
the correlation time
\begin{equation*}
    {{\tau }_{cor}}=C_{\Xi \Xi }^{-1}(0)\int_{0}^{+\infty }{\left| C_{\Xi \Xi }(u) \right|du}.
\end{equation*}
The autocovariance function of $\tilde{\Xi }(\tilde{t};\theta )$ takes the form
$C_{\tilde{\Xi }\tilde{\Xi }}(\tilde{t},\tilde{s})=$ $\sigma_{\tilde{\Xi }}^{2}C_{{{{\tilde{\Xi }}}_{0}}{{{\tilde{\Xi }}}_{0}}(\cdot)}(\tilde{t},\tilde{s})$, 
with variance $\sigma_{\tilde{\Xi }}^{2}=\Pi_{\Xi}^{2}$.
Note that the potential function corresponding to the oscillator (\ref{eq:6.2a}) has two stable minima at $x=\pm1$

\subsection{Corresponding pdf-evolution models and numerical results}\label{sec6.2}

Using the state-vector $\bfv{X}=({{X}_{1}},{{X}_{2}})=(\tilde{X},\dot{\tilde{X}})$, the system of RDEs corresponding to 
equation (\ref{eq:6.2a}) takes the form of system (\ref{eqs:2.1}) with $N=2$. For this case, the functions ${{h}_{1,2}}$,
in equation (\hyperref[eq:2.1a]{2.1a}), are specified as ${{h}_{1}}(\bfv{X})={{X}_{2}}$ and ${{h}_{2}}(\bfv{X})=-X_{1}^{3}+{{X}_{1}}-2\zeta{{X}_{2}}$, 
while the excitation vector is $\Xi =(0,\tilde{\Xi }).$The Jacobian matrix of the vector $\bfv{h}$, takes the form
\begin{equation}
    {{\mathbf{J}}^{\bfv{h}}}(\bfv{X}(t;\theta ))=\left[
        \begin{matrix}
            0                       & 1       \\
            1-3X_{1}^{2}(t,\theta ) & -2\zeta \\
        \end{matrix}
        \right].\label{eq:6.5}
\end{equation}
Both SCT equation and novel genFPKE, corresponding to the oscillator under study, take the form of the general equation 
(\ref{eq:2.6}), differing only by the structure of their diffusion coefficients $\mathcal{D}_{vn}^{{{X}^{0}}\Xi (\cdot )}$ 
and $\mathcal{D}_{vn}^{\Xi \Xi }$, given by equations (\ref{eqs:4.17}) and (\ref{eqs:5.19}), respectively. Assuming, for 
simplicity, that the initial value is uncorrelated to the excitation, $\mathbf{C}_{{{\bfv{X}}^{0}}\tilde{\Xi }}=0$, the 
diffusion coefficients $\mathcal{D}_{vn}^{{{X}^{0}}\Xi }$ become zero. Further, the coefficients $\mathcal{D}_{v1}^{\Xi \Xi }$ 
are also zero, since ${{\Xi }_{1}}=0$, by definition. For both cases, the diffusion coefficients $\mathcal{D}_{v2}^{\Xi \Xi }$ 
are of the following integral form
\begin{equation}
    \mathcal{D}_{v2}^{\Xi \Xi }\left[{{f}_{\bfv{X}(\cdot )}}(
        \cdot );\bfv{x},t \right]=\int\limits_{{{t}_{0}}}^{t}{C_{\tilde{\Xi }\tilde{\Xi }}(t,s) \left\{ \begin{aligned}&D_{\nu 2}^{\operatorname{ngFPK}}[{{f}_{\bfv{X}(\cdot )}}(\cdot );\bfv{x},t,s] \\&D_{\nu 2}^{\operatorname{SCT}}(\bfv{x},t,s)\end{aligned} \right\}}ds
        , \label{eq:6.6}
\end{equation}
The components $D_{\nu 2}^{\operatorname{ngFPK}}$, corresponding to the novel genFPKE, are given by the formulae
\begin{subequations}
    \begin{align}
         & D_{12}^{\operatorname{ngFPK}}({{f}_{\bfv{X}(\cdot )}}(\cdot );\bfv{x},t,s)={{\Phi }_{12}}[\mathbf{R}[{{f}_{\bfv{X}(\cdot )}}(\cdot ),\cdot ]](t;s), \label{eq:6.7a}                           \\
         & D_{22}^{\operatorname{ngFPK}}({{f}_{\bfv{X}(\cdot )}}(\cdot );\bfv{x},t,s)=\nonumber                                                                                                          \\
         & =3\left({{m}_{X_{1}^{2}}}(t)-x_{1}^{2}\right){{\Phi }_{12}}[\mathbf{R}[{{f}_{\bfv{X}(\cdot )}}(\cdot ),\cdot ]](t;s)+{{\Phi }_{22}}[\mathbf{R}[{{f}_{\bfv{X}(\cdot )}}(\cdot ),\cdot ]](t;s).
    \end{align}
    \label{eq:6.7b}
\end{subequations}
where the matrix
\begin{equation*}
    \mathbf{R}(t):=\mathbf{R}[{{f}_{\bfv{X}(\cdot )}}(\cdot ),t]=\left[
        \begin{matrix}
            0                       & 1       \\
            1-3{{m}_{X_{1}^{2}}}(t) & -2\zeta \\
        \end{matrix}
        \right],
\end{equation*}
is the expected value of ${{\mathbf{J}}^{\bfv{h}}}(\bfv{X}(t;\theta ))$, equation (\ref{eq:6.5}), and the function 
${{m}_{X_{1}^{2}}}(t)$ is given by
\begin{equation}
    {{m}_{X_{1}^{2}}}(t):={{m}_{X_{1}^{2}}}({{f}_{\bfv{X}(t)}}(\cdot );t)=\mathbb{E}^{\theta }\left[ X_{1}^{2}(t;\theta ) \right]. \label{eq:6.8}
\end{equation}
The components $D_{\nu 2}^{\operatorname{SCT}}$, corresponding to the SCT approximation, are given by
\addtocounter{equation}{1}
\begin{equation}
    D_{12}^{\operatorname{SCT}}(\bfv{x},t,s)=(t-s), \ \ \ D_{22}^{\operatorname{SCT}}(\bfv{x},t,s)=1-2\zeta (t-s).\tag{6.9a,b}\label{eq:6.9a,b}
\end{equation}

Concerning the computation of the novel genFPKE, note that the function ${{m}_{X_{1}^{2}}}(t)$ is actually a moment of 
the unknown response, realizing the dependence of the matrix $\mathbf{R}$ on the density ${{f}_{\bfv{X}(\cdot )}}(\cdot )$. 
This somewhat peculiar term (\ref{eq:6.8}) calls for a special numerical treatment since, at each time instant, depends 
on the unknown density at that time. This peculiarity is treated by an iterative scheme as follows: the current-time value 
${{m}_{X_{1}^{2}}}(t)$, needed for the calculation of diffusion coefficients (\ref{eq:6.6}), is first estimated by 
extrapolation based on the three previous time steps (after the second step), and then improved by iterations at the current 
time. That is, for $\varepsilon >0$ and $m_{X_{1}^{2}}^{(0)}(t)=m_{X_{1}^{2}}(t)$,\\

\noindent
i) Calculate $\mathcal{D}_{v2}^{\Xi\Xi}$ by using the (initially extrapolated) $m_{X_{1}^{2}}^{(i)}(t)$ and obtain the 
PUFEM approximation $f_{\bfv{X}(t)}^{(i)}(\bfv{x})$,\\\noindent
ii) Calculate $m_{X_{1}^{2}}^{(i+1)}(t)$ from $f_{\bfv{X}(t)}^{(i)}(\bfv{x})$ and check $d=|m_{X_{1}^{2}}^{(i+1)}(t)-m_{X_{1}^{2}}^{(i)}(t)|<\varepsilon$,\\
iii.1) if $d<\varepsilon$, set $m_{X_{1}^{2}}(t)=m_{X_{1}^{2}}^{(i)}(t)$, $f_{\bfv{X}(t)}(\bfv{x})=f_{\bfv{X}(t)}^{(i)}(\bfv{x})$, 
and continue to the next time,\\
iii.2) if $d>\varepsilon$, return to step i) replacing $m_{X_{1}^{2}}^{(i)}(t)$ by $m_{X_{1}^{2}}^{(i+1)}(t)$.\\

\noindent
Usually, one or two iterations suffice to achieve convergence. Last, the state-transition matrix $\Phi [\mathbf{R}](t;s)$ 
is approximated via a Magnus expansion with two terms, namely
\begin{equation*}
    \Phi [\mathbf{R}](t;s)\approx \exp \left[ \int_{s}^{t}{\mathbf{R}({{u}_{1}})d{{u}_{1}}}
    +\frac{1}{2}\int_{s}^{t}{\int_{s}^{{{u}_{1}}}{[\mathbf{R}({{u}_{1}}),\mathbf{R}({{u}_{1}})]d{{u}_{2}}}d{{u}_{1}}}
    \right],
\end{equation*}
and the resulting matrix exponential is computed via explicit closed form expressions, see \cite{Bernstein_1993}.

We consider the oscillator (\ref{eq:6.1a}) with dimensional parameters $m=1\operatorname{kg}$, $b=1\operatorname{kg}{{\sec }^{-1}}$,
${{\eta }_{1}}=-1\operatorname{kg}{{\sec }^{-2}}$, ${{\eta }_{3}}=0.36\operatorname{kg}{{m}^{-2}}{{\sec }^{-2}}$ and 
${{\xi }_{0}}=1\operatorname{kg}m{{\sec }^{-2}}$. This example corresponds to equation (\ref{eq:6.2a}) with $\zeta =0.5$, 
subjected to a GF excitation $\tilde{\Xi }(\tilde{t};\theta )$ with intensity $\Pi_{\Xi }^{2}=0.36$. That is, we study a 
case with intermediate noise intensity or equivalently (for our normalization) an intermediately nonlinear one. Results 
are presented with respect to the normalized correlation time $\tilde{\tau }={{\tau }_{cor}}/{{\tau }_{relax}}$, called 
relative correlation time, where ${{\tau }_{relax}}$ denotes the relaxation time of the unforced harmonic oscillator. 
In all figures presented below, the final time is chosen to be in the long-time stationary regime, to check the validity 
of genFPKE in both the transient and the stationary regimes. The MC simulations are constructed from a sample of ${{10}^{5}}$ 
numerical experiments for each tested case.

The numerical results presented concern the non-dimensional oscillator (\ref{eq:6.2a}) with initial value and 
excitation parameter values given in \hyperref[tab:1]{Table 1} or \hyperref[tab:2]{2}. From now on, tilda is discarded from the notation, except 
for the relative correlation time, $\tilde{\tau}$.

\begin{figure} 
    \centering
    \includegraphics[width=0.8\textwidth]{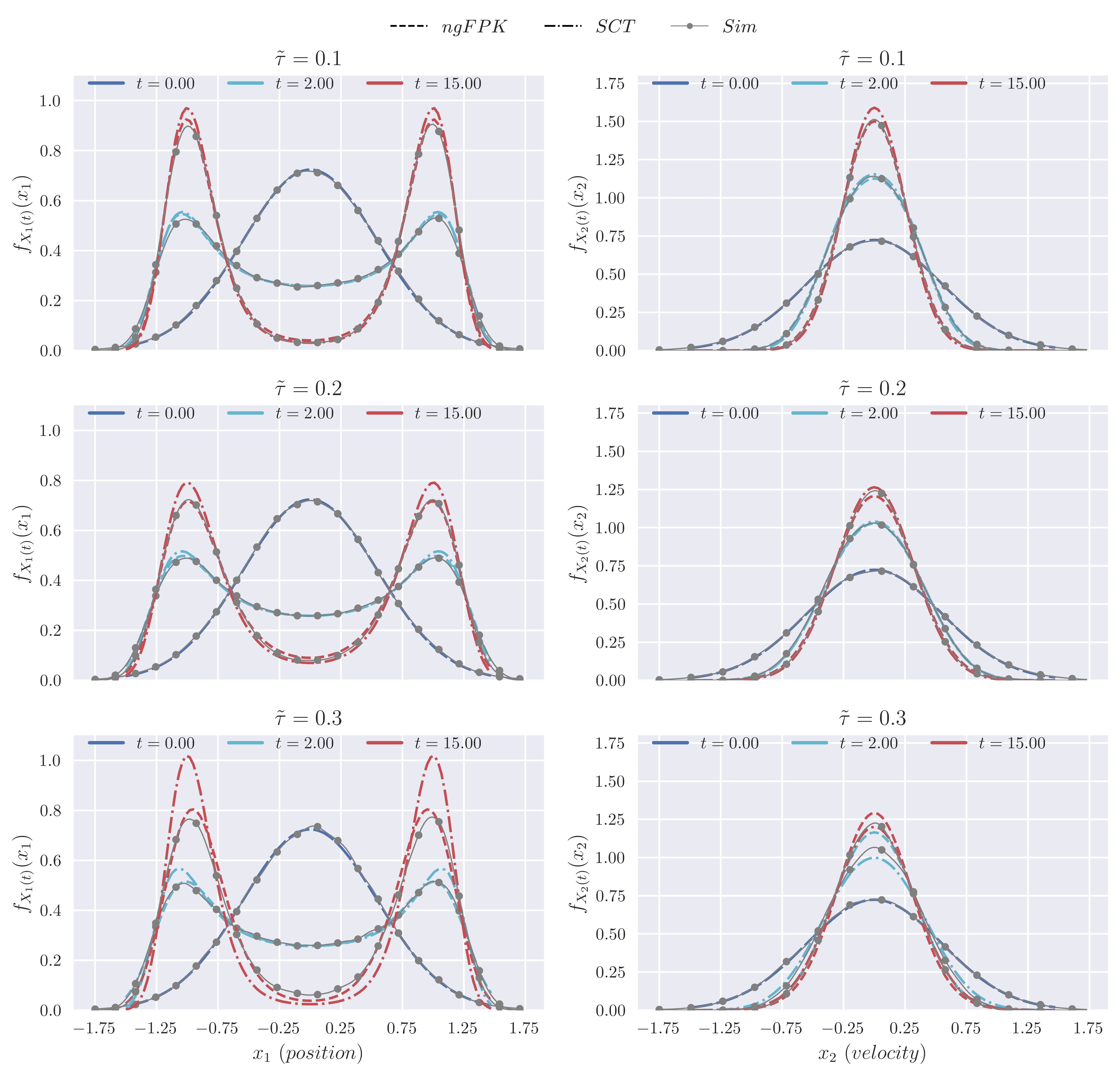}
    \caption{Evolution of response marginal pdfs for the oscillator (\hyperref[eq:6.2a]{6.2}),
        configured as described in \hyperref[tab:1]{Table 1} [$\bfv{m}_{\bfv{X}^0} =\bfv{0} , m_{\Xi}(t) = 0$], for three values of relative correlation time. 
        Marginal pdfs obtained via the PUFEM solutions to the SCT approximation (dashed-dotted lines) and novel genFPKE 
        (dashed lines) are compared to MC simulations (marked continuous lines), in different time instances. 
        The greatest time corresponds to the long-time steady-state regime.}
    \label{fig:3}
\end{figure}

\begin{table}[h] \label{tab:1}
    \centering
    \caption{Numerical values of the parameters corresponding to the oscillator under numerical investigation in \hyperref[fig:3]{Figures 3}, \ref{fig:5} and \ref{fig:6}}
    \renewcommand{\arraystretch}{1.5}
    \begin{tabular}{c c c c c c}
        \multicolumn{6}{c}{\textbf{Oscillator/Excitation}} \\
        \toprule 
        Parameters  & $\zeta$ & $m_{\Xi}$ & $\sigma^{2}_{\Pi_\Xi}$ & $\omega_{\Xi}$ & $\tilde{\tau}$      \\
        Values      & 0.5     & 0       & 0.36                 & 2.5          & $\{0.1, 0.2, 0.3\}$ \\
        \midrule
        \multicolumn{6}{c}{\textbf{Initial Value}} \\
        \toprule
        Parameters  & \multicolumn{2}{c}{$\bfv{m}_{\bfv{X}^0}$} & \multicolumn{3}{c}{$\bfv{C}_{\bfv{X}^0 \bfv{X}^0}$} \\
        Values      & \multicolumn{2}{c}{(0,0)}    & \multicolumn{3}{c}{0.3$\bfv{I}$}         \\
        \bottomrule
    \end{tabular}
\end{table}
In \hyperref[fig:3]{Figure 3}, we compare the 1D marginal distributions, ${{f}_{{{X}_{1}}(t)}}({{x}_{1}})$ (position) and
${{f}_{{{X}_{2}}(t)}}({{x}_{2}})$ (velocity), as obtained via the PUFEM solution of the SCT equation and novel genFPKE, as well 
as via MC simulation. In this test, we consider that initial value and excitation have zero mean. We find that for $\tilde{\tau }
=0.1$ both equations perform well enough, almost being in full agreement with MC simulations. However, as the relative 
correlation time $\tilde{\tau }$ increases, SCT approximation overestimates the peak values of the position density, especially in the 
steady state. This situation becomes more profound for $\tilde{\tau }=0.3$, where the overestimation becomes more than 22\%. 
The pdf of the velocity, ${{f}_{{{X}_{2}}(t)}}({{x}_{2}})$, is well approximated by both equations. A small overastimation 
of the pick value by the SCT approximation is again present.

\begin{table}[h] \label{tab:2}
    \centering
    \caption{Numerical values of the parameters corresponding to the oscillator under numerical investigation in 
                \hyperref[fig:4]{figures/Figures 4} and \ref{fig:7}}
    \renewcommand{\arraystretch}{1.5}
    \begin{tabular}{c c c c c c}
        \multicolumn{6}{c}{\textbf{Oscillator/Excitation}} \\
        \toprule 
        Parameters  & $\zeta$ & $m_{\Xi}$ & $\sigma^{2}_{\Pi_\Xi}$ & $\omega_{\Xi}$ & $\tilde{\tau}$ \\
        Values      & 0.5     & $\frac{0.5{{e}^{6(t-1)}}}{1+{{e}^{6(t-1)}}} \to 0.5$ & 0.36 & 2.5 & 0.3 \\
        \midrule
        \multicolumn{6}{c}{\textbf{Initial Value}} \\
        \toprule
        Parameters  & \multicolumn{2}{c}{$\bfv{m}_{\bfv{X}^0}$} & \multicolumn{3}{c}{$\bfv{C}_{\bfv{X}^0\bfv{X}^0}$} \\
        Values      & \multicolumn{2}{c}{(-0.5,-0.5)} & \multicolumn{3}{c}{0.3$\bfv{I}$} \\
        \bottomrule
    \end{tabular}
\end{table}    

\begin{figure} 
    \centering
    \includegraphics[width=0.8\textwidth]{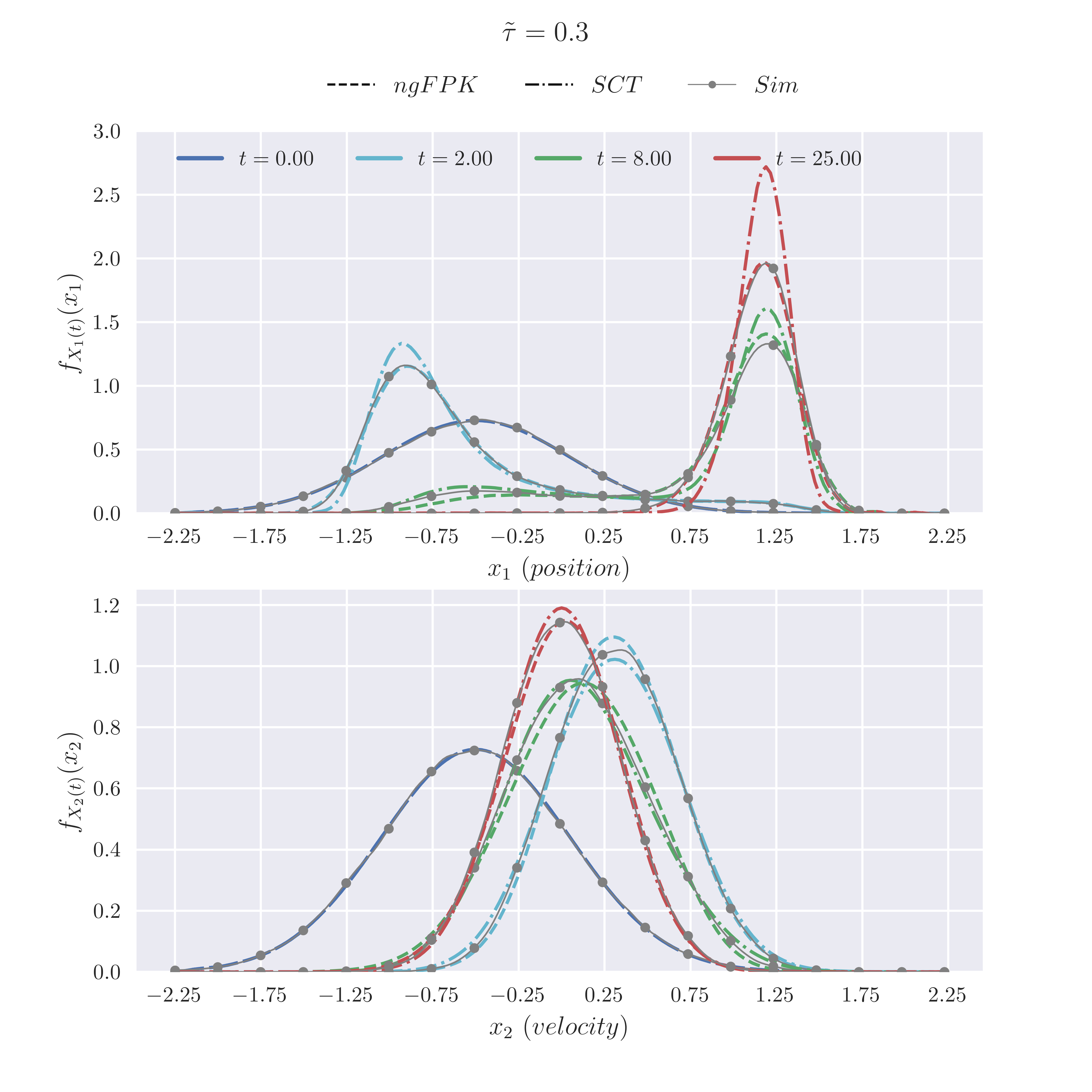}
    \caption{Evolution of response marginal pdfs for the oscillator (\hyperref[eq:6.2a]{6.2}),
        configured as described in \hyperref[tab:2]{Table 2} [$m_{X_1^0} , m_{X_2^0}, m_{\Xi}(t \to \infty) \neq 0$]. 
        Marginal pdfs obtained via the PUFEM solutions to the SCT approximation (dashed-dotted lines) and novel genFPKE (dashed
        lines) are compared to MC simulations (marked continuous lines), in different time instances. The greatest time corresponds to the long-time
        steady-state regime}
    \label{fig:4}
\end{figure}

In \hyperref[fig:4]{Figure 4}, we demonstrate the evolution of the marginal pdfs of the same oscillator, now with initial value and 
excitation of non-zero mean, as described in \hyperref[tab:2]{Table 2}. To interpret the results shown in \hyperref[fig:4]{Figure 4}, we recall
that the local minima of the oscillator's  potential well are located at $x_{1}=\pm1$. The initial position density 
$f_{X_{1}(t_0)}(x_1)$ is centered at $x_{1}=-0.5$, on the right of the first minimum (-1), while the long-time mean of the excitation 
is located at $x_{1}=+0.5$,  near the second minimum (+1). Having these facts in mind, we understand the evolution shown in 
\hyperref[fig:4]{Figure 4} as follows: at short times, $t=2$, most of the probability mass lies near the first minium of 
the potential well (-1), exhibiting a reasonable asymmetry due to the combined effect of the initial density distribution 
and the attraction at (-1). At later times, $t=8$ and $t=20$, the probability mass moves to the right minimum of the potential 
well (+1), again displaying a small asymmetry, now influenced by the long-time excitation (centered at +0.5) and the attraction 
at (+1). This evolution is realized by all solutions (SCT equation, ngFPKE and MC simulations). However,  the SCT model systematically 
overestimates the peak values, whereas the ngFPKE align closely with the MC simulation. The overestimation of the long-time 
mean value by the SCT model is approximately 35\%. A clearer visualization of this probability mass transition from the 
first potential well (-1) to the second (+1), realized by the evolution of the 3D graph of the joint pdf $f_{X_{1}(t)X_{2}(t)}(x_1,x_2)$,
is \href{https://www.dropbox.com/scl/fo/qmp7wzx0tdcpvvqhf5j8o/AHuDTbDD8i3NHZ-yfq2mGu0?rlkey=i6o38dgci3b0i5tjby03s9sbc&st=7bdui2ss&dl=0}{available in a video online}.
Note that, if the long-time mean value of the excitation remains sufficiently close to zero (\text{\small{e.g.}} $\scriptstyle m_{\Xi}(t\to\infty)=0.15$), 
the long-time pdf $f_{X_{1}(t_0)}(x)$ retains a bi-modal structure, albeit assymetric, as expected.

\begin{figure} 
    \centering
    \includegraphics[width=0.8\textwidth]{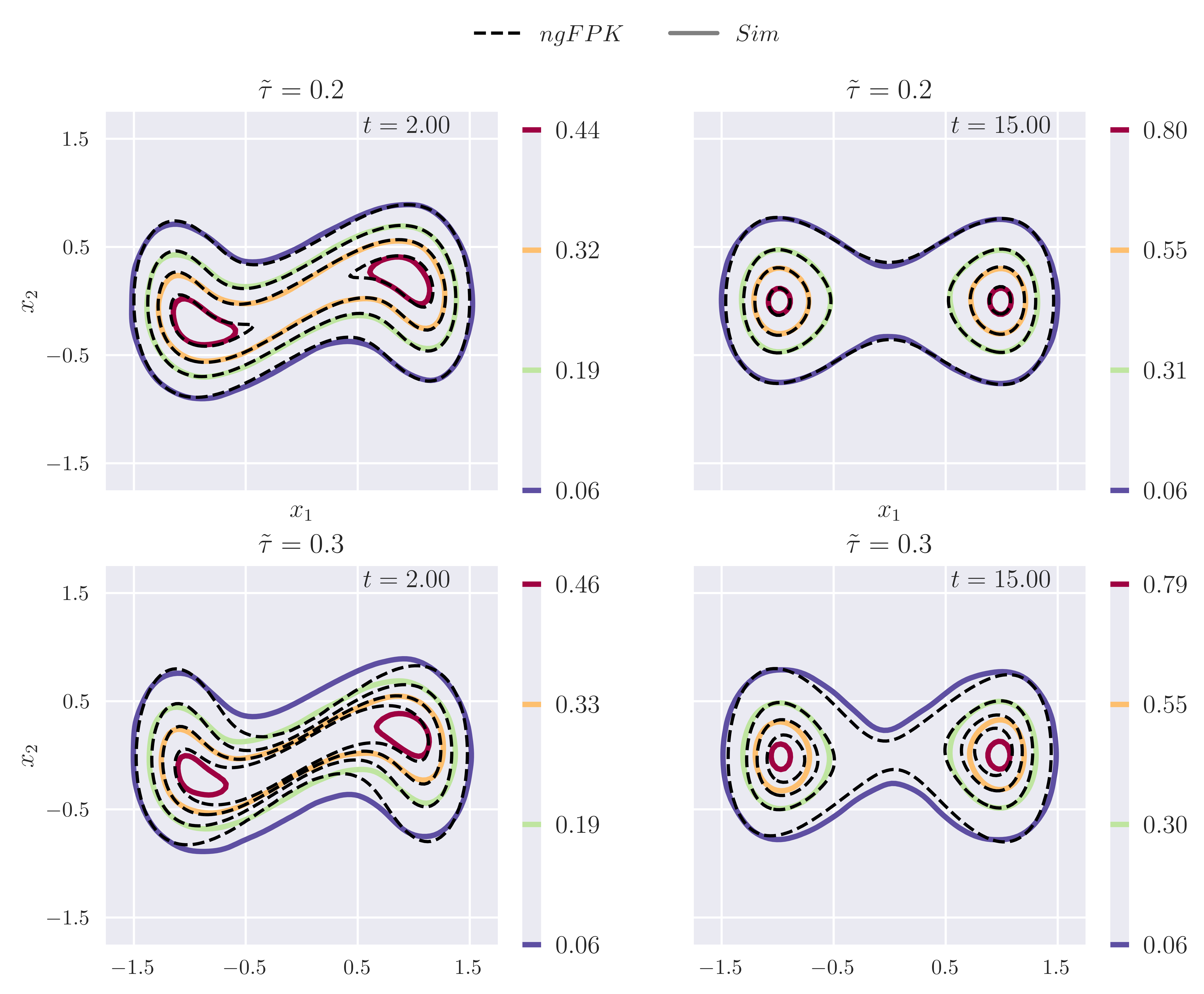}
    \caption{Contours of response pdfs $f_{\bfv{X}(t)}^{ngFPK}(\bfv{x})$, as obtained via the PUFEM solutions of the 
        novel genFPKE (dashed lines) and MC simulations (colored continuous lines), for the oscillator (\hyperref[eq:6.2a]{6.2}),
        configured as described in \hyperref[tab:1]{Table 1}. The pdfs are shown at two time instances 
        (transient and steady state), for two values of relative correlation time.}
    \label{fig:5}
\end{figure}

\begin{figure} 
    \centering
    \includegraphics[width=0.8\textwidth]{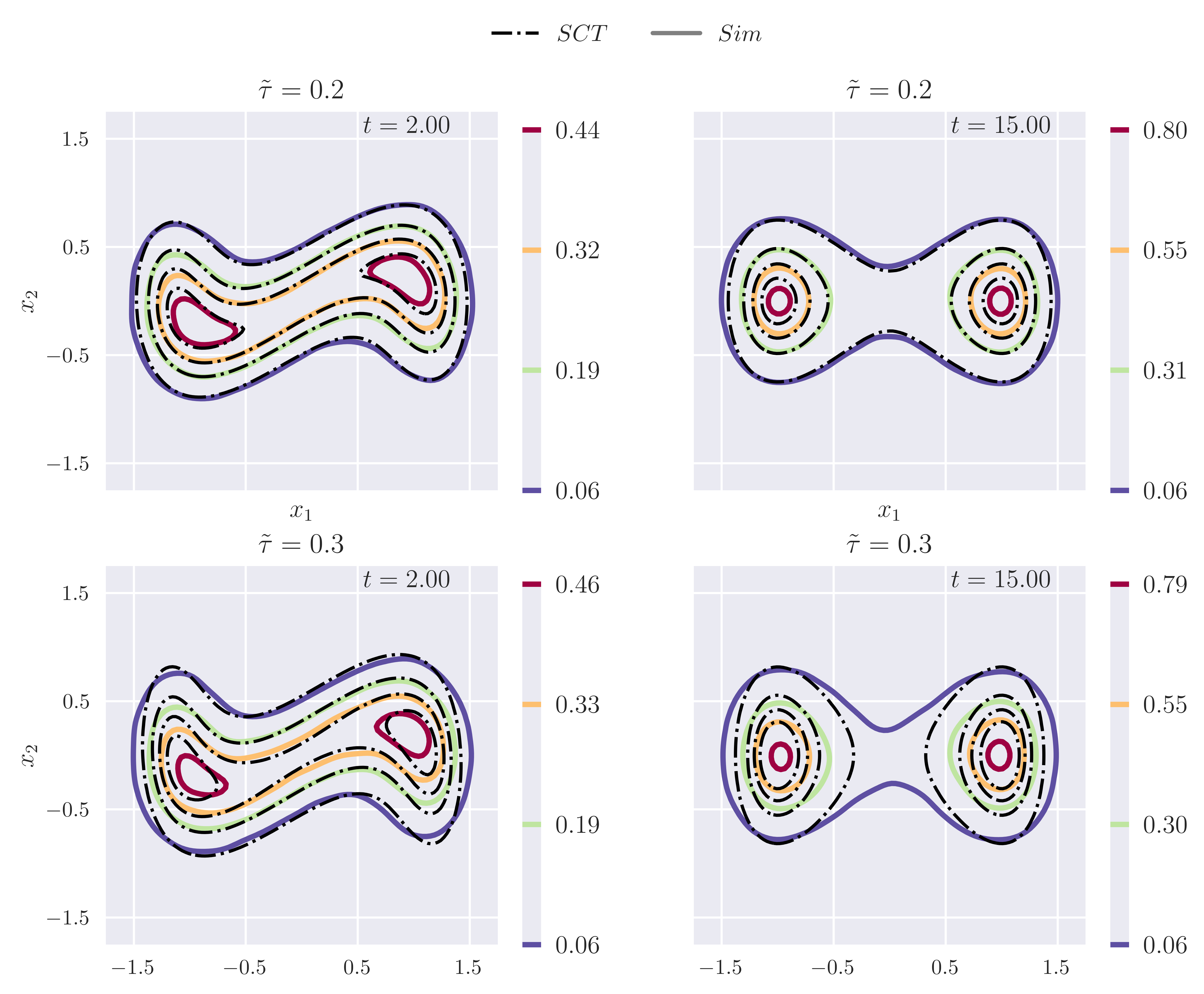}
    \caption{Contours of response pdfs $f_{\bfv{X}(t)}^{SCT}(\bfv{x})$, as obtained via the PUFEM solutions of the SCT equation 
        (dashed-dotted lines) and MC simulations (colored continuous lines), for the oscillator (\hyperref[eq:6.2a]{6.2}), 
        configured as described in \hyperref[tab:1]{Table 1}. The pdfs are shown at two time instances 
        (transient and steady state), for two values of relative correlation time.}
    \label{fig:6}
\end{figure}

In \hyperref[fig:5]{Figure 5}, we compare contour lines of the 2D response pdf for the same oscillator with parameter 
values given in \hyperref[tab:1]{Table 1}, as obtained via PUFEM solution of the novel genFPKE, denoted $f_{\bfv{X}(t)}^{ngFPK}(\bfv{x})$, 
with MC simulations. We observe that, for relative correlation time $\tilde{\tau }=0.3$, in the steady state ($t=15$), the
$f_{\bfv{X}(t)}^{ngFPK}(\bfv{x})$ contours are in good agreement with the MC simulations, except from the highest-level 
curve, close to the density’s picks. For the same oscillator, in \hyperref[fig:6]{Figure 6}, we perform the same comparison
for contour lines of the 2D response pdf obtained by solving the SCT equation, denoted $f_{\bfv{X}(t)}^{SCT}(\bfv{x})$, using PUFEM. 
We see that for relative correlation time $\tilde{\tau }=0.3$, the SCT approximation performs poorly at steady state ($t=15$), 
since the contours of $f_{\bfv{X}(t)}^{SCT}(\bfv{x})$ fails in capturing the geometric characteristics of the corresponding
MC simulation. Further, it is worth noticing that the SCT equation seems to perform better in the early transient state, 
$t=2$, although its long-time steady state limit is a low-quality approximation, much worse than the one obtained by
the novel genFPKE.


\begin{figure} 
    \centering
    \includegraphics[width=\textwidth]{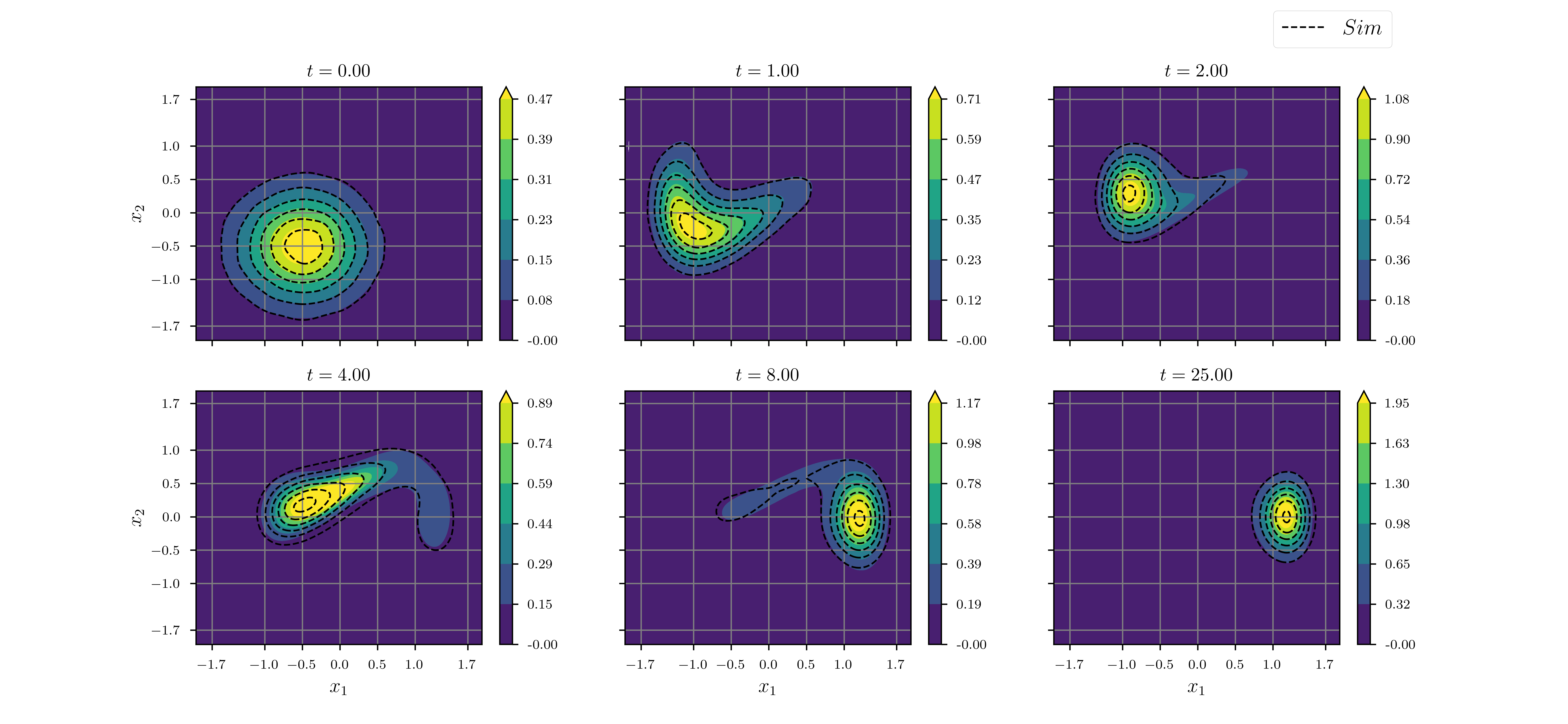}
    \caption{Contour plots of the 2D response pdf of the oscillator (\hyperref[eq:6.2a](6.2)),
        configured as described in \hyperref[tab:2]{Table 2}. Response pdfs $f_{\bfv{X}(t)}^{ngFPK}(\bfv{x})$ , 
        obtained via the PUFEM solutions of the novel genFPKE are demonstrated in different time
        instances. Contour lines of corresponding MC simulations are denoted with
        black dashed-dotted lines. }
    \label{fig:7}
\end{figure}

In \hyperref[fig:7]{Figure 7}, we demonstrate the evolution of the 2D response pdf of the oscillator described by 
\hyperref[tab:2]{Table 2}, for the case $\tilde{\tau} = 0.3$. Contour plots obtained via the solution of the novel 
genFPKE are compared with MC simulations. It is clear that the numerical solution of the ngFPKE is in very good 
agreement with the MC simulation results, during the whole evolution, from the initial time up to the long-time steady 
state.

    \clearpage


\clearpage

\section{Discussion and conclusions} \label{sec7}

The novel genFPKE derived in the present work, exhibits two salient features which, to the best of our knowledge, do not 
occur in any other existing approach. First, it separates the mean value of the response from its random fluctuations, 
and keeps intact (without any simplifications) the effect of the mean-value evolution, see equations (\hyperref[eq:5.13a]{5.13}). 
Second, and related with the first, it is nonlinear and nonlocal in time through the dependence of the diffusion coefficients 
on the history of a generalized moment of the response. This fact reflects the non-Markovian character of the response, 
and extends the range of validity of the novel genFPKE to higher values of the relative correlation time.

The novel genFPKE is able to reproduce various existing model equations, under appropriate simplifications, showcasing its 
consistency with existing theories. The numerical results obtained for the joint pdf $f_{X_1(t)X_2(t)}(x_1,x_2)$ by 
exploiting the novel genFPKE are in good agreement with corresponding results obtained by using MC simulations. 
This agreement pertains throughout the whole evolution of the response, from the initial time up to the long-time steady 
state. Excitation is not restricted to an Ornstein-Uhlenbeck process, as usually happens in similar studies. It may be 
any regular (smoothly correlated) Gaussian process, specified by means of its mean value and its covariance matrix, both 
arbitrarily evolving in time. 

The separation of the mean value of the response from its fluctuation around it (called subsequently the \textit{separation technique}) 
is a nontrivial generalization to the multidimensional case of the corresponding separation developed by the same group 
\cite{Mamis_2019} for the scalar case. A precursor of this technique, in the scalar case, can be traced back to H{\"a}nggi’s \textit{decoupling approximation} 
(or \textit{H{\"a}nggi’s ansatz}); see \cite{Hanggi_1985,Hanggi_1994}. H{\"a}nggi’s implementation corresponds to replacing 
$h'(X(t;\theta))$ (in our notation) by its mean value, neglecting the random fluctuation about it. Besides, to avoid nonlocality 
in time, H{\"a}nggi replaces $\mathbb{E}\left[h'(X(t;\theta))\right]$ by its long-time stationary value (assuming time-invariant RDE), 
which invalidates the resulting equation in the transient regime. Further, the ad hoc character of H{\"a}nggi's decoupling approximation, 
led him to the conclusion that “…the decoupling approximation is not suitable to approximate multidimensional features such 
as multidimensional probabilities that may exhibit colored dependent correlation among the state variables” \cite{Hanggi_1994}, 
p. 273. Our analysis, \hyperref[sec5]{Section 5}, shows that the \textit{separation technique}, as presented in this paper, 
can be implemented systematically either in one-dimensional or in multidimensional cases, retaining the effect of random 
fluctuations around the mean value of the response, leading to an approximate genFPKE which is valid both in the transient 
and in the long-time regime. 

A simpler pdf-evolution equation, the SCT approximation, has been also derived and studied. This equation is also new in 
the sense that it cannot be found in the literature, although can be derived as a special case from some complicated equations 
attacking directly the multiplicative case; see \cite{Dekker_1982,Fox_1983,Garrido_1982} and the discussion at the end of 
\hyperref[eqs:4.17]{Section 4.3}. Numerical evidence reveals that the novel genFPKE outperforms the SCT approximation, 
especially in the long-time steady-state regime.

The novel genFPKE has been derived by the SLE (which is exact but non-closed) after applying two main approximations: 
i) the use of a first-order Magnus expansion for the representation of the state-transition matrix $\Phi[\mathbf{B}]$; 
see equation (\ref{eq:5.15a}) and ii) the use of a first-order Taylor expansion for the exponentiated integral (\ref{eq:5.15b}), 
which can be considered as a current-time approximation. Both approximations can be improved, e.g. by retaining the corresponding 
second-order terms. More straightforward is the inclusion of the second-order term in the Taylor expansion, utilized to 
simplify the exponentiated integral (\ref{eq:5.15b}). A preliminary analysis of the structure of this approximation shows 
that richer, yet significantly more complicated, expressions are obtained for the diffusion coefficients. This approach 
might be promising, extending further the range of applicability to higher values of the relative correlation time. 
Nevertheless, we have not yet carried out any in-depth study of this more complicated equation. The present methodology 
can be extended to the multiplicative excitation case, working with the joint response-excitation pdf instead of the 
response pdf. This extension will be presented in a forthcoming paper.

    \subsection*{Data Availability:} The data that support the findings of this study are
    available from the authors upon reasonable request.
    
    \subsection*{Acknowledgments}
        The authors would like to thank Zaxarias Kapelonis for instructive discussions on the numerical implementation in the 
        initial phase of this work, as well as for the development of the MC simulation code prototype.

    \subsection*{Author Contributions}
        Conceptualization [Gerassimos A. Athanassoulis, Konstantinos Mamis]; 
        Formal analysis [Gerassimos A. Athanassoulis, Nikolaos P. Nikoletatos-Kekatos]; 
        Methodology [Gerassimos A. Athanassoulis, Konstantinos Mamis, Nikolaos P. Nikoletatos-Kekatos]; 
        Supervision [Gerassimos A. Athanassoulis]; 
        Writing–original draft [Gerassimos A. Athanassoulis, Nikolaos P. Nikoletatos-Kekatos]; 
        Writing–review \& editing [Gerassimos A. Athanassoulis, Nikolaos P. Nikoletatos-Kekatos]; 
        Software [Nikolaos P. Nikoletatos-Kekatos]; 
        Validation [Nikolaos P. Nikoletatos-Kekatos]; 
        Visualization [Nikolaos P. Nikoletatos-Kekatos];



    \appendix  

    \phantomsection
    \addcontentsline{toc}{section}{Appendix A: List of Abbreviations (in alphabetical order)}  
    \section*{Appendix A: List of Abbreviations (in alphabetical order)} \label{secA1}
    
    \begin{itemize}
        \item $\operatorname{FF}\ell$: Function Functional
        \item FPK:  Fokker-Plank-Kolmogorov
        \item FPKE: Fokker-Plank-Kolmogorov Equation
        \item genFPKE: generalized FPKE
        \item GF: Gaussian Filter
        \item IVP(s): Initial Value Problem(s)
        \item MC: Monte Carlo
        \item NF: Novikov-Furutsu
        \item ngFPKE: Novel generalized FPK
        \item ODE(s): Ordinary Differential Equation(s)
        \item PDE(s): Partial Differential Equation(s)
        \item pdf(s): probability density function(s)
        \item PUFEM: Partition of Unity Finite Element Method
        \item RDE(s): Random Differential Equation(s)
        \item SCT: Small Correlation Time
        \item SLE: Stochastic Liouville Equation
    \end{itemize}


    \phantomsection
    \addcontentsline{toc}{section}{Appendix B: On the State-Transition Matrix and its Approximations}  
    \section*{Appendix B: On the State-Transition Matrix and its Approximations} \label{secB1}

    \renewcommand{\theequation}{B.\arabic{equation}}
    \renewcommand{\thefigure}{\arabic{figure}}  
    \setcounter{figure}{7}

    Explicit forms of the state-transition matrix of a general (time-varying) linear system are
    provided by the Peano \cite{Brockett_1970,DaCunha_2005,Baake_2011} and Magnus \cite{Blanes_2009,Magnus_1954} series
    expansions. We present herewith the standard formulation of both expansions, to fix the notation and facilitate the
    reader to follow the flow of calculations in \hyperref[sec4]{Sections 4} and \ref{sec5}.

    Consider the linear, time-varying IVP
    \begin{subequations} \label{eq:B1}
        \begin{align}
            & \mathbf{\dot{y}}(t)=\mathbf{A}(t)\mathbf{y}(t),                     \\
            & \mathbf{y}({{t}_{0}})={{\mathbf{y}}^{0}}, \ \ \ t\in [{{t}_{0}},T].
        \end{align}
    \end{subequations}
    The corresponding state-transition matrix $\Phi (t;{{t}_{0}})$ realizes the solution of (\ref{eq:B1}) in the form
    $\mathbf{y}(t)=\Phi (t;{{t}_{0}}){{\mathbf{y}}^{0}}$ ($\Phi ({{t}_{0}};{{t}_{0}})=\bfv{I}$). In fact, for any
    given state $\mathbf{y}(s)$, $s\in [{{t}_{0}},T]$, it holds that $\mathbf{y}(t)=\Phi (t;s)\mathbf{y}(s)$,
    $\forall t\ge s$. Further, a state-transition matrix is never singular, and it is ${{C}^{k+1}}$ when
    $\mathbf{A}(t)$ is ${{C}^{k}}$. Besides ${{\Phi }^{-1}}(t;s)=\Phi (s;t)$, $\forall s\in [{{t}_{0}},T]$, and
    $\Phi (t;s)$ satisfies the matrix differential equation, $\dot{\Phi }(t;s)=\mathbf{A}(t)\Phi (t;s)$, with
    initial value $\Phi (s;s)=\bfv{I}$. Given the fundamental matrix $\mathbf{U}$ of (\ref{eq:B1}), $\Phi $ is explicitly
    constructed as $\Phi (t;s)=$ $\mathbf{U}(t){{\mathbf{U}}^{-1}}(s)$. In some cases, we are obliged to consider
    simultaneously the state-transition matrices of two different linear time-varying systems (see, e.g., \hyperref[sec5.2]{Section 5.2}).
    Then, use is made of the more complicated notation $\Phi [\mathbf{A}](t;s)$, where $\mathbf{A}$ is the system matrix.

    For a general time-varying linear system, the state-transition matrix $\Phi (t;s)$ cannot be expressed in closed
    form. There are, however, series expansions of $\Phi (t;s)$, in terms of the system matrix $\mathbf{A}(t)$. One such
    series expansion is the Peano-Baker series, obtained directly from Picard iterations, which has the form
    \begin{align} \label{eq:B2}
        & \Phi (t;s)=\Phi [\mathbf{A}](t;s)=\bfv{I}+\int_{s}^{t}{\mathbf{A}(u)}du+\int_{s}^{t}{\mathbf{A}({{u}_{1}})\int_{s}^{{{u}_{1}}}{\mathbf{A}({{u}_{2}})}d{{u}_{2}}}d{{u}_{1}}+ \nonumber \\
        & +\int_{s}^{t}{\mathbf{A}({{u}_{1}})\int_{s}^{{{u}_{1}}}{\mathbf{A}({{u}_{2}})}\int_{s}^{{{u}_{2}}}{\mathbf{A}({{u}_{3}})}d{{u}_{3}}d{{u}_{2}}}d{{u}_{1}}+\cdots .
    \end{align}
    This series is uniformly convergent, see \cite{Brockett_1970} Sec. 3, under the assumption of continuous $\mathbf{A}$; see also \cite{Frazer_1938} Sec. 2.11 and \cite{Baake_2011}. In classical statistical mechanics, the above formula is usually identified as the definition of the ordered exponential, also expressed by the notation
    \begin{equation} \label{eq:B3}
        {{\exp }_{\operatorname{ord}}}\left( \int_{s}^{t}{\mathbf{A}(u)du} \right)=\bfv{I}+\sum\limits_{n=1}^{+\infty }{\frac{1}{n!}\int_{s}^{t}{\cdots \int_{s}^{t}{O[\mathbf{A}({{u}_{1}})\cdots \mathbf{A}({{u}_{n}})]}d{{u}_{n}}}\cdots d{{u}_{1}}},
    \end{equation}
    where the operator $O[\cdot ]$ ensures the time ordering. In connection to the derivation of genFPKE, the ordered
    exponential is the central mathematical entity of the ordered cumulant expansion approach, see Sec. 1.2 for
    references. The series (\ref{eq:B2}) is closely related to the Dyson series used in quantum mechanics, \cite{Reed_1975}
    Sec. X.12, whose terms formally differs from the corresponding ones in Peano-Baker series by a factor${{(-i)}^{n}}$.
    An alternative to the Peano-Baker series is the Magnus series expansion. The latter originate in the monumental work
    of Magnus \cite{Magnus_1954} who followed the approach of expressing the solution of the non-autonomous matrix
    differential equation $\mathbf{\dot{Y}}(t)=\mathbf{A}(t)\mathbf{Y}(t)$, $\mathbf{Y}(s)=\bfv{I}$ as a matrix exponential
    \begin{equation*}
        \mathbf{Y}(t)=\Phi [\mathbf{A}](t;s)\mathbf{Y}(s)=\exp \left( \bfv{\Omega} (t;s) \right).
    \end{equation*}
    Then, the matrix $\bfv{\Omega} (t;s)$ is expressed via the series
    \begin{equation} \label{eq:B4}
        \bfv{\Omega} (t;s):=\sum\limits_{i=1}^{+\infty }{{{\bfv{\Omega} }_{i}}(t;s)},
    \end{equation}
    where, by denoting $\left[ \mathbf{A},\mathbf{B} \right]=\mathbf{A}\mathbf{B}-\mathbf{B}\mathbf{A}$ the matrix
    commutator, its first three terms read
    \begin{align*}
        & {{\bfv{\Omega} }_{1}}(t;s)=\int_{s}^{t}{\mathbf{A}(u)du},\ \ \ \ {{\bfv{\Omega} }_{2}}(t;s)=\frac{1}{2}\int_{s}^{t}{\int_{s}^{{{u}_{1}}}{\left[ \mathbf{A}({{u}_{1}}),\mathbf{A}({{u}_{2}}) \right]d{{u}_{2}}d{{u}_{1}}}},                                                                                                             \\
        & {{\bfv{\Omega} }_{3}}(t;s)=\frac{1}{6}\int_{s}^{t}{ \int_{s}^{{{u}_{1}}}{ \int_{s}^{{{u}_{2}}}{\left[ \mathbf{A}({{u}_{1}}),\left[ \mathbf{A}({{u}_{2}}),\mathbf{A}({{u}_{3}}) \right] \right]+\left[ \mathbf{A}({{u}_{3}}),\left[ \mathbf{A}({{u}_{2}}),\mathbf{A}({{u}_{1}}) \right] \right]}d{{u}_{3}}d{{u}_{2}}d{{u}_{1}}}}.
    \end{align*}

    In general, the Magnus expansion is a handy tool in many situations, since it directly provides an exponential
    representation of the state-transition matrix. Conditions under which (\ref{eq:B3}) exists are discussed in \cite{Blanes_2009}
    Sec. 2.7.1, and the convergence of the series (\ref{eq:B4}) has been extensively studied by many authors, see \cite{Casas_2007},
    \cite{Blanes_2009} Sec. 2.7.2 and references therein. Despite that the Magnus expansion usually has a small radius of
    convergence, it is a precious tool for approximations in small time intervals $t-s$, since it is well behaved under
    truncation of any order, maintaining qualitative and often geometric characteristics of the solution, see \cite{Blanes_2009}
    Sec. 1.1. These features make Magnus expansion a more suitable tool, both for analytic and numerical approximations,
    even though it may seem easier to perform calculations/computations with the formally simpler Peano-Baker series.
    Both above representations of $\Phi $ are utilized in \hyperref[sec4]{Sections 4}  and \ref{sec5}, respectively.

    Last, it is worth mentioning that there are two special cases for which the series (\ref{eq:B2}) can be summed, obtaining a
    closed-form expression for $\Phi $. The first is for constant $\mathbf{A}$ (time-invariant systems), and the second
    is the case where $\mathbf{A}(\tau )$ commutes with its integral (special time-varying structure) \cite{Brockett_1970},
    Section 1.5. In these cases, $\Phi $ attains the form of the matrix exponential
    $\Phi =exp(\int_{{{t}_{0}}}^{t}{\mathbf{A}(\tau )d\tau })$. In the above cases, all commutators in the Magnus series
    (\ref{eq:B4}) become zero.

\begin{figure} 
        \centering
        \includegraphics[width=\textwidth]{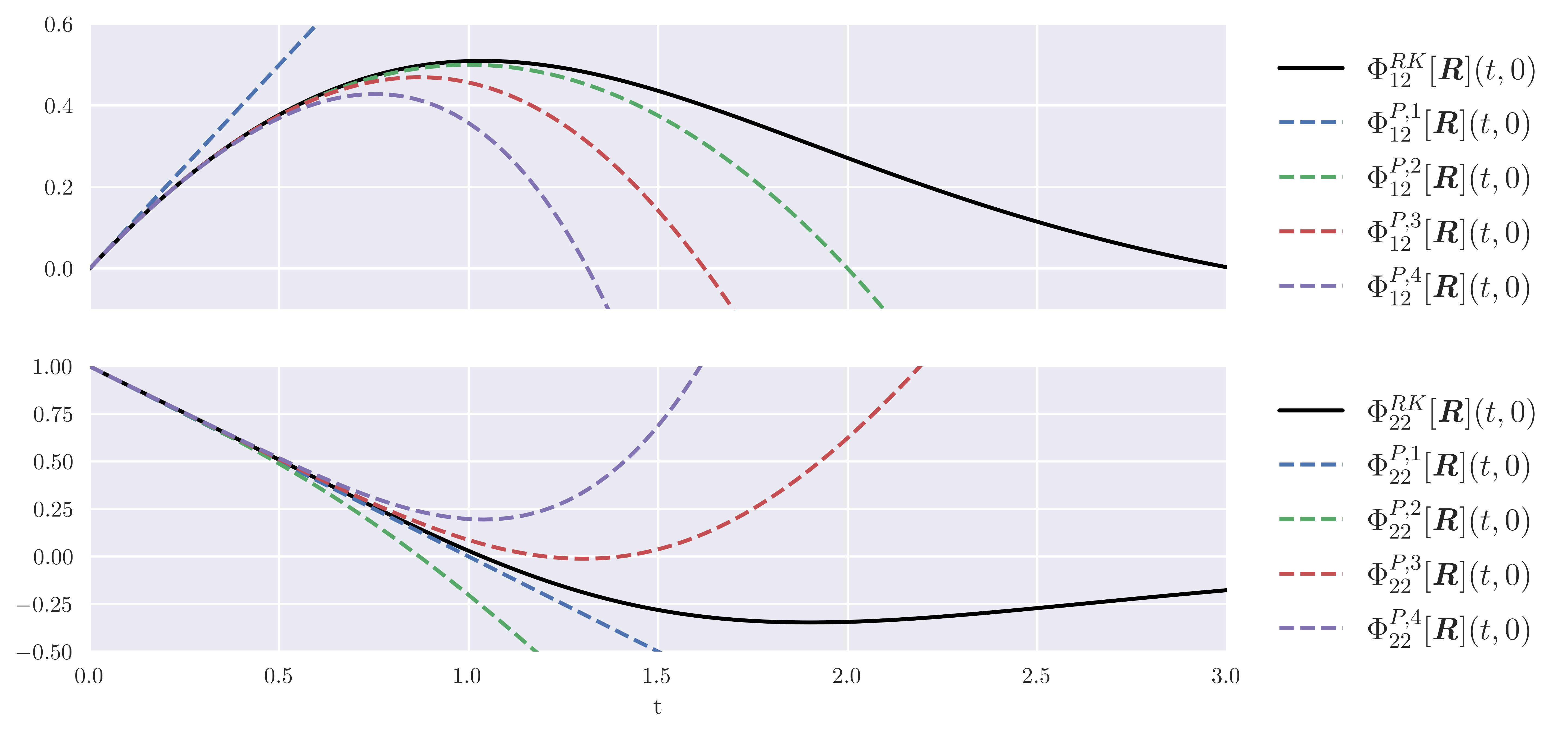}
        \caption{Comparison of Peano Baker approximation of $\Phi _{i2}[\mathbf{R}](t;s=0)$, up to the fourth order 
        ($\Phi _{i2}^{P,n}[\mathbf{R}](t;s=0)$, dashed lines) and corresponding RK solution ($\Phi _{i2}^{RK}[\mathbf{R}](t;s=0)$, 
        continuous lines).}
        \label{fig:9}
    \end{figure}

\begin{figure} 
        \centering
        \includegraphics[width=\textwidth]{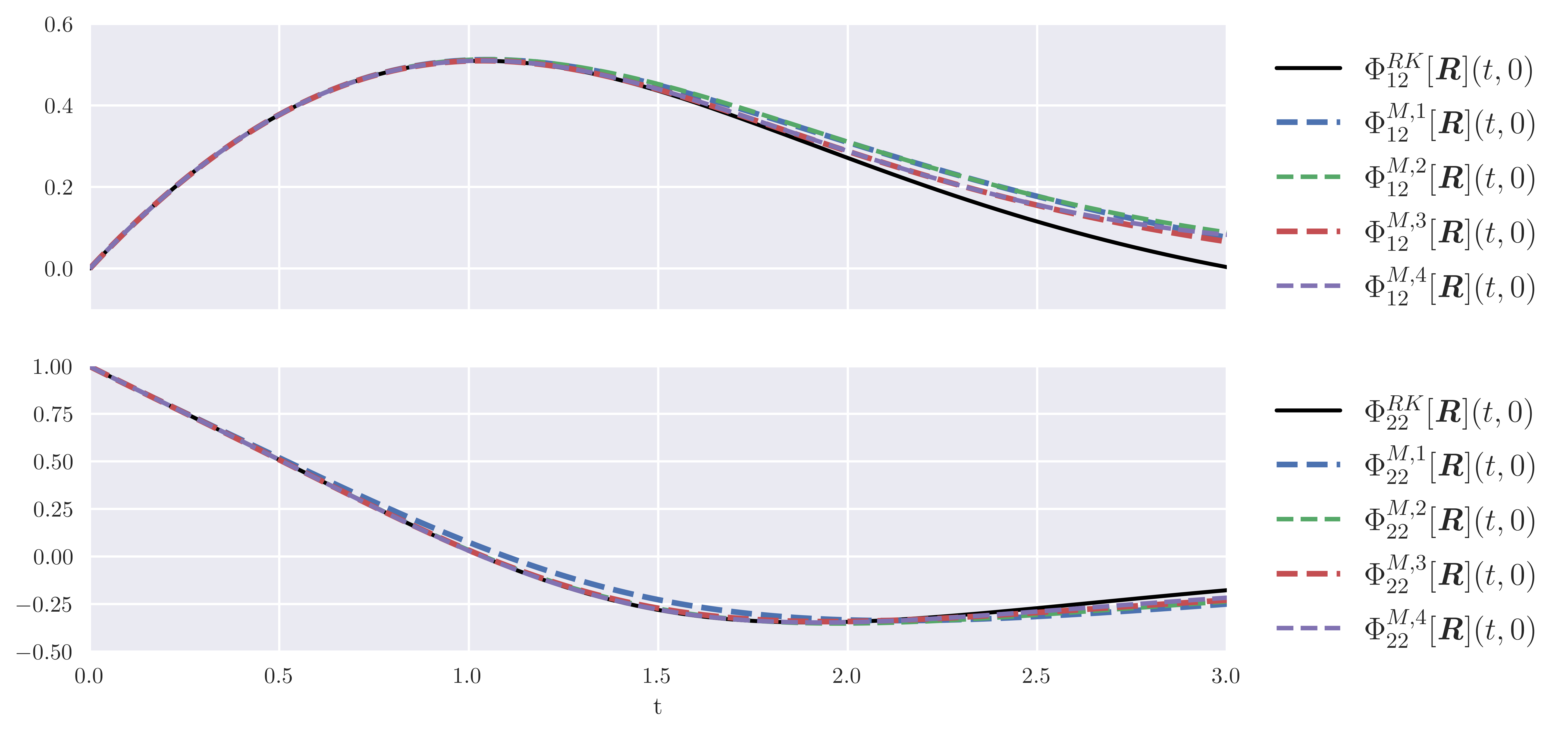}
        \caption{Comparison of Magnus approximation of $\Phi _{i2}[\mathbf{R}](t;s=0)$, up to the fourth order 
        ($\Phi _{i2}^{M,n}[\mathbf{R}](t;s=0)$, dashed lines) and corresponding RK solution ($\Phi _{i2}^{RK}[\mathbf{R}](t;s=0)$, 
        continuous lines).}
        \label{fig:10}
    \end{figure}

    We close this Appendix by constructing an example to compare the efficiency of the Peano-Baker and Magnus expansions.
    The example is motivated from \hyperref[sec6]{Section 6}, where we solve the novel genFPKE (\ref{eq:2.6}), corresponding to the Duffing oscillator
    (\ref{eq:6.2a}). In this case, the Magnus expansion is used for the determination of the components
    ${{\Phi }_{12}}$ and ${{\Phi }_{22}}$, of the state-transition matrix
    $\Phi [\mathbf{R}[{{f}_{\bfv{X}(\cdot )}}(\cdot ),\cdot ]](t;s)$, required for the computation of the diffusion
    coefficients (\ref{eq:6.6}), Sec. 6. The matrix $\mathbf{R}(t):=\mathbf{R}[{{f}_{\bfv{X}(\cdot )}}(\cdot ),t]$ has the form
    \begin{equation*}
        \mathbf{R}(t)=\left[ \begin{matrix}
                0                       & 1       \\
                1-3{{m}_{X_{1}^{2}}}(t) & -2\zeta \\
            \end{matrix} \right],
    \end{equation*}
    where the second moment ${{m}_{X_{1}^{2}}}(t)$ is unknown since it depends on the unknown density
    ${{f}_{\bfv{X}(t)}}(\bfv{x}).$ However, for a specific oscillator (i.e. oscillator (\ref{eq:6.2a}), with parameter values as
    in \hyperref[tab:2]{Table 2}), we can retrieve ${{m}_{X_{1}^{2}}}(t)$ from the data of the MC simulation, making the matrix
    $\mathbf{R}$ a priori known for a given time interval $[s,T]$. Then, the state-transition matrix
    $\Phi [\mathbf{R}](t;s)$ is computed by solving the IVPs
    \begin{equation*}
        {{\mathbf{\dot{y}}}_{i}}(t)=\mathbf{R}(t){{\mathbf{y}}_{i}}(t), \ \ \ \mathbf{y}(s)={{\mathbf{e}}_{i}}, \ \ \ i=1,2,
    \end{equation*}
    via a fourth order Runge-Kutta (RK) method. That is, we approximate $\Phi [\mathbf{R}](t;s)$ via 
    ${{\Phi }^{RK}}[\mathbf{R}](t;s)=({{\mathbf{y}}_{1}}(t),{{\mathbf{y}}_{2}}(t))$ (${{\mathbf{y}}_{1,2}}$ are column vectors), 
    which is considered as reference solution. In \hyperref[fig:9]{Figure 9} and \ref{fig:10}, we compare the RK 
    solution $\Phi _{i2}^{RK}[\mathbf{R}](t;s)$, $i=1,2$, with the Peano-Baker approximation $\Phi _{i2}^{P,n}[\mathbf{R}](t;s)$, 
    obtained via the truncated series (\ref{eq:B2}) with $n=1,2,3,4$ terms, and the Magnus approximation 
    $\Phi _{i2}^{M,n}[\mathbf{R}](t;s)$, obtained via the exponentiation of the truncated series (\ref{eq:B4}) with $n=1,2,3,4$ terms.

    The superiority of the Magnus approximation over the Peano approximation along the full range $[0,T=3]$ is clearly
    seen in the above example. This behavior has been observed in many other examples we have investigated numerically.

    \newpage
    \addcontentsline{toc}{section}{References}  
    \printbibliography

    \clearpage
    \phantomsection
\addcontentsline{toc}{section}{Supplementary Material}

\begin{refsection} 

\renewcommand{\theequation}{\arabic{equation}}
\renewcommand{\thefigure}{\arabic{figure}}  
\setcounter{figure}{0}
\setcounter{equation}{0}
\setcounter{table}{0}

\section*{\large \centering Supplementary Material \\ to \\ A systematic path to non-Markovian
dynamics II: Probabilistic response of nonlinear multidimensional
systems to Gaussian colored noise excitation}

\subsection*{Application to the underdamped linear oscillator and corresponding
numerical results}

In continuation of \hyperref[sec4.1]{Section 4.1} of the main paper, here we elaborate on equation (\ref{eq:4.4}) for the case of an underdamped
harmonic oscillator, driven by a colored Gaussian excitation. This case is both trivial and interesting. On the one hand,
its solution is known in closed form; on the other hand, it can be used as a benchmark problem to check the sufficiency of 
the Monte-Carlo simulations, as well as some features of the numerical solution scheme developed for solving the general 
equation (\ref{eq:2.6}); see \hyperref[sec2.2]{Section 2.2} and \hyperref[sec5.2]{5.2}. of the main paper. The corresponding dynamical equation reads
\begin{equation}     \phantomsection        \label{eqsmm:1}
\ddot{X}(t;\theta )+2\zeta{{\omega }_{0}}\dot{X}(t;\theta )+\omega_{0}^{2}
X(t;\theta )=\Xi(t;\theta ),
\end{equation} 
where $0<\zeta<1$ denotes the dimensionless damping coefficient and ${{\omega }_{0}}$ is the natural frequency of the
undamped oscillator. Introducing the state-vector $\bfv{X} = {{({{X}_{1}},{{X}_{2}})}^{\operatorname{T}}}={{(X,\dot{X})}^{\operatorname{T}}}$,
the oscillator is expressed in the form of the RDE system (\ref{eqs:2.1}) of the main paper, with $N=2$. The ${{h}_{n}}$ functions
are specified in the forms:
\addtocounter{equation}{1}
\begin{equation} \phantomsection \label{eqsmm:2.1a,b}
{{h}_{1}}(\bfv{X}(t;\theta ))={{X}_{2}}(t;\theta ),\ \ \ {{h}_{2}}(\bfv
{X}(t;\theta ))=-\omega_{0}^{2}{{X}_{1}}(t;\theta )-2\zeta{{\omega }_{0}}
{{X}_{2}}(t;\theta ). \tag{2.1a,b} 
\end{equation}
According to the above, the state-space representation of equation (\ref{eqsmm:1}), takes the form of the following system
of linear RDEs
\addtocounter{equation}{1}
\begin{equation} \phantomsection \label{eqs_sm:3}
\bfv{\dot{X}}(t;\theta )={\mathbf{J}}^{\bfv{h}} + \bfv{\Xi}(t;\theta ), \ \ \ \ \bfv{X}({{t}_{0}};\theta )=(X_{1}^{0}(\theta ),X_{2}^{0}(\theta
)),   \tag{3a,b}
\end{equation}
where the system matrix ${\mathbf{J}}^{\bfv{h}}$ has the form
\begin{equation*}
{{\mathbf{J}}^{\bfv{h}}}=\nabla ({{h}_{1}},{{h}_{2}})=\left[
    \begin{matrix}
    0               & 1                      \\
    -\omega_{0}^{2} & -2\zeta{{\omega }_{0}} \\
    \end{matrix}
    \right],
\end{equation*}
and the excitation vector reads as
$\bfv{\Xi} (t;\theta )={{(\Xi_{1},\Xi_{2})}^{\operatorname{T}}}={{(0,\Xi(t;\theta ))}^{\operatorname{T}}}$.
The excitation vector is not a full vector since the first equation of the system is unforced, $\Xi_{1}=0$.
Regarding the excitation $\Xi_{2}=\Xi$, we considering it as colored Gaussian of non-zero mean ${{m}_{\Xi }}(t)\ne 0$,
which for simplicity, it is further assumed uncorrelated to the initial value ${{\bfv{X}}^{0}}(\theta )$. Therefore, the
cross-covariance matrix $C_{\bfv{X^{0}}{\bfv{\Xi}}}$ and the autocovariances $C_{{{\Xi }_{n}}{{\Xi }_{1}}}$, for $n=1,2$,
vanish. This results in the zeroing of the coefficients $\mathcal{D}_{\nu n}^{{{X}_{0}}\mathbf{\Xi}}$, for $\nu,n =1,2$ and,
$\mathcal{D}_{\nu 1}^{\mathbf{\Xi} \mathbf{\Xi} }$, for $\nu =1,2$, of the general genFPKE (\ref{eq:4.4}). To specify the remaining 
diffusion coefficients, $\mathcal{D}_{\nu 2}^{\mathbf{\Xi} \mathbf{\Xi} }$, for $\nu =1,2$ (see equation (\ref{eq:4.5b}) of the main paper), 
we calculate the state-transition matrix $\Phi^{\mathbf{\Xi} }$, equation (\hyperref[eqs:4.6a,b]{4.6b}). Taking into account the system
matrix of the variational problems (\ref{eqs:3.14}), ${\mathbf{J}}^{\bfv{h}}$, we obtain (see e.g. \cite{Bernstein_1993}, \cite{Mamis_2018} Sec. 4,
or \cite{Perko_2001})
\phantomsection
\label{eqsmm:4}
\begin{align} 
    \phantomsection
    {{\Phi }^{\mathbf{\Xi} }}(t,s) & = {{e}^{\left( t-s \right){{\mathbf{J}}^{\bfv{h}}}}} \nonumber \\
    \phantomsection
    & = {{e}^{-a(t-s)}}\sin (\gamma (t-s))\left[
    \begin{matrix}
    \cot (\gamma (t-s))+a/\gamma & 1/\gamma \\
    \phantomsection
    -\omega_{0}^{2}/\gamma & \cot (\gamma (t-s))-a/\gamma \\
    \phantomsection
    \end{matrix}
    \right], \phantomsection 
\end{align}
where $a=\zeta{{\omega }_{0}}$ and $\gamma ={{\omega }_{0}}{{(1-{{\zeta }^{2}})}^{1\mathbf{/}2}}$.
Using equation (\hyperref[eqsmm:4]{4}), the diffusion coefficients (\ref{eq:4.5b}) of the main paper are completely determined for the case of an underdamped
oscillator, taking the form:
\begin{subequations} \phantomsection \label{eqs_sm:5}
    \begin{align}
        \mathcal{D}_{12}^{\mathbf{\Xi} \mathbf{\Xi} }(t) & = \int\limits_{{{t}_{0}}}^{t}{C_{\Xi_{2} \Xi_{2} }(t,s)\Phi _{12}^{\mathbf{\Xi} }(t,s)ds}=  \nonumber                                                              \\
                                                        & =\frac{1}{\gamma}\int\limits_{{{t}_{0}}}^{t}{C_{\Xi (\cdot )\Xi (\cdot)}^{{}}(t,s){{e}^{-a(t-s)}}\sin (\gamma (t-s))ds},                                           \\
        \mathcal{D}_{22}^{\mathbf{\Xi} \mathbf{\Xi} }(t) & = \int\limits_{{{t}_{0}}}^{t}{C_{\Xi_{2} \Xi_{2} }(t,s)\Phi _{22}^{\mathbf{\Xi} }(t,s)ds}=  \nonumber                                                              \\
                                                        & =\frac{1}{\gamma}\int\limits_{{{t}_{0}}}^{t}{C_{\Xi (\cdot )\Xi (\cdot)}^{{}}(t,s){{e}^{-a(t-s)}}\left( \gamma \cos(\gamma (t-s))-a\sin (\gamma (t-s)) \right)ds}.
    \end{align}
\end{subequations}
The genFPKE for this case reads as 
\begin{equation} \phantomsection \label{eqsmm:6}
\frac{\partial {{f}_{\bfv{X}(t)}}(\bfv{x})}{\partial t}+\sum\limits_{n=1}^{N}{\frac{\partial }{\partial {{x}_{n}}}\left[ \left( {{h}_{n}}(\bfv{x})+{{m}_{{{\Xi }_{n}}}}(t) \right){{f}_{\bfv{X}(t)}}(\bfv{x}) \right]}
= \sum\limits_{\nu =1}^{2}{\mathcal{D}_{\nu 2}^{\mathbf{\Xi} \mathbf{\Xi} }(t){{\partial }_{{{x}_{2}}{{x}_{\nu }}}}{{f}_{\bfv{X}(t)}}(\bfv{x})}.
\end{equation}
This equation is linear and can be rewritten in the form
$\partial{{f}_{\bfv{X}(t)}}(\bfv{x})/\partial t+L[{{f}_{\bfv{X}(t)}}(\bfv{x}
)]=0$,
where the second order, linear operator $L[\boldsymbol{\cdot}]$ is given by 
\begin{align} \phantomsection \label{eqsmm:7}
L[\boldsymbol{\cdot}]=-\mathcal{D}_{12}^{\mathbf{\Xi} \mathbf{\Xi} }(t){{\partial }_{{{x}_{2}}{{x}_{1}}}}
\boldsymbol{\cdot} & -\mathcal{D}_{22}^{\mathbf{\Xi} \mathbf{\Xi} }(t){{\partial }_{{{x}_{2}}{{x}_{2}}}}
\boldsymbol{\cdot}+{{x}_{2}}{{\partial }_{{{x}_{1}}}}\boldsymbol{\cdot}+\nonumber                        \\
                    & +
\left({{m}_{\Xi}}(t)-\omega_{0}^{2}{{x}_{1}}-2\zeta{{\omega }_{0}}
{{x}_{2}}\right){{\partial }_{{{x}_{2}}}}\boldsymbol{\cdot}-2\zeta{{\omega }_{0}}
\boldsymbol{\cdot}\, .
\end{align}
For comparison purposes, we note that, in the case of white noise excitation, equation (\ref{eqsmm:6}) becomes the standard
FPKE, and the operator $L[\cdot ]$ takes the form
\begin{equation*}
{{L}_{\text{FPK}}}[\boldsymbol{\cdot}]=-D(t){{\partial }_{{{x}_{2}}{{x}_{2}}}}
\boldsymbol{\cdot}+{{x}_{2}}{{\partial }_{{{x}_{1}}}}\boldsymbol{\cdot}+
\left(-\omega_{0}^{2}{{x}_{1}}-2\zeta{{\omega }_{0}}{{x}_{2}}\right){{\partial }_{{{x}_{2}}}}
\boldsymbol{\cdot}-2\zeta{{\omega }_{0}}\boldsymbol{\cdot},
\end{equation*}
where $D(t)$ is the noise intensity of the excitation
$\Xi(t;\theta )$.
\begin{remark}
The operator $L[\boldsymbol{\cdot}]$ is not a standard parabolic operator, in the sense that it is asymmetric with
respect to the second-order derivatives (the pure diffusive part). More precisely, one of the second and one of the
mixed partial derivatives are lacking in the right-hand side of equation (\ref{eqsmm:7}).
\end{remark}

The analytic solution of the genFPK (\ref{eqsmm:6}) is easily found; see e.g. \cite{Mamis_phd}. It is the bivariate Gaussian density with mean value vector
\begin{equation*}
{{\bfv{m}}_{\bfv{X}}}(t)={{\Phi
    }^{{{\bfv{X}}^{0}}}}(t,{{t}_{0}}){{\bfv{m}}_{{{\bfv{X}}^{0}}}}+\int_{{{t}_{0}}}^{t}{{{\Phi
        }^{\mathbf{\Xi} }}(t,s){{\bfv{m}}_{\mathbf{\Xi} }}(s)}ds,
\end{equation*}
and autocovariance matrix
\begin{align*}
{{\bfv{C}}_{\bfv{X}\bfv{X}}}(t) & = {{\mathbb{E}}^{\theta
    }}\left[ \left( \bfv{X}(t,\theta
    ){{\bfv{X}}^{\operatorname{T}}}(t,\theta ) \right)
\right]-{{m}_{\bfv{X}}}(t)m_{\bfv{X}}^{\operatorname{T}}(t) =                                                                  \\
                                & ={{\Phi}^{{{\bfv{X}}^{0}}}}(t,{{t}_{0}}){{\bfv{C}}_{{{\bfv{X}}^{0}}{{\bfv{X}}^{0}}}}{{\left(
{{\Phi }^{{{\bfv{X}}^{0}}}}(t,{{t}_{0}})
\right)}^{\operatorname{T}}}+                                                                                                  \\
                                & \quad \quad \quad + \int_{{{t}_{0}}}^{t}{\int_{{{t}_{0}}}^{t}{{{\Phi
        }^{\mathbf{\Xi} }}(t,{{s}_{1}}){{\bfv{C}}_{\mathbf{\Xi\Xi}
        }}({{s}_{1}},s){{\left( {{\Phi }^{\mathbf{\Xi} }}(t,s)
        \right)}^{\operatorname{T}}}d{{s}_{1}}}ds}.
\end{align*}
The matrix ${{\Phi }^{\mathbf{\Xi} }}$ is given by equation (\hyperref[eqsmm:4]{4}), and the matrix ${{\Phi }^{{{\bfv{X}}^{0}}}}$ 
results by substituting $s={{t}_{0}}$ in the latter.

\begin{figure} 
\centering
\includegraphics[width=0.8\textwidth]{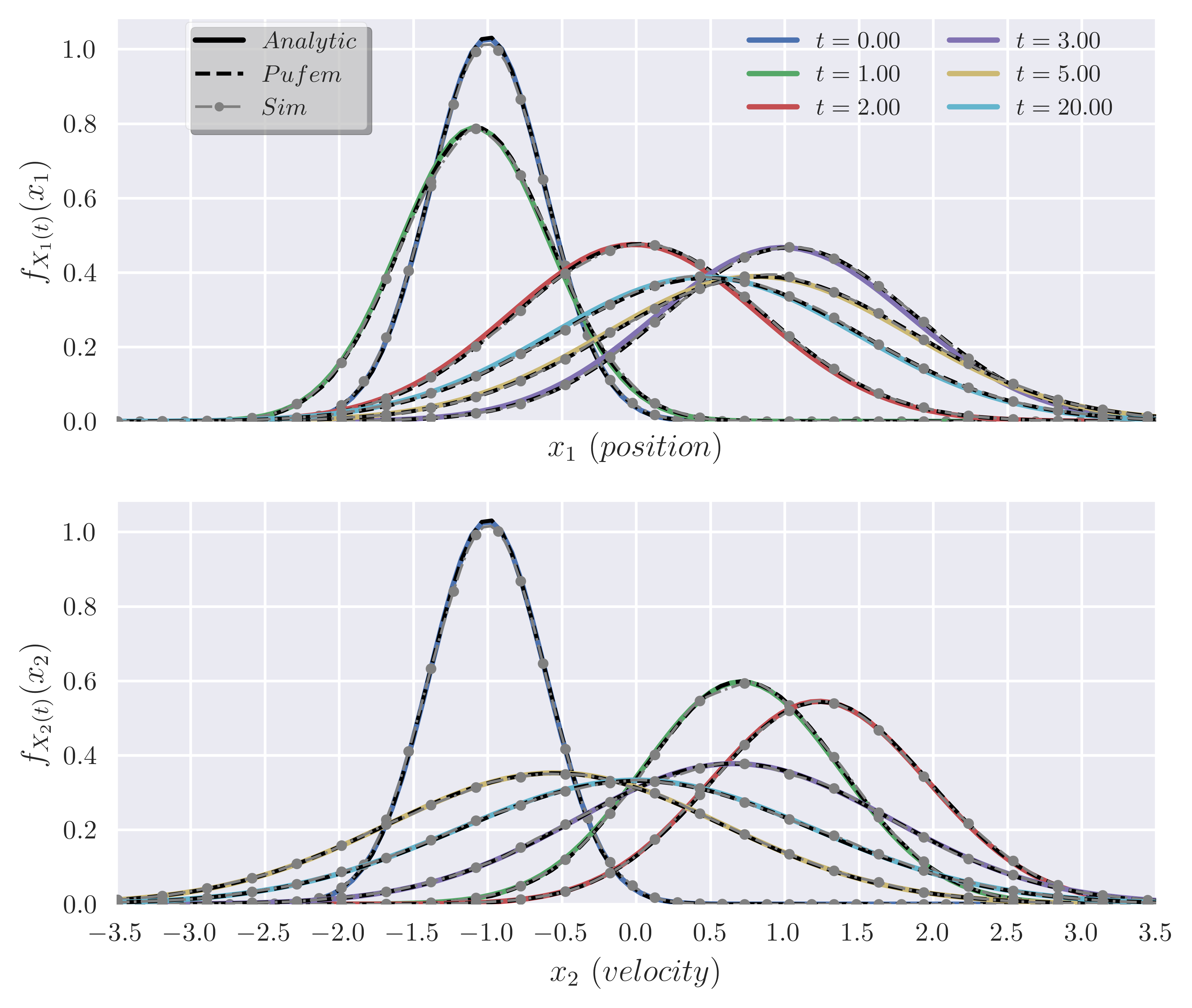}
\caption{Evolution of response marginal pdfs for the oscillator (\ref{eqsmm:1}) configured as described in \hyperref[tab_sm:1]{Table 1}. Analytic
    marginal pdfs (continuous colored lines) are compared to corresponding ones obtained via the PUFEM solution of the genFPKE
    (dashed black lines) and MC simulation (marked grey lines), at different times. }
\label{fig_sm:1}
\end{figure}

\begin{figure} 
\centering
\includegraphics[width=0.8\textwidth]{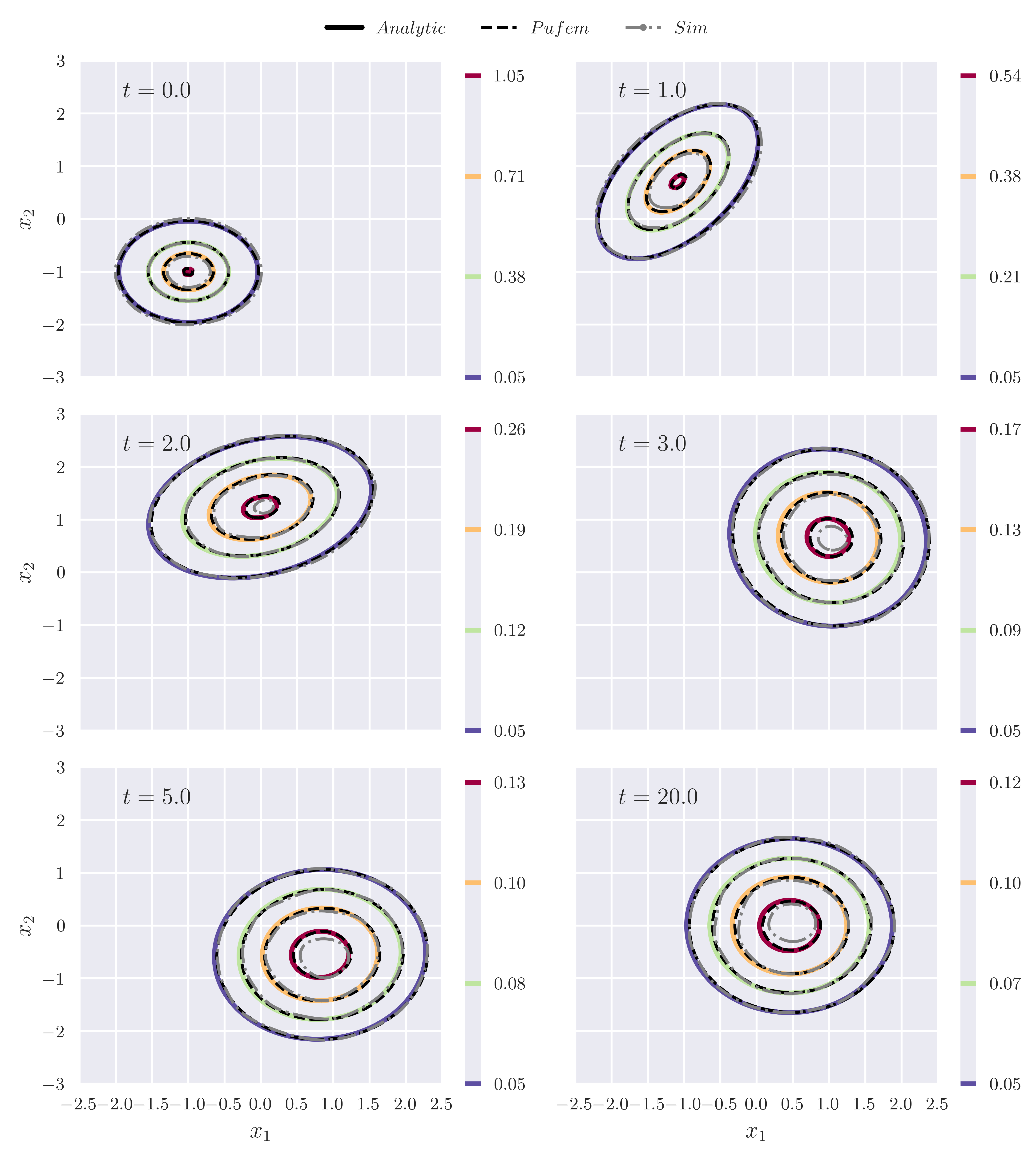}
\caption{Evolution contours of 2D response pdf for the for the oscillator (\ref{eqsmm:1}) configured as described in \hyperref[tab_sm:1]{Table 1}.
    Contour lines of the analytic pdf (continuous colored lines) are compared to corresponding PUFEM approximations
    (dashed black lines) and MC simulation (dashed dotted grey lines), at different time instances. }
\label{fig_sm:2}
\end{figure}

The excitation of the linear oscillator (\ref{eqsmm:1}) is the same as the excitation used in the presentation of the Duffing 
oscillator in \hyperref[sec6]{Section 6} of the main paper; see equations (\hyperref[eq:6.3b,c]{6.3}) and (\ref{eq:6.4}). The numerical results to be presented subsequently concern 
the oscillator (\ref{eqsmm:1}) with parameter values given in \hyperref[tab_sm:1]{Table 1}. In all figures presented below, the 
largest time for which results are shown is chosen to be well-within the long-time stationary regime, in order to check the 
validity of genFPKE in both the transient and the stationary regimes. The presented MC simulations are constructed from a 
sample of ${{10}^{5}}$ numerical experiments. In \hyperref[fig_sm:1]{Figure 1}, the marginal pdfs of the response are demonstrated 
in different time instances. In \hyperref[fig_sm:2]{Figure 2}, the 2D response pdf is demonstrated at different times. In 
both figures, PUFEM approximations of the densities are favorably compared with analytic ones and MC simulations.

\begin{table}[h] \label{tab_sm:1}
\centering
\caption{Configuration of the problem under numerical investigation.}
\renewcommand{\arraystretch}{1.5}
\begin{tabular}{c c c c c c}
    \multicolumn{6}{c}{\textbf{Oscillator Parameters}}                                                                                                                \\
    \toprule
    Parameters & \multicolumn{2}{c}{$\zeta$}                   & \multicolumn{3}{c}{$\omega_{0}$}                                                                     \\
    Values     & \multicolumn{2}{c}{0.25}                      & \multicolumn{3}{c}{$1{{\sec }^{-1}}$}                                                                \\
    \midrule
    \multicolumn{6}{c}{\textbf{Initial Value}}                                                                                                                        \\
    \toprule
    Parameters & \multicolumn{2}{c}{$\bfv{m}_{\bfv{X^0}}$}     & \multicolumn{3}{c}{$\bfv{C}_{\bfv{X^0 X^0}}$}                                                              \\
    Values     & \multicolumn{2}{c}{$(-1m,-1m{{\sec }^{-1}})$} & \multicolumn{3}{c}{$0.15\bfv{I}{{m}^{2}}{{\sec }^{-2}}$}                                             \\
    \midrule
    \multicolumn{6}{c}{\textbf{Excitation Parameters}}                                                                                                                \\
    \toprule
    Parameters & ${{m}_{\Xi }}$                                & $\sigma _{\Xi }^{2}$                                     & ${{\omega }_{\Xi }}$ & ${{\tau }_{corr}}$ \\
    Values     & $0.5m{{\sec }^{-1}}$                          & $1m{{\sec }^{-2}}$                                       & $1.5{{\sec }^{-1}}$  & $2\sec $           \\
    \bottomrule
\end{tabular}
\end{table}

The results suggest that the PUFEM approximation, as well as the MC simulation are in good agreement with the analytical
solution of the linear genFPKE both in the transient and long-steady-state. Further, considering a case with constant 
non-zero mean excitation, we find that the long time, $t>15$, mean value of the position ${{m}_{{{\mathbf{X}}_{1}}}}(t)=
\mathbb{E}_{{}}^{\theta }\left[{{X}_{1}}(t;\theta ) \right]$, becomes equal to ${{m}_{\Xi }}/\omega^{2}_{0}$, as expected.

In \hyperref[fig:3]{Figure 3} (\href{https://www.dropbox.com/scl/fo/qmp7wzx0tdcpvvqhf5j8o/AHuDTbDD8i3NHZ-yfq2mGu0?rlkey=i6o38dgci3b0i5tjby03s9sbc&dl=0}{Multimedia available online})
the evolution of the probabilistic solution, from the initial time up to the long-time stationary state, is presented. 
PUFEM approximations of the response 1D and 2D pdfs are nicely compared with corresponding analytic densities.
\clearpage
\begin{samepage}
\noindent  \textbf{Conclusion.} Numerical results presented herein show that, in the simple case of a linear damped oscillator, the MC 
simulations and the numerical scheme for solving the genFPKE are able to grasp all features of the corresponding 
analytical solution. Note that, the pdf-evolution equation for nonlinear RDEs contains additional terms, which disappear 
in the linear case.     
\end{samepage}

\begin{figure} 
\centering
\includegraphics[width=0.8\textwidth]{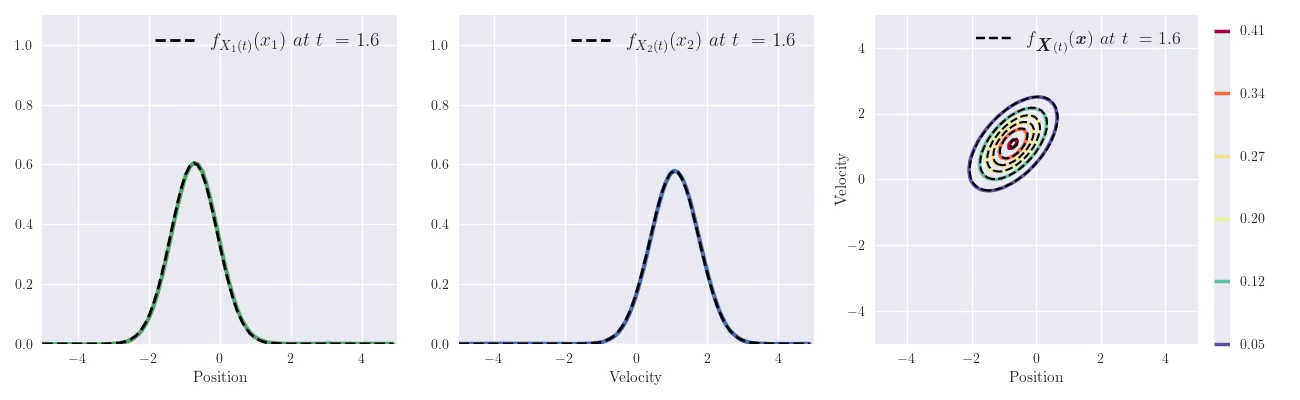}
\caption{Response pdfs of the linear RDE (\ref{eqsmm:1}), configured as described in \hyperref[tab_sm:1]{Table 1}, at an intermediate time. In left and middle
    panel, analytic marginal pdfs (continuous colored lines) are compared to PUFEM approximations (dashed black lines).
    In the right panel, the 2D pdf is demonstrated. Contour lines of the analytic pdf (continuous colored lines) are
    compared to corresponding PUFEM ap-proximations (dashed black lines). A video of the complete evolution from the initial state, up to the steady state, 
    is \href{https://www.dropbox.com/scl/fo/qmp7wzx0tdcpvvqhf5j8o/AHuDTbDD8i3NHZ-yfq2mGu0?rlkey=i6o38dgci3b0i5tjby03s9sbc&dl=0}{available online}}.
\label{fig_sm:3}
\end{figure}

\printbibliography[heading=subbibliography] 

\end{refsection}

\end{document}